\documentclass[preprint,12pt,authoryear]{elsarticle}
\usepackage[T1]{fontenc}
\usepackage{lmodern}
\input{glyphtounicode}\pdfgentounicode=1
\usepackage{amsmath,amssymb,amsthm,mathtools,bm}
\usepackage{graphicx,booktabs,multirow,array}
\usepackage{algorithm,algpseudocode}
\usepackage[hidelinks]{hyperref}
\usepackage[capitalise,nameinlink]{cleveref}
\crefname{appendix}{}{}
\Crefname{appendix}{}{}
\usepackage{xcolor}
\usepackage[section]{placeins}

\usepackage[margin=2.5cm]{geometry}
\newtheorem{lemma}{Lemma}
\newtheorem{proposition}{Proposition}
\theoremstyle{remark}
\newtheorem{remark}{Remark}
\journal{Computer Methods in Applied Mechanics and Engineering}

\newcommand{\OedGradErr}{$4.1\cdot10^{-8}$}
\newcommand{\OedRed}{2}
\newcommand{\OedRedPts}{10}
\newcommand{\OedRhoOpt}{0.16, 1.61, 25.00}
\newcommand{\OedStdBase}{0.0085}
\newcommand{\OedStdOpt}{0.0084}
\newcommand{\OedStdPts}{0.0094}

\newcommand{\OedTGradErr}{$6.5\cdot10^{-8}$}
\newcommand{\OedTRed}{15}
\newcommand{\OedTRedPts}{54}

\newcommand{\OedTStdBase}{0.0172}
\newcommand{\OedTStdOpt}{0.0145}
\newcommand{\OedTStdPts}{0.0313}
\newcommand{\OedTWopt}{0.80, 0.80, 0.80}
\newcommand{\OedWopt}{0.80, 2.54, 10.00}

\newcommand{\PMIaRmse}{0.080}

\newcommand{\PMIaSref}{1.79}
\newcommand{\PMIaSsel}{3}

\newcommand{\PMIaahplus}{0.89}

\newcommand{\PMIad}{1.00}
\newcommand{\PMIak}{1.35}

\newcommand{\PMIbResD}{11.0}

\newcommand{\PMIbRmse}{0.054}

\newcommand{\PMIbSref}{2.75}
\newcommand{\PMIbSsel}{7}
\newcommand{\PMIbUnconvCV}{0.044}
\newcommand{\PMIbUnconvNC}{45.1\,\%}
\newcommand{\PMIbUnconvRE}{3.9\,\%}
\newcommand{\PMIbahminus}{0.91}
\newcommand{\PMIbahplus}{0.41}
\newcommand{\PMIbahplusHi}{0.43}
\newcommand{\PMIbahplusLo}{0.41}
\newcommand{\PMIbbD}{0.77}
\newcommand{\PMIbbZ}{0.55}
\newcommand{\PMIbd}{0.98}
\newcommand{\PMIbk}{1.42}

\newcommand{\StrFineXJErrT}{0.012}
\newcommand{\StrFineXN}{4}

\newcommand{\StrFineXNErrT}{0.113}

\newcommand{\StrFineXPErrT}{0.012}
\newcommand{\StrIoneMax}{1.09}

\newcommand{\StrJAhRMSE}{0.003}

\newcommand{\StrJErrT}{0.012}

\newcommand{\StrNAhRMSE}{0.014}

\newcommand{\StrNErrT}{0.115}
\newcommand{\StrNseed}{8}

\newcommand{\StrPAhRMSE}{0.006}

\newcommand{\StrPErrT}{0.010}
\newcommand{\StrTriaxMax}{0.65}
\newcommand{\StrTriaxMin}{-0.05}
\newcommand{\StrTruthAh}{0.43}
\newcommand{\StrTruthRmse}{0.066}

\newcommand{\StrTubeJAhRMSE}{0.014}

\newcommand{\StrTubeJErrT}{0.017}
\newcommand{\StrTubeN}{6}

\newcommand{\StrTubeNAhRMSE}{0.051}

\newcommand{\StrTubeNErrT}{0.263}

\newcommand{\StrTubePAhRMSE}{0.031}

\newcommand{\StrTubePErrT}{0.024}
\newcommand{\SynAAdm}{75.0\,\%}

\newcommand{\SynCAdm}{100.0\,\%}

\newcommand{\SynN}{24}
\newcommand{\SynUAdm}{45.8\,\%}

\newcommand{\VPMIaMonoCV}{0.081}

\newcommand{\VPMIaChebCV}{0.072}

\newcommand{\VPMIaPbpCV}{0.070}

\newcommand{\VPMIaPbpd}{0.96}

\newcommand{\VPMIaPbpbeta}{0.74}
\newcommand{\VPMIaPbpgamma}{0.451}

\newcommand{\VPMIbMonoCV}{0.060}

\newcommand{\VPMIbMonoahpLo}{0.40}
\newcommand{\VPMIbMonoahpHi}{0.43}

\newcommand{\VPMIbExtRmse}{0.051}
\newcommand{\VPMIbExtCV}{0.058}

\newcommand{\VPMIbSmaxCV}{0.061}

\newcommand{\VPMIbSmaxahp}{0.37}

\newcommand{\VPMIbSmaxp}{1.6}
\newcommand{\VPMIbSmaxahpLo}{0.37}
\newcommand{\VPMIbSmaxahpHi}{0.39}

\newcommand{\VPMIbGateCV}{0.046}

\newcommand{\VPMIbGateNC}{39.0}

\newcommand{\VPMIbGatePenCV}{1.160}

\newcommand{\VPMIbGatePenNC}{3.8}

\newcommand{\VPAsixMonoRmse}{0.045}
\newcommand{\VPAsixMonoCV}{0.047}

\newcommand{\VPAsixPbpRmse}{0.036}
\newcommand{\VPAsixPbpCV}{0.043}

\newcommand{\VPAsixPbpd}{1.09}
\newcommand{\VPAsixPbpk}{1.00}

\newcommand{\VPAsixPbpbeta}{0.72}
\newcommand{\VPAsixPbpgamma}{0.734}

\newcommand{\VPAsixSmaxCV}{0.044}

\newcommand{\VPAsixSmaxbZ}{1.30}
\newcommand{\VPAsixSmaxbD}{0.64}
\newcommand{\VPAsixSmaxahp}{1.36}

\newcommand{\VPAsixGateCV}{0.059}

\newcommand{\VPAsixGateNC}{71.1}

\newcommand{\VPAsixGatePenCV}{0.188}

\newcommand{\VConcMonoRmse}{0.137}
\newcommand{\VConcMonoCV}{0.194}

\newcommand{\VConcMonod}{7.45}

\newcommand{\VConcPbpRmse}{0.033}
\newcommand{\VConcPbpCV}{0.044}

\newcommand{\VConcPbpd}{10.12}
\newcommand{\VConcPbpk}{1.62}
\newcommand{\VConcPbpbZ}{0.99}
\newcommand{\VConcPbpbD}{10.91}
\newcommand{\VConcPbpahp}{1.10}

\newcommand{\VConcExtahpLo}{1.09}
\newcommand{\VConcExtahpHi}{1.13}
\newcommand{\VConcSmaxRmse}{0.038}
\newcommand{\VConcSmaxCV}{0.062}

\newcommand{\VConcSmaxahp}{0.97}

\newcommand{\VConcSmaxahpLo}{0.82}
\newcommand{\VConcSmaxahpHi}{1.00}

\newcommand{\VConcGateCV}{0.874}

\newcommand{\VConcGateNC}{64.9}

\newcommand{\VConcGatePenCV}{0.052}

\newcommand{\VTSSAMonoRmse}{0.093}
\newcommand{\VTSSAMonoCV}{0.094}

\newcommand{\VTSSAMonoNC}{1.0}

\newcommand{\VTSSAPbpRmse}{0.081}
\newcommand{\VTSSAPbpCV}{0.086}

\newcommand{\VTSSAPbpeET}{1.87}
\newcommand{\VTSSAPbpePS}{1.15}

\newcommand{\VTSSAPbpbeta}{0.01}
\newcommand{\VTSSAPbpgamma}{0.995}

\newcommand{\VTSSASmaxCV}{0.089}

\newcommand{\VTSSAGateRmse}{0.079}

\newcommand{\VTSSAGateNC}{36.7}

\newcommand{\TSSAWratio}{3.2}
\newcommand{\TSSAWratioUT}{2.6}
\newcommand{\TSSAlmaxMin}{2.57}
\newcommand{\TSSAlmaxMax}{2.85}

\newcommand{\TrelRelRms}{4.2}
\newcommand{\TrelAlphas}{-1, 2, 7}
\newcommand{\TSSAhypRelRms}{16.2}
\newcommand{\TSSAhypAlphas}{-6, 2, 3}

\newcommand{\SynTThreeMonoEn}{0.003}

\newcommand{\SynTThreeExtEn}{0.003}

\newcommand{\SynTFourMonoEn}{0.066}

\newcommand{\SynTFourExtEn}{0.005}

\newcommand{\SynTFourSmaxEn}{0.026}

\newcommand{\SynTFiveMonoEn}{0.053}

\newcommand{\SynTFiveExtEn}{0.003}

\newcommand{\SynTFiveSmaxEn}{0.007}

\newcommand{\SynTSixSmaxEn}{0.237}

\newcommand{\SpreadMeas}{4}
\newcommand{\SpreadExtr}{12}

\begin{document}
\begin{frontmatter}
\title{Sparse, admissible and interpretable: discovery of isotropic limit surfaces with extended invariant libraries, physically gated mixtures of experts, and end-to-end structural identification}

\author[tud]{Michael A. Kraus\corref{cor}}
\ead{kraus@ismd.tu-darmstadt.de}
\cortext[cor]{Corresponding author}
\affiliation[tud]{organization={Institute of Structural Mechanics and Design (ISM+D), Technical University of Darmstadt},
            city={Darmstadt}, country={Germany}}

\begin{abstract}
Choosing and fitting a limit surface, a yield or failure criterion, to a few dozen multiaxial strength tests is a model-selection problem with physical side conditions. The equivalent stress concept, as systematised by Kolupaev, provides a large catalogue of criteria together with admissibility conditions: convexity, a unique definition of the surface without spurious outer contours, explicit solvability for the equivalent stress, and bounds on inelastic Poisson ratios. We recast the task as sparse, constrained and differentiable model discovery and compare three representations. The first is a library of invariant monomials $I_1^aq^mr_3^c$, which contains the classical criteria as sparse members. The second extends this library by trigonometric Lode-angle terms, which isotropy restricts to cosine modes, i.e.\ rational invariants with negative powers of $q$, and by Podg\'orski--Bigoni--Piccolroaz atoms with learnable shape parameters. The third is a mixture of convex gauge experts, either with a competition (smooth-maximum) gate or with a conventional direction gate, with hard-concrete and Gumbel-softmax sparsity. A ray-wise first-root definition of the surface gives explicit solvability and an exactly differentiable objective. Convexity and the absence of re-entrant branches are differentiable penalties for the libraries and hold by construction for the competition gate. That gate acts as a physically motivated regime selector between mechanisms. Under stated assumptions its responsibilities are, in addition, the share of failure probability in competing-risk Weibull statistics, with the exponent as Weibull modulus, the share of dissipation in Koiter's multi-surface flow rule, and a Boltzmann weight in the Gibbs variational principle. A direction gate has none of these meanings. In a synthetic benchmark the extended library recovers out-of-library truths such as Mohr--Coulomb and PBP surfaces with errors below one per cent, whereas direction-gated mixtures are non-convex on a third to two thirds of all directions. We analyse five experimental data sets: two PMI foams, polyamide PA6, concrete, and the structural silicone TSSA in Hencky-strain space, where uniaxial, planar and equibiaxial failure lie on the three characteristic meridians. A sharp Lode function matters more than many terms. With PBP atoms the concrete data, which the monomial library could not resolve, are fitted with a fourfold lower cross-validated error, and the measured strength ratios are reproduced within a few per cent. For TSSA the selected surface is a maximum-principal-stretch criterion with a cavitation term. Stored-energy contours, from Ogden models of the TSSA curves and of Treloar's natural rubber, contradict a critical-energy reading. The competition gate separates tensile cell-wall fracture from cell collapse in the foams. The fitted exponents, however, do not match the Weibull moduli of the replicate scatter, so the quantitative statistical reading is a hypothesis that these data do not support. Values that the tests do not constrain, above all the hydrostatic tensile strength, often differ between admissible model classes by several bootstrap widths. Model-form uncertainty therefore has to be reported. End-to-end identification from notched-bar failure loads through a differentiable finite element model removes the bias of nominal-stress mapping, and gradient-based notch design reduces the predicted uncertainty of the hydrostatic tensile strength.

\end{abstract}

\end{frontmatter}

\section{Introduction}
\label{sec:intro}
Limit surfaces, the yield or failure surfaces in stress space, are among the most widely used constitutive ingredients in structural analysis. They underlie stress-based design checks of polymers, foams, adhesives, concrete and rock, they serve as yield functions in plasticity, and they are the basis of the equivalent stress concept that reduces multiaxial states to a scalar comparison with a uniaxial strength \citep{Yu2002,Altenbach2014,Kolupaev2018}. More than a century of work has produced a large catalogue of criteria: von Mises, Tresca, Drucker--Prager, Mohr--Coulomb, the Burzy\'nski--Yagn family, the unified strength theory of Yu, the criteria of Podg\'orski and Bigoni--Piccolroaz, and many generalisations \citep{Podgorski1985,Bigoni2004,Yu2002}. The monograph of \citet{Kolupaev2018} systematises this catalogue in the invariants $(I_1, I_2', I_3')$. It formulates necessary conditions and plausibility assumptions for admissible criteria (convexity, a unique definition of the surface without additional outer contours, bounds on inelastic Poisson ratios, a minimal number of parameters), discusses fitting strategies and multiaxial test programmes, and applies them to metals, polymers, foams and concrete.

Two observations from that body of work motivate the present paper. First, the choice of the criterion is in practice a \emph{model selection} problem: Kolupaev compares dozens of candidate forms for a given data set and states that there is no ``unique'' fitting, because the result depends on the weighting of the data, the chosen criterion and its number of degrees of freedom, the objective function, the constraints, and the final selection among optimal fits \citep[Sect.~12.5]{Kolupaev2018}. Second, the admissibility conditions are checked largely by hand: spurious branches of high-power criteria ``can not be avoided by introduction of new constraints'', results ``must be controlled visually within each step'', and physically meaningful objectives such as the distance along the ray from the origin are considered too expensive for parameter identification \citep[Sect.~12.2, 12.5]{Kolupaev2018}.

Data-driven constitutive modelling offers tools for exactly this kind of problem \citep{Fuhg2024review}. Neural-network yield functions have been trained with Sobolev losses \citep{Vlassis2021} or with input-convex architectures that guarantee convexity by construction \citep{Fuhg2022yield}. Constitutive artificial neural networks \citep{Linka2023} and sparse regression in the spirit of SINDy \citep{Brunton2016,Rudy2017,Kaheman2020} produce interpretable models from function libraries. The EUCLID framework discovers hyperelastic, plastic and generalised standard material models from full-field displacements and reaction forces without stress labels \citep{Flaschel2021,Flaschel2022,Flaschel2023}. Most recently it has been extended to non-associated pressure-sensitive plasticity \citep{Xu2025}. That work represents the yield surface by a Fourier series in the Lode angle combined with a linear pressure term and an exponential cap, promotes sparsity by an $\ell_1$-type penalty, and enforces convexity of the deviatoric section at a finite set of Lode angles. It is demonstrated on synthetic full-field data for perfectly plastic materials. Our Chebyshev library is closely related to this Fourier representation. Neither EUCLID nor, to our knowledge, any other discovery framework treats the second admissibility condition emphasised by Kolupaev, the unique definition of the surface without spurious outer contours. Without it, sparse polynomial criteria routinely acquire such contours, as Section~\ref{sec:results} shows.

Two further questions arise once a library is in use. The first concerns its content. Polynomial invariant libraries tie the resolution in the Lode angle to the stress degree. Sharp deviatoric shapes, from the Tresca hexagon to the Rankine triangle, therefore cannot be represented sparsely. Kolupaev's catalogue and the elastomer literature, by contrast, contain compact non-polynomial forms such as the Podg\'orski--Bigoni--Piccolroaz (PBP) Lode function \citep{Podgorski1985,Bigoni2004,Rosendahl2019}. Which non-polynomial terms are admissible and useful, and are trigonometric terms in the Lode angle among them? The second question is whether a library is needed at all. Mixtures of experts \citep{Jacobs1991,Shazeer2017}, with a learnable gate that blends a few simple criteria, are the standard machine-learning answer to regime-dependent behaviour and have been used in physics-informed learning \citep{Bischof2022moe}. Multi-surface criteria \citep[Ch.~11]{Kolupaev2018} and the coupled distortional--dilatational criterion for silicone adhesives \citep{Drass2020coupled} are mixtures with a hard gate. Whether such gates can be interpreted physically, rather than only fitted, has to our knowledge not been asked.

Strength data for the materials treated by \citet{Kolupaev2018}, such as structural foams, polymers, concrete and adhesives, look different. They consist of a few dozen failure stress states from tubular, cube, disc and cruciform specimens, often with large scatter and with some regions of stress space (hydrostatic tension) practically inaccessible. They are typically reported as tables of principal stresses at failure. For such data, the relevant questions are: which compact, admissible surface does the data support; how uncertain are extrapolated values such as the hydrostatic tensile strength; and which additional, preferably simple, tests would reduce that uncertainty most?

This paper answers these questions with a framework that merges Kolupaev's equivalent stress concept with differentiable programming and sparse model discovery. The contributions are:
\begin{enumerate}
\item \emph{Invariant libraries} of increasing expressiveness: monomials $I_1^a q^m r_3^c$ with $q=\sqrt{3I_2'}$ and $r_3=q^3\cos3\theta=\tfrac{27}{2}I_3'$, which contain the classical criteria including Kolupaev's cosine ansatz as sparse members (Section~\ref{sec:library}); Lode-angle Fourier modes, which are shown to reduce to Chebyshev polynomials of $\cos3\theta$ and hence to rational invariants with negative powers of $q$, since isotropy excludes all sine modes (Proposition~\ref{prop:cos}); and parametric PBP atoms with learnable shape parameters (Section~\ref{sec:atoms}).
\item A \emph{ray-wise} definition of the surface by the first root along rays from the origin. It gives explicit solvability for $\sigma_\mathrm{eq}$ and homogeneity for every library member and turns Kolupaev's ray-based quality measure into a cheap, exactly differentiable identification objective. Implicit sensitivities follow from a single Newton correction (Lemma~\ref{lem:implicit}).
\item \emph{Differentiable admissibility penalties} for convexity (tangential curvature in principal-stress space, justified by Lewis' theorem), for the unique definition of the surface (a no-re-entrance condition, Proposition~\ref{prop:star}) and for inelastic Poisson ratios. Together they replace visual control by quantitative audits.
\item \emph{Mixture-of-experts baselines} with convex gauge experts: a competition gate (smooth maximum) that is convex and unique by construction, and a conventional direction gate with and without a convexity penalty, both with sparsity-enforcing relaxations (hard-concrete $L_0$, Gumbel-softmax) (Section~\ref{sec:moe}). We show that the competition gate has a threefold physical meaning: competing-risk fracture statistics with the exponent as Weibull modulus, Koiter's multi-surface flow rule, and the Gibbs variational principle. A direction gate has none of these meanings (Section~\ref{sec:gate_meaning}).
\item A \emph{selection algorithm} that combines exhaustive and beam-searched screening with constrained refits, leakage-free cross-validation and bootstrap uncertainty. It is applied to five experimental data sets: the PMI foams Rohacell 71IG and 110IG, polyamide PA6 and concrete from \citet{Kolupaev2018}, and the transparent structural silicone adhesive TSSA in Hencky-strain space from the test programme of \citet{Drass2020coupled}. For TSSA it is complemented by an energetic analysis with hyperelastic models identified from the TSSA curves and from Treloar's natural-rubber data \citep{Treloar1944} (Section~\ref{sec:hencky}).
\item \emph{End-to-end identification from structural failure loads} of notched round bars under superposed pressure, through a differentiable axisymmetric FE model with exact polynomial load-path roots (Proposition~\ref{prop:affine}). This is complemented by \emph{gradient-based optimal design} of the notch geometry to minimise the posterior variance of the hydrostatic tensile strength.
\item A \emph{cross-material analysis} of all examples. It separates what the data determine from what the model class decides, and it tests the gate interpretation against the replicate scatter (Section~\ref{sec:overall}).
\end{enumerate}
The paper is organised as follows. Section~\ref{sec:background} recalls invariants, characteristic values and Kolupaev's admissibility conditions. Section~\ref{sec:method} presents the method. Section~\ref{sec:results} reports a synthetic benchmark, the five experimental case studies, the structural identification, the design study and the analysis across all examples. Section~\ref{sec:disc} discusses limitations and Section~\ref{sec:concl} concludes. Code and transcribed data are provided as supplementary material.

\section{Equivalent stress concept and admissibility of limit surfaces}
\label{sec:background}
\subsection{Invariants and the Burzy\'nski plane}
Throughout, $\boldsymbol\sigma$ denotes the Cauchy stress, $\sigma_\mathrm{I}\ge\sigma_\mathrm{II}\ge\sigma_\mathrm{III}$ its principal values and $\mathbf s=\boldsymbol\sigma-\tfrac13 I_1\mathbf 1$ its deviator. Following \citet{Kolupaev2018} we use the axiatoric--deviatoric invariants
\begin{equation}
  I_1=\operatorname{tr}\boldsymbol\sigma,\qquad I_2'=\tfrac12\,\mathbf s:\mathbf s,\qquad I_3'=\det\mathbf s,\qquad
  \cos3\theta=\frac{3\sqrt3}{2}\,\frac{I_3'}{I_2'^{3/2}},
  \label{eq:invariants}
\end{equation}
where $\theta\in[0,\pi/3]$ is the stress angle, $\theta=0$ on the tensile meridian (uniaxial tension, $Z$) and $\theta=\pi/3$ on the compressive meridian (uniaxial compression, $D$). We abbreviate the von Mises measure and a cubic Lode invariant as
\begin{equation}
  q=\sqrt{3I_2'},\qquad r_3 = q^3\cos3\theta=\tfrac{27}{2}\,I_3'. \label{eq:q_r3}
\end{equation}
The identity $q^3\cos3\theta=\tfrac{27}{2}I_3'$ follows directly from \eqref{eq:invariants}; it is the reason why Lode-angle dependence can be expressed by \emph{polynomials} in the stress. Stresses are normalised by a reference strength $\sigma_\mathrm{ref}$ (here the mean uniaxial tensile strength, or the normalisation of the original data source), and all invariants below refer to $\bar{\boldsymbol\sigma}=\boldsymbol\sigma/\sigma_\mathrm{ref}$; the bar is dropped where no confusion can arise. The Burzy\'nski plane $(I_1/\sigma_+,\ \sqrt{3I_2'}/\sigma_+)$ shows meridian sections of a limit surface; the $\pi$-plane shows sections $I_1=\mathrm{const}$ \citep{Kolupaev2018}.

\subsection{Characteristic values}
A limit surface is conveniently summarised by the dimensionless values of \citet[Chap.~5]{Kolupaev2018}, all normalised by the model's own uniaxial tensile strength $\sigma_+$:
$d=\sigma_-/\sigma_+$ (uniaxial compression, point $D$), $k=\sqrt3\,\tau_*/\sigma_+$ (torsion, $K$), $b_Z$ and $b_D$ (equibiaxial tension/compression, $B_Z$, $B_D$), $i_Z$ and $u_D$ (plane stress with $\sigma_\mathrm{II}=\sigma_\mathrm I/2$, $I_Z$, $U_D$), $a^\mathrm{hyd}_\pm$ (hydrostatic tension/compression, $A_Z$, $A_D$, defined per principal stress), and the inelastic Poisson ratios
\begin{equation}
  \nu^\mathrm{in}_+ = -\left.\frac{\partial\Phi/\partial\sigma_\mathrm{II}}{\partial\Phi/\partial\sigma_\mathrm{I}}\right|_{Z},\qquad
  \nu^\mathrm{in}_- = -\left.\frac{\partial\Phi/\partial\sigma_\mathrm{II}}{\partial\Phi/\partial\sigma_\mathrm{I}}\right|_{D},
  \label{eq:nu_in}
\end{equation}
which follow from an associated flow rule at $Z$ and $D$. For pressure-insensitive surfaces $\nu^\mathrm{in}_\pm=1/2$. For brittle failure surfaces, which do not govern plastic flow, $\nu^\mathrm{in}_\pm$ should be read as a geometric measure of the surface's slope at $Z$ and $D$, as in \citet[Sect.~5.4]{Kolupaev2018}.

\subsection{Admissibility conditions}
\label{sec:kolupaev_conditions}
\citet[Sect.~8.2]{Kolupaev2018} lists necessary conditions and plausibility assumptions for a generalised criterion. Those relevant here are: (i) convexity of the limit surface (Drucker's postulate); (ii) a unique definition of the surface, i.e.\ no additional outer contours enclosing the physically meaningful one; (iii) explicit solvability for the equivalent stress $\sigma_\mathrm{eq}$; (iv) homogeneity in the stresses; (v) dependence on all three invariants, through rational functions of $I_1$ and $I_3'$; (vi) a maximal stress power $n\le 12$; (vii) a minimal number of dimensionless parameters; (viii) continuous differentiability, including the hydrostatic nodes; and (ix) bounds on the inelastic Poisson ratios, in particular $\nu^\mathrm{in}_+<1/2$ \citep[Eq.~(12.18)]{Kolupaev2018}. The same monograph states that ``there is no `unique' fitting of the surface to the measured data'' \citep[Sect.~12.5]{Kolupaev2018}, that higher powers produce branches outside the validity region which ``can not be avoided by introduction of new constraints'', and that optimisation results ``must be controlled visually within each step''. Table~\ref{tab:conditions} summarises how the present framework turns each condition into either a property of the representation or a differentiable penalty; the remainder of this section and Section~\ref{sec:method} give the details.

\begin{table}[!htbp]
\centering\small
\caption{Admissibility conditions of \citet[Sect.~8.2, 12.4]{Kolupaev2018} and their treatment in the present framework.}
\label{tab:conditions}
\begin{tabular}{p{0.24\linewidth}p{0.39\linewidth}p{0.27\linewidth}}
\toprule
Condition & Library models (Sects.~\ref{sec:library}--\ref{sec:atoms}) & Competition MoE (Sect.~\ref{sec:moe}) \\
\midrule
Convexity & Differentiable penalty on the tangential curvature, Eq.~\eqref{eq:Pconv}, on 240 directions in principal-stress space (Lewis' theorem); numerical audit on 2000 directions, no continuum certificate & By construction (Prop.~\ref{prop:expert}) \\
Unique definition, no outer contours & First-root convention plus no-re-entrance penalty, Eq.~\eqref{eq:Puniq}, within $t\le t_\mathrm{max}$; Proposition~\ref{prop:star}; numerical audit & By construction (gauge) \\
Explicit solvability for $\sigma_\mathrm{eq}$; homogeneity & By construction: $\sigma_\mathrm{eq}$ is the gauge of the star-shaped admissible set, Eq.~\eqref{eq:gauge} & By construction \\
All three invariants, rational in $I_1$, $I_3'$ & Monomial and Chebyshev terms are rational; PBP atoms are not & PBP Lode function (not rational) \\
Maximal power $n\le12$ & Library degree $n=6$ & Degree one \\
Continuous differentiability, incl.\ hydrostatic nodes & Not enforced; terms linear in $q$ (also $m=1$ Chebyshev and PBP terms) give a conical node (Remark~\ref{rem:axis}) & Not enforced; conical node if an active expert has $b_e=0$ \\
Minimal number of parameters & Sparse selection with cross-validation (Section~\ref{sec:selection}) & Number of experts by cross-validation; $L_0$ gates \\
Dimensionless parameters & Normalisation by $\sigma_\mathrm{ref}$ & same \\
Bounds on $\nu^\mathrm{in}_\pm$ & Differentiable penalty, Eq.~\eqref{eq:Pnu} & Not enforced (reported) \\
$\sigma_\mathrm{eq}=\sigma_+$ (optional) & Anchor penalty, Eq.~\eqref{eq:Panchor}; discussed in Section~\ref{sec:disc} & not used \\
\bottomrule
\end{tabular}
\end{table}

\section{Method}
\label{sec:method}
\begin{figure}[!htbp]
\centering
\includegraphics[width=\linewidth]{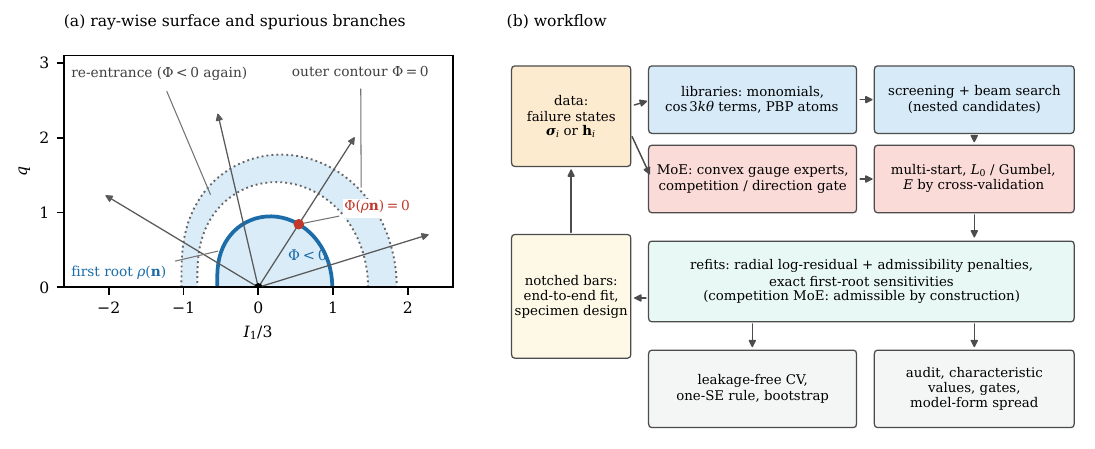}
\caption{(a) Ray-wise definition of the surface: along each ray from the origin, the first sign change of $\Phi$ from negative to positive defines the radius $\rho(\mathbf n)$ (thick). A polynomial $\Phi$ can become negative again further out (re-entrance) and possess outer contours of its zero set (dotted), which violate Kolupaev's unique-definition condition and are penalised. (b) Workflow of the identification.}
\label{fig:method}
\end{figure}

\Cref{fig:method} summarises the method. A candidate representation, either a sparse library model or a mixture of gauge experts, defines the surface by the first root along rays from the origin. Admissibility is imposed by differentiable penalties or holds by construction. Supports, shape parameters and the number of experts are selected by screening, constrained refits and leakage-free cross-validation.

\subsection{Invariant library}
\label{sec:library}
We represent the limit function as a sparse linear combination of invariant monomials,
\begin{equation}
  \Phi(\boldsymbol\sigma;\boldsymbol\xi)=\sum_{k=1}^{K}\xi_k\,m_k(\bar{\boldsymbol\sigma})-1,\qquad
  m_k = I_1^{a_k}\,q^{m_k}\,r_3^{c_k},\qquad 1\le n_k:=a_k+m_k+3c_k\le n,
  \label{eq:library}
\end{equation}
with non-negative integers $(a_k,m_k,c_k)$ and $n=6$, which gives $K=38$ candidate terms. The constant $-1$ fixes the scale of $\boldsymbol\xi$ and places the stress-free state inside the admissible set, $\Phi(\mathbf 0)=-1<0$. Each monomial is isotropic and positively homogeneous of degree $n_k$, $m_k(t\boldsymbol\sigma)=t^{n_k}m_k(\boldsymbol\sigma)$ for $t>0$. Terms with even $m_k$ are polynomials in the stress components, because $q^2=3I_2'$ and $r_3=\tfrac{27}{2}I_3'$; terms with odd $m_k$ are smooth everywhere except on the hydrostatic axis, where they produce cone-type vertices (as in the Drucker--Prager criterion).

Many classical criteria are \emph{sparse} members of \eqref{eq:library} (Table~\ref{tab:classical}). In particular, the cosine ansatz of power six \citep[Eq.~(9.25)]{Kolupaev2018}, $(3I_2')^3(1+c_3\cos3\theta+c_6\cos^23\theta)/(1+c_3+c_6)=\sigma_\mathrm{eq}^6$, is the three-term combination $\{q^6,\ q^3r_3,\ r_3^2\}$, since $q^6\cos3\theta=q^3r_3$ and $q^6\cos^23\theta=r_3^2$. Its combination with a polynomial $I_1$-substitution adds powers of $I_1$.

\begin{table}[!htbp]
\centering\small
\caption{Classical criteria as members of the library \eqref{eq:library} (normalised by $\sigma_+$; $d=\sigma_-/\sigma_+$).}
\label{tab:classical}
\begin{tabular}{lll}
\toprule
Criterion & $\Phi+1$ & Terms \\
\midrule
von Mises & $q^2$ (equivalently $q$) & 1 \\
Drucker--Prager cone & $\frac{d+1}{2d}\,q+\frac{d-1}{2d}\,I_1$ & 2 \\
Burzy\'nski--Yagn paraboloid & $\frac1d\,q^2+\frac{d-1}{d}\,I_1$ & 2 \\
Burzy\'nski--Yagn ellipsoid/hyperboloid & $\xi_1 q^2+\xi_2 I_1+\xi_3 I_1^2$ & 3 \\
Cosine ansatz, power 6 \citep[Eq.~(9.25)]{Kolupaev2018} & $\frac{q^6+c_3\,q^3r_3+c_6\,r_3^2}{1+c_3+c_6}$ & 3 \\
Tresca (polynomial form, \citealp{Hill1950}) & $\frac{4}{27}(q^6-r_3^2)-q^4+2q^2$ & 4 \\
\bottomrule
\end{tabular}
\end{table}

\begin{remark}[Non-uniqueness of the representation]
The zero set of $\Phi$ is unchanged under $\Phi\mapsto h\circ(\Phi+1)-1$ for any increasing $h$ with $h(1)=1$; for example, $q$ and $q^2$ describe the same von Mises cylinder. Supports therefore cannot be compared one-to-one with a ``true'' support. We assess identified models by the geometry of the surface they define (Section~\ref{sec:synthetic}).
\end{remark}

\begin{remark}[Polynomial forms of piecewise-linear criteria]
The polynomial form of the Tresca criterion in Table~\ref{tab:classical} equals, up to a positive factor, $-\prod_{i<j}[(\sigma_i-\sigma_j)^2-\sigma_+^2]$. Along the tensile ray it reduces to $-(\sigma^2-1)^2\le0$: it touches zero at $Z$ without changing sign and becomes negative again outside the hexagon. Such representations violate the unique-definition condition, and the first-root convention below does not recover the hexagon at its vertices. Piecewise-linear surfaces (Tresca, Mohr--Coulomb, Rankine) are therefore only approximated by smooth library members. Section~\ref{sec:synthetic} quantifies this for Mohr--Coulomb.
\end{remark}

\subsection{Beyond monomials I: Lode-angle Fourier modes and rational invariants}
\label{sec:cheb}
The monomial library \eqref{eq:library} couples two things that have nothing to do with each other physically: the stress degree of a term and its resolution in the Lode angle. Since $r_3^c=q^{3c}\cos^c3\theta$, a monomial of degree $n_k$ can carry Lode harmonics up to $\cos(3c\,\theta)$ with $3c\le n_k\le6$ only. Sharp deviatoric features at low stress degree, such as the nearly hexagonal or triangular sections of rubbers and of granular and frictional materials, then need many terms or cannot be represented at all (\cref{fig:concept}b). The question is which non-polynomial Lode functions are admissible.
\begin{proposition}[Isotropy admits only cosine modes]\label{prop:cos}
Let $\Phi$ be an isotropic function of the stress, written as $\Phi=\hat\Phi(I_1,q,\theta)$ with the Lode angle $\theta\in[0,\pi/3]$ continued to $\mathbb R$ through the principal-stress representation $\sigma_i=I_1/3+\tfrac23q\cos(\theta-2\pi(i-1)/3)$. Then $\hat\Phi$ is even and $2\pi/3$-periodic in $\theta$. Its Fourier series therefore contains only the modes $\cos(3k\theta)=T_k(\cos3\theta)$, $k\in\mathbb N_0$, where $T_k$ is the Chebyshev polynomial of the first kind.
\end{proposition}
\begin{proof}
A shift $\theta\mapsto\theta+2\pi/3$ permutes the principal stresses cyclically, and $\theta\mapsto-\theta$ exchanges $\sigma_2$ and $\sigma_3$. An isotropic function is a symmetric function of the principal stresses and is invariant under both operations. Evenness removes the sine modes, and periodicity leaves only the multiples of $3$.
\end{proof}
Sine modes such as $\sin3\theta$ would make the criterion depend on the \emph{ordering} of the principal stresses. They are therefore excluded, and a ``trigonometric'' library is a Chebyshev library in $\cos3\theta=r_3/q^3$. We add the terms
\begin{equation}
  m_{a,m,k}=I_1^{a}\,q^{m}\,T_k\!\left(\frac{r_3}{q^3}\right)=I_1^a\,q^m\cos(3k\theta),\qquad m\ge1,\quad 1\le a+m\le6,\quad 1\le k\le3,
  \label{eq:cheb}
\end{equation}
to the monomials. Expanding $T_k$ shows that $m_{a,m,k}$ is a sum of terms $I_1^a r_3^j q^{m-3j}$, $j\le k$, i.e.\ a \emph{rational} invariant function with negative powers of $q$ whenever $3k>m$. Each term is still isotropic and positively homogeneous of degree $a+m$. The ray polynomial \eqref{eq:raypoly} and all sensitivities are therefore unchanged. The restriction $m\ge1$ keeps the terms continuous on the hydrostatic axis, where the Lode angle is undefined: $|q^m T_k|\le q^m\to0$. Terms with $k=1$ and $m\ge3$ coincide with monomials and are omitted. This leaves $53$ new terms; the union with the monomials (``Chebyshev library'') has $K=91$ members.

\begin{remark}[Regularity on the hydrostatic axis]\label{rem:axis}
Continuity on the axis does not imply differentiability. Near $q=0$ a term $I_1^aq^m\cos3k\theta$ behaves like $q^m$ times a bounded function of the deviatoric direction. It is therefore $C^{m-1}$, but in general not $C^m$, at the axis. For $m=1$ its gradient has no limit there, and a surface on which such a term is active has a conical node on the hydrostatic axis, like a Drucker--Prager cone. This is not new to the extended libraries: the monomial $q$ itself, the Drucker--Prager term, is not differentiable on the axis, and the PBP atoms $I_1^a(q\Omega)^m$ behave in the same way. Kolupaev's condition (viii), continuous differentiability at the hydrostatic nodes, is therefore not enforced but reported (Table~\ref{tab:conditions}). For pressure-sensitive brittle materials a conical node is often physically acceptable; where it is not, the library can be restricted to $m\ge2$ in all Lode-dependent terms. The reduced regularity has no effect on the identification. (a) Lemma~\ref{lem:implicit} needs only the derivatives of $G(t,\boldsymbol\xi)=\Phi(t\mathbf n;\boldsymbol\xi)$ along a fixed ray. For every direction, $G$ is a polynomial in $t$ (Eq.~\eqref{eq:raypoly}) whose coefficients are the feature values $m_k(\mathbf n)$, and it is linear in the library coefficients. On the axis itself all terms with $m\ge1$ vanish identically, so $G$ reduces to the pure $I_1$ part and remains a smooth function of $t$ and $\boldsymbol\xi$. The derivatives with respect to the PBP parameters $(\beta,\gamma)$ are smooth for $\gamma<1$, which is why we bound $\gamma\le0.995$. (b) Only the surface normal and the curvature at the node are undefined. They enter the convexity penalty and the audit, whose directions exclude a $4^\circ$ (penalty) or $2^\circ$ (audit) cone around the axis, and the inelastic Poisson ratios, which are evaluated at $Z$ and $D$ away from the axis. (c) Numerically, $\cos3\theta=r_3/q^3$ is evaluated with $q^2$ regularised by $10^{-12}$ and clipped to $[-1,1]$ (\cref{app:impl2}). The convexity of the node itself is not tested directly. Near the node, however, the surface is approximately a cone with the local deviatoric section as directrix, and a re-entrant (non-convex) node would make that section concave. This shows up as negative tangential curvature in the neighbouring directions just outside the excluded cone, which the penalty and the audit do check.
\end{remark}

\subsection{Beyond monomials II: parametric atoms}
\label{sec:atoms}
\begin{figure}[!htbp]
\centering
\includegraphics[width=\linewidth]{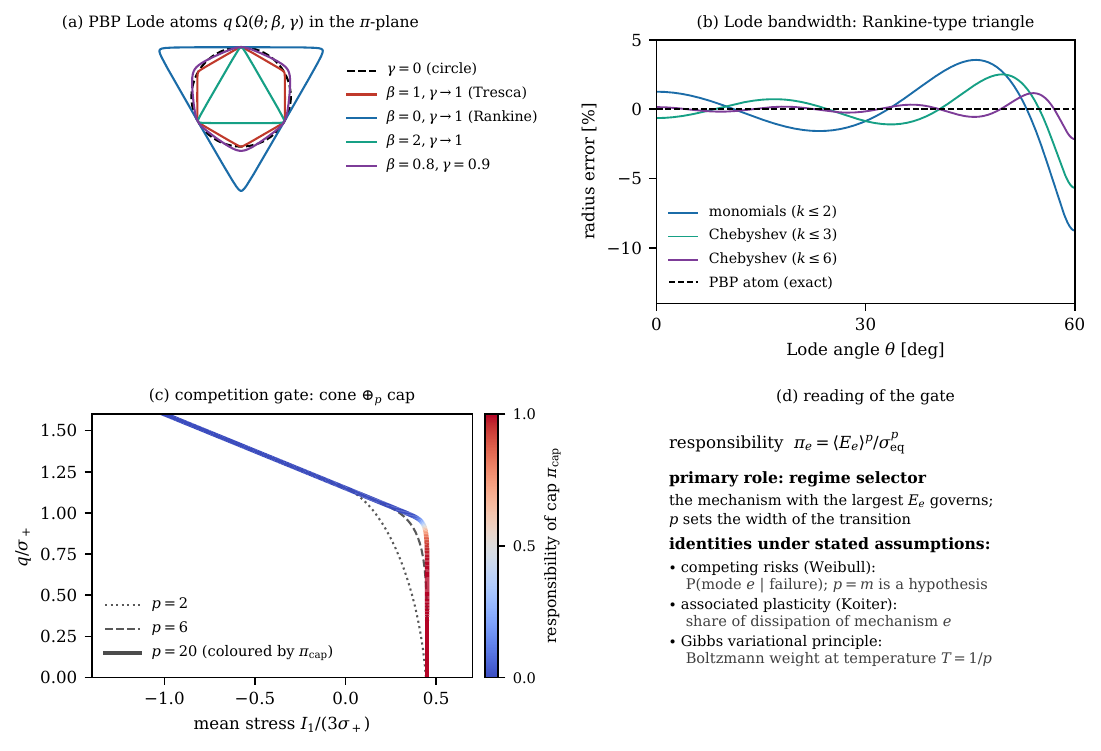}
\caption{Extended representations. (a) Deviatoric sections of the PBP atom \eqref{eq:pbp} for several $(\beta,\gamma)$. (b) Radius error of Lode-bandwidth-limited Fourier (Chebyshev) approximations of a Rankine-type triangle; monomials of degree six resolve only $k\le2$, a single PBP atom is exact. (c) Competition gate \eqref{eq:smax} of a Drucker--Prager cone and a mean-stress cap in the Burzy\'nski plane for $p=2,6,20$; the curve for $p=20$ is coloured by the responsibility of the cap. (d) Primary reading as regime selector and the three mechanistic readings of the responsibilities, each exact under its own assumptions (Section~\ref{sec:gate_meaning}).}
\label{fig:concept}
\end{figure}

Several established deviatoric shape functions are not finite Fourier sums. The most versatile one is due to \citet{Podgorski1985} and \citet{Bigoni2004} (PBP),
\begin{equation}
  \Omega(\theta;\beta,\gamma)=\frac{\cos\!\left[\beta\frac{\pi}{6}-\frac13\arccos(\gamma\cos3\theta)\right]}{\cos\!\left[\beta\frac{\pi}{6}-\frac13\arccos\gamma\right]},\qquad \beta\in[0,2],\ \gamma\in[0,1],
  \label{eq:pbp}
\end{equation}
normalised to $\Omega=1$ on the tension meridian. The deviatoric section $q\,\Omega(\theta)=\mathrm{const}$ is convex for all $(\beta,\gamma)$ in this range. It contains the circle ($\gamma=0$), the Tresca hexagon ($\beta=1,\gamma\to1$), and the two Rankine-type triangles ($\beta=0$ or $\beta=2$, $\gamma\to1$), with smooth transitions in between \citep{Bigoni2004} (\cref{fig:concept}a). Its corners for $\gamma\to1$ are exactly the features that polynomial libraries cannot resolve. We use it as a \emph{parametric atom}: $q$ is replaced by $q\Omega$ in the $q$-containing monomials without $r_3$,
\begin{equation}
  m^{\Omega}_{a,m}=I_1^a\,\bigl(q\,\Omega(\theta;\beta,\gamma)\bigr)^m,\qquad m\ge1,\quad 1\le a+m\le6,
  \label{eq:atoms}
\end{equation}
with one pair $(\beta,\gamma)$ shared by all atoms of a model. The $21$ atoms are homogeneous of degree $a+m$. They are added to the monomials (``PBP library'', $K=59$) or to the monomials and Chebyshev terms (``extended library'', $K=112$). The shape parameters are fitted jointly with the active coefficients by bounded L-BFGS-B. We bound $\gamma\le0.995$ because $\partial_\gamma\arccos(\gamma\cos3\theta)$ is unbounded at $\gamma\cos3\theta=\pm1$. Screening (Algorithm~\ref{alg:beam}) needs fixed features, so it uses the $(\beta,\gamma)$ of a preliminary fit of the core $\{q\Omega,(q\Omega)^2,I_1\}$, and every refit then frees them again. In the information criteria the two shape parameters count as parameters whenever an atom is active. Other Lode functions, such as those of \citet{Altenbach2014} or \citet{Hamdi2006}, enter in the same way. Non-homogeneous meridian functions, such as the \citet{Bigoni2004} cap, would need the generic first-root search instead of the ray polynomial and are not included.

\begin{remark}[Sparsity depends on the basis]
The three libraries span nested function spaces (monomials $\subset$ Chebyshev, and monomials $\subset$ PBP via $\gamma=0$). Sparsity, however, is a property of a representation, not of a surface. A Tresca-like or triangular deviatoric section is a single PBP atom, but has no finite monomial or Chebyshev representation. Conversely, the trigonal criterion with $\{q^3,r_3,I_1\}$ (truth T3 in Section~\ref{sec:synthetic}) is not a finite combination of PBP atoms. A union library lets the selection choose the sparsest description. The price is collinearity between near-equivalent terms, which makes the selected \emph{support} less stable than the selected \emph{surface}.
\end{remark}

\subsection{Ray-wise definition of the surface and the equivalent stress}
For a unit direction $\mathbf n$ in principal-stress space let
\begin{equation}
  \rho(\mathbf n;\boldsymbol\xi)=\min\{t>0:\ \Phi(t\mathbf n;\boldsymbol\xi)=0,\ \Phi \text{ changes sign from} -\text{to}+\}
  \label{eq:firstroot}
\end{equation}
be the radius of the limit surface along $\mathbf n$ (first root), with $\rho=\infty$ if no root exists (open surface in that direction). Because the monomials are homogeneous,
\begin{equation}
  \Phi(t\mathbf n;\boldsymbol\xi)=\sum_{p=1}^{n}A_p(\mathbf n;\boldsymbol\xi)\,t^p-1,\qquad A_p=\sum_{k:\,n_k=p}\xi_k\,m_k(\mathbf n),
  \label{eq:raypoly}
\end{equation}
is a polynomial in $t$ whose coefficients are linear in $\boldsymbol\xi$. The monomials $m_k(\mathbf n)$ of fixed directions (data directions, constraint directions) are evaluated once, so the root search reduces to evaluating a degree-six polynomial. The equivalent stress is the gauge (Minkowski functional) of the admissible set, scaled to uniaxial tension,
\begin{equation}
  \sigma_\mathrm{eq}(\boldsymbol\sigma)=\sigma_+\,\frac{\|\boldsymbol\sigma\|}{\rho(\boldsymbol\sigma/\|\boldsymbol\sigma\|)},\qquad \sigma_+=\rho(\mathbf e_Z),
  \label{eq:gauge}
\end{equation}
which is positively homogeneous of degree one and satisfies $\sigma_\mathrm{eq}=\sigma_\mathrm I$ in uniaxial tension. Explicit solvability for $\sigma_\mathrm{eq}$ and homogeneity, conditions (iii) and (iv), therefore hold by construction for every member of the library, including non-homogeneous combinations of terms of different degree.

\paragraph{Root finding and exact sensitivities}
The root is bracketed on a fixed grid $0=t_0<t_1<\dots<t_G=t_\mathrm{max}$ ($G=256$, quadratic spacing) and refined by bisection with detached parameters. Exact first derivatives with respect to $\boldsymbol\xi$ are recovered by one implicit Newton correction.
\begin{lemma}\label{lem:implicit}
Let $G(t,\boldsymbol\xi)=\Phi(t\mathbf n;\boldsymbol\xi)$, let $t_s$ be a root with $G_{,t}(t_s,\boldsymbol\xi)\neq0$, and define $\hat t(\boldsymbol\xi)=t_s-G(t_s,\boldsymbol\xi)/G_{,t}(t_s,\boldsymbol\xi)$ with $t_s$ treated as a constant. Then $\partial\hat t/\partial\boldsymbol\xi=-G_{,\boldsymbol\xi}/G_{,t}$, which is the implicit-function-theorem derivative of the root.
\end{lemma}
\begin{proof}
Differentiating $\hat t$ gives $-G_{,\boldsymbol\xi}/G_{,t}+G\,G_{,t\boldsymbol\xi}/G_{,t}^2$, and the second term vanishes because $G(t_s,\boldsymbol\xi)=0$.
\end{proof}
Second derivatives, which the design problem of Section~\ref{sec:oed} needs, are obtained with two un-detached Newton corrections, since each Newton step doubles the order of derivative consistency. All derivatives are computed by automatic differentiation in JAX \citep{jax2018}.

\subsection{Differentiable admissibility penalties}
\label{sec:penalties}
\paragraph{Unique definition}
\begin{proposition}\label{prop:star}
Let $\Phi$ be continuous with $\Phi(\mathbf 0)<0$ and assume that along every ray $\Phi(t\mathbf n)\ge0$ for all $t\ge\rho(\mathbf n)$. Then $\{\Phi<0\}=\{t\mathbf n:0\le t<\rho(\mathbf n)\}$ is star-shaped with respect to the origin, its boundary is the first-root surface, and $\{\Phi<0\}$ has no further components. In particular, there is no outer contour in the sense of \citet[Sect.~8.2.2]{Kolupaev2018}.
\end{proposition}
\begin{proof}
Every $\boldsymbol\sigma=t\mathbf n$ with $t\ge\rho(\mathbf n)$ satisfies $\Phi\ge0$ by assumption, and every point with $t<\rho(\mathbf n)$ satisfies $\Phi<0$ by definition of the first root.
\end{proof}
We enforce the assumption of Proposition~\ref{prop:star} with a small margin $\epsilon=0.02$, which also excludes near-tangent outer branches. The penalty is evaluated on $N_d=240$ quasi-uniform directions (Fibonacci sphere, excluding a $4^\circ$ cone around the hydrostatic axis) and $n_o=64$ geometrically spaced radii $t_j\in[1.1\rho,\ t_\mathrm{max}]$:
\begin{equation}
  P_\mathrm{uniq}(\boldsymbol\xi)=\frac{1}{N_dn_o}\sum_{d,j}\mathrm{ReLU}\big(\epsilon-\Phi(t_j\mathbf n_d;\boldsymbol\xi)\big)^2 .
  \label{eq:Puniq}
\end{equation}
\begin{remark}[Validity region]
If the leading homogeneous form of a polynomial $\Phi$ is indefinite, $\Phi(t\mathbf n)$ becomes negative again for large $t$ along some rays. Uniqueness can therefore only be required within a bounded validity region. We use $t\le t_\mathrm{max}=3\max_ir_i$, three times the largest tested stress. This makes explicit the validity region of \citet[Sect.~12.5]{Kolupaev2018}, outside of which spurious branches are tolerated. The identified criteria must not be evaluated beyond it.
\end{remark}

\paragraph{Convexity}
For a level set with $\nabla\Phi\neq\mathbf 0$, local convexity at a surface point is equivalent to positive semi-definiteness of the Hessian restricted to the tangent space, $\mathbf v\cdot\nabla^2\Phi\,\mathbf v\ge0$ for all $\mathbf v\perp\nabla\Phi$. A closed, connected set whose boundary is locally convex everywhere is convex (Tietze--Nakajima theorem). For isotropic $\Phi$, convexity of the admissible set in the six-dimensional stress space is equivalent to convexity of its image in principal-stress space \citep{Lewis1996}, so three-dimensional checks suffice. With the projector $\mathbf P=\mathbf 1-\hat{\mathbf g}\otimes\hat{\mathbf g}$, $\hat{\mathbf g}=\nabla\Phi/\|\nabla\Phi\|$ evaluated at $\boldsymbol\sigma_d=\rho(\mathbf n_d)\mathbf n_d$, we penalise negative eigenvalues of the dimensionless curvature matrix
\begin{equation}
  \boldsymbol\kappa_d=\frac{\rho(\mathbf n_d)}{\|\nabla\Phi\|}\,\mathbf P\,\nabla^2\Phi\,\mathbf P,\qquad
  P_\mathrm{conv}(\boldsymbol\xi)=\frac{1}{N_d}\sum_{d}\sum_{i=1}^3\mathrm{ReLU}\big(-\lambda_i(\boldsymbol\kappa_d)\big)^2 .
  \label{eq:Pconv}
\end{equation}
The normal direction contributes a zero eigenvalue and is therefore not penalised.

\paragraph{Numerical audit versus certificate}
\label{sec:convex_audit}
The penalties \eqref{eq:Puniq} and \eqref{eq:Pconv} act on $240$ directions. After fitting, every model is audited on a denser, independent set of $2000$ Fibonacci directions (excluding a $2^\circ$ cone around the hydrostatic axis). A direction counts as non-convex if the smallest scaled tangential curvature is below $-10^{-3}$, and as re-entrant if $\Phi<-10^{-6}$ at any of $200$ geometrically spaced radii between $1.01\rho$ and $t_\mathrm{max}$ (\cref{app:impl}). The audit is a numerical check on finitely many directions and radii. It cannot exclude violations between the sampled directions, on the excluded cone, or beyond $t_\mathrm{max}$. When we report that a library model has no or few audit violations, we therefore mean the result of this discrete audit, not a certificate of convexity or uniqueness on the continuum. Exact certificates would require, for example, sum-of-squares relaxations of the polynomial sub-library (Section~\ref{sec:disc}). Only the properties stated in Propositions~\ref{prop:star}, \ref{prop:expert} and in Section~\ref{sec:moe} hold for every direction; the audit fractions of the competition MoE are zero by construction and serve as a check of the implementation.

\paragraph{Inelastic Poisson ratios}
Following \citet[Sect.~5.4, Eq.~(12.18)]{Kolupaev2018} we require $-1\le\nu^\mathrm{in}_+\le1/2$ and $\nu^\mathrm{in}_-\ge-1$:
\begin{equation}
  P_\nu=\mathrm{ReLU}(\nu^\mathrm{in}_+-\tfrac12)^2+\mathrm{ReLU}(-1-\nu^\mathrm{in}_+)^2+\mathrm{ReLU}(-1-\nu^\mathrm{in}_-)^2 .
  \label{eq:Pnu}
\end{equation}
We do not impose $\nu^\mathrm{in}_-\le1/2$. Strongly pressure-sensitive materials with open surfaces, such as concrete, violate it (values around 5.5 are reported for the concrete data of Section~\ref{sec:real}; \citealp[Table~14.15]{Kolupaev2018}), and the Burzy\'nski--Yagn paraboloid with $d>1$ also gives $\nu^\mathrm{in}_->1/2$.

\paragraph{Optional anchor}
\citet[Sect.~12.4]{Kolupaev2018} recommends fixing $\sigma_\mathrm{eq}=\sigma_+$ to the uniaxial tensile strength, because the tension test is the most reliable one. This can be added as
\begin{equation}
  P_\mathrm{anchor}=\big(\ln\rho(\mathbf e_Z)-\ln\bar\sigma_+^\mathrm{meas}\big)^2 .
  \label{eq:Panchor}
\end{equation}
It is switched off by default; Section~\ref{sec:disc} discusses why.

\subsection{Identification from failure stress states}
\label{sec:objective}
Given $N$ measured failure states $\boldsymbol\sigma_i$ with directions $\mathbf n_i$ and radii $r_i=\|\boldsymbol\sigma_i\|$, we minimise
\begin{equation}
  J(\boldsymbol\xi)=\frac1N\sum_{i=1}^N\Big[\ln\frac{\rho(\mathbf n_i;\boldsymbol\xi)}{r_i}\Big]^2
  +\mu\big(\mu_cP_\mathrm{conv}+\mu_uP_\mathrm{uniq}+\mu_\nu P_\nu\big)+\alpha\|\tilde{\boldsymbol\xi}\|^2,
  \label{eq:objective}
\end{equation}
with $\tilde\xi_k=\xi_k s_k$ the coefficients of scaled features ($s_k$ the root-mean-square of $m_k$ over the data), $\mu_c=\mu_u=\mu_\nu=10$ and a small ridge $\alpha=10^{-6}$. The penalty factor $\mu$ is continued over $\{1,100\}$. The radial log-residual is the relative error of the predicted failure load along a proportional load path through the data point. It is the logarithmic counterpart of the ray-based and equivalent-stress-based quality measures $f_\mathrm{ray}$ and $f_\mathrm{eq}$ of \citet[Sect.~12.2]{Kolupaev2018}. Kolupaev considered these ``physical'' objectives too expensive for parameter identification and used them only to compare fits. With \eqref{eq:raypoly} and Lemma~\ref{lem:implicit} they are cheap and exactly differentiable, so we use them directly for identification. If a data direction has no root ($\rho=\infty$), its residual is replaced by $\ln(t_\mathrm{max}/r_i)+10\,\mathrm{ReLU}(-\Phi(\boldsymbol\sigma_i))$, which keeps a gradient towards closing the surface. $J$ is minimised with L-BFGS-B \citep{Byrd1995} using exact gradients.

\subsection{Sparse selection and model choice}
\label{sec:selection}
Regression on all $K=38$ terms of a strongly collinear library from a few dozen points is ill-posed. Thresholding schemes in the spirit of SINDy \citep{Brunton2016,Rudy2017} and greedy forward selection both failed in preliminary tests: both discard or never reach the correct support because the first selected term (typically $q$) masks the others. We therefore combine best-subset screening with constrained radial refits (Algorithm~\ref{alg:beam}). The screening starts from the implicit, linear SINDy-type problem $\sum_k\xi_km_k(\boldsymbol\sigma_i)=1$ \citep[cf.][]{Kaheman2020}. Its algebraic residual is a poor proxy for the radial error, so we linearise the radial residual instead. By Euler's theorem for homogeneous functions, $t\,\partial_t\Phi(t\mathbf n)=\sum_k n_k\xi_km_k$, hence
\begin{equation}
  \ln\frac{\rho(\mathbf n_i)}{r_i}\approx-\frac{\Phi(\boldsymbol\sigma_i)}{\sum_kn_k\xi_km_k(\boldsymbol\sigma_i)} ,
  \label{eq:euler}
\end{equation}
and we minimise this first-order radial residual by two iteratively reweighted least-squares updates. Neither score is uniformly better: the Euler-linearised score is essential for the trigonal truth of Section~\ref{sec:synthetic}, while the plain algebraic score works better for the concrete data, whose radii span more than one order of magnitude. Candidates are therefore taken from the union of the best supports under both scores. Supports for which the denominator in \eqref{eq:euler} is non-positive at a data point are penalised, since there $\Phi$ decreases outward and the point cannot lie on a first root. The screening is exhaustive for $s\le3$ (9177 supports for the monomial and 234\,248 for the extended library) and uses beam search for $s>3$; each candidate requires only small least-squares solves. In a preliminary run of the benchmark of Section~\ref{sec:synthetic} with plain algebraic ranking, the constrained pipeline missed the three-term truth T3 in five of six realisations and returned surfaces with $6$--$14\%$ radial error. With the Euler-linearised score it recovers the true support (\cref{sec:synthetic}). Only the best candidates, together with warm-started extensions of the previously selected support, are refitted with the full objective \eqref{eq:objective}. For the union libraries the selection paths of the nested smaller libraries are added as candidates. A larger library thus never returns a worse support of a given size than a library it contains, unless the refit fails. With PBP atoms the screening uses the shape parameters of a preliminary core fit, and the refits free them again (Section~\ref{sec:atoms}). The final support size is chosen by $k$-fold cross-validation of the radial residual using the one-standard-error rule \citep{Hastie2009}. Every fold is refitted from the algebraic initialisation on its training data and not warm-started from the full-data fit, which would leak information (\cref{app:impl2}). Parameter uncertainty is quantified by a non-parametric bootstrap with the support held fixed \citep{Efron1993}.

\begin{algorithm}[t]
\caption{Screened beam search with constrained radial refits.}\label{alg:beam}
\begin{algorithmic}[1]
\Require data $\{\boldsymbol\sigma_i\}$, library $\{m_k\}$, $s_\mathrm{max}$, beam width $B=60$, refits $R=8$
\State $\mathbf A_{ik}\gets m_k(\boldsymbol\sigma_i)/s_k$
\For{$s=1,\dots,s_\mathrm{max}$}
  \State candidate supports $\mathcal C_s\gets$ all $s$-subsets ($s\le3$) or one-term extensions of the beam (else)
  \State rank $\mathcal C_s$ by the Euler-linearised radial residual \eqref{eq:euler} (IRLS) and by the algebraic residual; beam $=$ best $B$ and $B/2$ supports, respectively
  \State refit the best $R$ supports under each score, the best $\max(R/2,2)$ extensions of $\mathcal S_{s-1}$ (warm start) and the size-$s$ supports of nested libraries with \eqref{eq:objective}
  \State $\mathcal S_s\gets$ support with the lowest penalised objective
\EndFor
\State choose $s^\star$ by leakage-free $k$-fold CV (one-standard-error rule); bootstrap on $\mathcal S_{s^\star}$
\end{algorithmic}
\end{algorithm}

\subsection{Mixtures of experts as baselines}
\label{sec:moe}
A library model is one equation with many terms. A mixture of experts (MoE) is a few simple equations, each describing one mechanism, combined by a gate \citep{Jacobs1991,Shazeer2017}. In the equivalent-stress setting it is natural to let each expert be a gauge, i.e.\ an equivalent stress of its own. We use the five-parameter family
\begin{equation}
  E_e(\boldsymbol\sigma)=\sqrt{a_e\,\bigl(q\,\Omega(\theta;\beta_e,\gamma_e)\bigr)^2+b_e\,I_1^2}+c_e\,I_1,\qquad a_e,b_e\ge0,\ \beta_e\in[0,2],\ \gamma_e\in[0,1),
  \label{eq:expert}
\end{equation}
which contains the von Mises cylinder, Drucker--Prager cones ($b_e=0$), the Rankine criterion ($b_e=0$, $\beta_e=0$, $\gamma_e\to1$, $a_e=4/9$, $c_e=1/3$), mean-stress (cavitation) caps ($a_e=b_e=0$, $c_e>0$) and shifted ellipsoids ($b_e>c_e^2$), each with a PBP deviatoric section.
\begin{proposition}\label{prop:expert}
Every expert \eqref{eq:expert} is convex and positively homogeneous of degree one in $\boldsymbol\sigma$.
\end{proposition}
\begin{proof}
$x(\boldsymbol\sigma)=q\,\Omega(\theta)$ is the gauge (Minkowski functional) of the convex prismatic set with the PBP section, hence convex, non-negative and homogeneous of degree one. The function $f(x,y)=\sqrt{a x^2+b y^2}$ is convex, homogeneous of degree one and non-decreasing in $x\ge0$. Composing it with the convex $x(\boldsymbol\sigma)$ and the linear $y=I_1$ preserves convexity, and adding the linear term $c_eI_1$ preserves both properties.
\end{proof}
We study two ways of combining the experts.

\paragraph{Competition gate (smooth maximum)}
\begin{equation}
  \sigma_\mathrm{eq}(\boldsymbol\sigma)=\Bigl(\sum_{e=1}^{E}\bigl\langle E_e(\boldsymbol\sigma)\bigr\rangle^p\Bigr)^{1/p},\qquad
  \pi_e(\boldsymbol\sigma)=\frac{\langle E_e\rangle^p}{\sigma_\mathrm{eq}^p},\qquad 1\le p\le p_\mathrm{max},
  \label{eq:smax}
\end{equation}
with $\langle x\rangle=\max(x,0)$, a learnable exponent $p$ ($p_\mathrm{max}=50$), and responsibilities $\pi_e\ge0$, $\sum_e\pi_e=1$. As a $p$-norm of non-negative convex functions, $\sigma_\mathrm{eq}$ is convex and homogeneous of degree one. The admissible set $\{\sigma_\mathrm{eq}\le1\}$ is therefore convex and star-shaped, and along every ray $\sigma_\mathrm{eq}(t\mathbf n)=t\,\sigma_\mathrm{eq}(\mathbf n)$ is strictly increasing. Conditions (i)--(iv) hold \emph{by construction} and no penalty is needed; directions in which all experts are non-positive are open. For $p\to\infty$, \eqref{eq:smax} becomes the non-smooth multi-surface criterion $\max_eE_e$ \citep[Ch.~11]{Kolupaev2018}. The coupled distortional--dilatational criterion of \citet{Drass2020coupled} for silicone adhesives is of this type, with two experts and $p\to\infty$.

\paragraph{Direction gate}
\begin{equation}
  \begin{gathered}
  \sigma_\mathrm{eq}(\boldsymbol\sigma)=\sum_{e=1}^{E}g_e(\mathbf n)\,E_e(\boldsymbol\sigma),\qquad \mathbf g=\mathrm{softmax}\bigl((\mathbf W\boldsymbol\phi(\mathbf n)+\mathbf b)/T\bigr),\\
  \boldsymbol\phi=\bigl(\cos3\theta,\cos6\theta,\eta,\eta^2,\eta\cos3\theta\bigr),
  \end{gathered}
  \label{eq:gate}
\end{equation}
with $\mathbf n=\boldsymbol\sigma/\|\boldsymbol\sigma\|$ and the direction triaxiality $\eta=I_1/(\sqrt3\|\boldsymbol\sigma\|)\in[-1,1]$. This is the standard MoE with a tunable soft gate. Because the gate depends on the direction only, $\sigma_\mathrm{eq}$ is still homogeneous of degree one, and the surface is unique by construction. It is \emph{not} convex in general, since the gate gradient enters the Hessian. We fit it without constraints (``gate'') and with the convexity penalty \eqref{eq:Pconv} (``gate+$P_c$'').

\paragraph{Sparsity and training}
The number of experts is the analogue of the support size. We choose $E\in\{1,2,3\}$ by cross-validation with the one-standard-error rule, as for the libraries. Two relaxations from machine learning enforce sparsity directly and serve as checks. (a) Hard-concrete $L_0$ gates $z_e\in[0,1]$ multiply the expert terms in \eqref{eq:smax}. Their expected number of non-zeros is penalised with weight $\lambda_0$, and they are trained with stochastic reparametrised gradients \citep{Louizos2018}. (b) For the direction gate, Gumbel-softmax noise with an annealed temperature $T:2\to0.1$ drives the routing towards hard assignments \citep{Jang2017,Maddison2017}. A symmetric Dirichlet prior with concentration below one on the gate weights would act similarly \citep{Blei2003}. All deterministic fits minimise the same radial log-residual as the libraries; for a gauge this is $\ln\rho(\mathbf n_i)-\ln r_i=-\ln\sigma_\mathrm{eq}(\bar{\boldsymbol\sigma}_i)$, so no root search is needed. We use L-BFGS on unconstrained reparametrisations (softplus for $a_e,b_e,$ sigmoids for $\beta_e,\gamma_e,p$) from $12$ random starts. The smooth-maximum model has $5E+1$ parameters ($5$ for $E=1$, where $p$ has no effect) and the direction-gated model $5E+6(E-1)$.

\subsection{Physical meaning of the gate}
\label{sec:gate_meaning}
A gate is useful beyond curve fitting only if it can be interpreted. We distinguish two levels. \emph{Phenomenologically}, the competition gate is a regime selector: each expert describes one failure or yield mechanism, the mechanism with the largest driving force $E_e$ governs, and $p$ controls the width of the transition between regimes ($p\to\infty$: sharp corner of a multi-surface criterion). This reading requires no further assumptions, and it is the one we use to interpret the fitted models in Section~\ref{sec:real}. \emph{Mechanistically}, the responsibilities $\pi_e$ admit three readings, which are exact identities under the assumptions stated for each. They give $p$ and $\pi_e$ a quantitative meaning, but whether the assumptions hold for a material is an empirical question. For the data of this paper they lead to testable hypotheses, which the available data do not confirm quantitatively (Section~\ref{sec:overall}).

\paragraph{Fracture statistics: competing risks}
Suppose that failure of a material element can be initiated by $E$ independent mechanisms with Weibull survival probabilities of a common modulus $m$, $P_{s,e}=\exp[-(E_e(\boldsymbol\sigma)/s_e)^m]$ \citep{Weibull1951}, and absorb the scales $s_e$ into $E_e$. Weakest-link survival is then $P_s=\prod_eP_{s,e}=\exp[-\sum_e E_e^m]=\exp[-\sigma_\mathrm{eq}^m]$ with \eqref{eq:smax} and $p=m$. Every quantile surface of the failure stress is a level set of $\sigma_\mathrm{eq}$. The probability that failure at $\boldsymbol\sigma$ is initiated by mechanism $e$ equals its share of the hazard rate, $E_e^m/\sum_jE_j^m=\pi_e$. Under these assumptions the exponent $p$ is the Weibull modulus, and $\pi_e$ is the predicted fraction of fracture surfaces of type $e$. Both can be checked against the scatter of replicate tests and against fractography. The assumptions are strong: the mechanisms must be statistically independent, share one modulus, and act on the same material volume. If the moduli differ, the system is no longer of the form \eqref{eq:smax}, and a fitted $p$ is an effective transition parameter rather than a modulus.

\paragraph{Plasticity: smoothed Koiter rule}
If $\sigma_\mathrm{eq}$ is used as a yield function with associated flow (an assumption that does not apply to brittle failure surfaces), then
\begin{equation}
  \dot{\boldsymbol\varepsilon}^\mathrm p=\dot\lambda\frac{\partial\sigma_\mathrm{eq}}{\partial\boldsymbol\sigma}=\sum_e\dot\lambda_e\frac{\partial E_e}{\partial\boldsymbol\sigma},\qquad \dot\lambda_e=\dot\lambda\Bigl(\frac{E_e}{\sigma_\mathrm{eq}}\Bigr)^{p-1},
\end{equation}
which is Koiter's rule for multi-surface plasticity \citep{Koiter1953} with smoothly distributed multipliers. By Euler's theorem the dissipation is $D=\boldsymbol\sigma:\dot{\boldsymbol\varepsilon}^\mathrm p=\dot\lambda\sigma_\mathrm{eq}$. The share carried by mechanism $e$ is $\dot\lambda_eE_e/D=\pi_e$. The gate therefore apportions the plastic dissipation among the mechanisms, and $p\to\infty$ recovers Koiter's corner rule.

\paragraph{Thermodynamics: Gibbs variational principle}
With $\psi_e=\ln E_e$, \eqref{eq:smax} reads
\begin{equation*}
  \ln\sigma_\mathrm{eq}=p^{-1}\ln\sum\nolimits_e\exp(p\,\psi_e),
\end{equation*}
a free energy at inverse temperature $p$. The Gibbs variational principle gives
\begin{equation}
  \ln\sigma_\mathrm{eq}=\max_{\boldsymbol\pi\in\Delta^{E-1}}\Bigl[\sum_e\pi_e\,\psi_e-\frac1p\sum_e\pi_e\ln\pi_e\Bigr],
  \label{eq:gibbs}
\end{equation}
and the maximiser is the Boltzmann distribution $\pi_e\propto\exp(p\psi_e)$, which is \eqref{eq:smax}. The gate is thus the entropy-regularised selection of the mechanism with the largest (logarithmic) driving force. The ``temperature'' $1/p$ measures the disorder of this selection, as $1/m$ does for the strength scatter in the competing-risk reading.

\paragraph{Direction gates have none of these meanings}
In \eqref{eq:gate} the gate is a free classifier of the loading direction. It does not depend on the experts' driving forces, and it does not result from a variational principle. Its gradient contributes $\sum_e E_e\,\partial g_e/\partial\boldsymbol\sigma$ to the normal of the surface, a term that no mechanism produces. It can be read only as a phenomenological map of where each expert is used, comparable to a deformation-mechanism map. A gate becomes physically interpretable once it is a function of the experts' own driving forces, and the competition gate \eqref{eq:smax} is the simplest such gate.

\subsection{Strain-space limit surfaces for rubber-like materials}
\label{sec:hencky}
Failure of nearly incompressible elastomers is usually described in deformation space. Stresses at failure depend on the hyperelastic model, and hydrostatic stresses are hardly measurable \citep{Rosendahl2019,Drass2020coupled}. The framework above uses only isotropy, homogeneity and the ray structure, so it applies unchanged to any isotropic symmetric tensor that vanishes in the reference state. We use the Hencky strain $\mathbf h=\ln\mathbf v$ with principal values $h_i=\ln\lambda_i$. In this space proportional log-strain paths are rays from the origin. The homogeneous tests on incompressible material lie in the deviatoric plane $\operatorname{tr}\mathbf h=\ln J=0$: uniaxial tension at $\theta=0$, planar (pure or simple) shear at $\theta=\pi/6$ and equibiaxial tension at $\theta=\pi/3$. Uniaxial compression has the same Lode angle as equibiaxial tension, so isotropy alone makes the two equivalent in this space. Cavitation is a critical volume change $J_c$ \citep{Drass2020coupled}, i.e.\ the plane $\operatorname{tr}\mathbf h=\ln J_c$. We represent it by its hydrostatic point $\mathbf h=\tfrac13\ln J_c\,(1,1,1)$. Data are normalised by $h_\mathrm{ref}$, the mean maximal principal log-strain at uniaxial failure. The Kolupaev characteristic values are replaced by the radius ratios $e_\mathrm{ET}$, $e_\mathrm{PS}$ and $e_\mathrm{HT}$ of the equibiaxial, planar and hydrostatic-tension directions to the uniaxial direction. The Poisson-ratio penalty does not apply.

\paragraph{Energetic reading}
A strain-space surface can be tested against energy-based failure hypotheses (critical stored energy, as in Beltrami's criterion and in tearing-energy approaches) if a strain-energy function $W$ is available. We identify incompressible Ogden models \citep{Ogden1972} $W=\sum_p\mu_p\alpha_p^{-1}(\lambda_1^{\alpha_p}+\lambda_2^{\alpha_p}+\lambda_3^{\alpha_p}-3)$, $\mu_p\alpha_p>0$, by exhaustive best-subset non-negative least squares on relative stress errors. The search runs over $17$ exponents between $-6$ and $10$ with at most three terms. We do this for Treloar's classical natural-rubber data in uniaxial, equibiaxial and pure-shear tension \citep{Treloar1944}, taken in digital form from \citet{HyperDataJl}. We also fit the engineering stress--strain curves of TSSA in uniaxial tension, equibiaxial (bulge) tension and simple shear of the same test programme. We then evaluate $W$ at the failure points. Treloar's data serve as a benchmark for the identification and as a contrasting material; they contain no failure information.

\subsection{Identification from structural failure loads}
\label{sec:structural}
Homogeneous multiaxial tests are expensive, and some regions of stress space, notably hydrostatic tension, are hardly accessible \citep[Sect.~13.2.8]{Kolupaev2018}. Notched specimens give access to triaxial tension, but their stress state is inhomogeneous. Mapping a measured failure load to a single ``nominal'' stress state then introduces a model error. We instead identify $\boldsymbol\xi$ end-to-end from failure loads by embedding the limit surface in a differentiable finite element model.

\paragraph{Specimen model}
We consider circumferentially notched round bars (outer radius $R=6$, notch depth $D_n=2$, net radius $a=4$, half-length $H=24$; arbitrary length unit) with the notch profile $r_\mathrm{out}(z)=R-D_n\exp(-(z/w)^2)$, whose root radius is $\rho_n=w^2/(2D_n)$. The specimens are loaded in tension, with an optional superposed hydrostatic pressure $p$ applied in a pressure chamber, as in the tube tests of \citet[Sect.~13.2]{Kolupaev2018}. Pre-failure behaviour is linear elastic ($\nu=0.3$). This is the main idealisation of the structural part. It makes the stress field affine in the load and enables the closed-form first root below; its limits for real foams are discussed in Section~\ref{sec:disc}. A mapped, graded mesh of $16\times36$ axisymmetric bilinear elements (half model) is generated directly from $w$, so that nodal coordinates, stiffness matrix, solution and Gauss-point stresses are all differentiable functions of $w$. For a homogeneous isotropic body, a pressure acting on the entire surface produces the exact field $-p\mathbf 1$, so the stress at Gauss point $x$ is
\begin{equation}
  \boldsymbol\sigma(x)=-p\,\mathbf 1+P\,\mathbf S(x;w),
\end{equation}
where $P$ is the net-section nominal axial stress and $\mathbf S$ the unit-load field.

\begin{proposition}\label{prop:affine}
A hydrostatic shift leaves $q$ and $r_3$ unchanged. Hence $\Phi(-p\mathbf 1+t\mathbf S)$ is a polynomial in $t$ of degree at most $n$ whose coefficients follow in closed form from the binomial expansion of $(I_1(\mathbf S)\,t-3p)^{a_k}$ multiplied by $q(\mathbf S)^{m_k}r_3(\mathbf S)^{c_k}t^{m_k+3c_k}$.
\end{proposition}
The predicted first-failure load is $P_f(\boldsymbol\xi,w,p)=\min_x t^\star(x)$, with $t^\star(x)$ the first root of the polynomial of Proposition~\ref{prop:affine} at Gauss point $x$. The critical location, and with it the critical stress triaxiality, is an output of the identified surface and not an assumption. The joint objective adds $\sum_j[\ln P_f(\boldsymbol\xi,w_j,p_j)-\ln P^\mathrm{obs}_j]^2$ to the point-data term of \eqref{eq:objective}, normalised by the total number of observations. For a linear elastic pre-failure response, the unit fields $\mathbf S(x;w_j)$ are computed once per specimen geometry. Gradients with respect to $\boldsymbol\xi$ pass through the min operator (to the critical Gauss point) and through the implicit root. We use a stress-based first-failure criterion evaluated at Gauss points. Non-local refinements such as the theory of critical distances \citep{Taylor2007} would enter as an averaging operator and are not needed for the moderate notches considered here.

\subsection{Gradient-based optimal specimen design}
\label{sec:oed}
The design problem asks which notch widths make the structural tests most informative about a quantity of interest. We take the hydrostatic tensile strength $a^\mathrm{hyd}_+$, which cannot be measured directly. With Gauss--Newton (Laplace) approximation of the posterior covariance of the active coefficients $\mathbf z$ \citep{Chaloner1995},
\begin{equation}
  \mathbf F(\mathbf w)=\frac{\mathbf J_\mathrm{pts}^\top\mathbf J_\mathrm{pts}}{s_\mathrm{pts}^2}+\sum_{j}\frac{\mathbf J_j(w_j)^\top\mathbf J_j(w_j)}{s_\mathrm{str}^2},\qquad
  \operatorname{Var}[a^\mathrm{hyd}_+]\approx\mathbf g^\top\mathbf F(\mathbf w)^{-1}\mathbf g,
  \label{eq:oed}
\end{equation}
where $\mathbf J_\mathrm{pts}=\partial\ln\rho(\mathbf n_i)/\partial\mathbf z$, $\mathbf J_j=\partial\ln P_f(\cdot,w_j,p)/\partial\mathbf z$ for the pressures $p\in\{0,\,0.3,\,0.6\}$, $\mathbf g=\partial a^\mathrm{hyd}_+/\partial\mathbf z$, and $s_\mathrm{pts}=s_\mathrm{str}=0.04$ are the assumed log-scatters. Minimising \eqref{eq:oed} over $\ln w_j$ requires mixed second derivatives $\partial^2\ln P_f/\partial\mathbf z\,\partial w$ through the FE solve, the mesh mapping, the min operator and the implicit root. We compute them by reverse-over-reverse automatic differentiation and verify them against central finite differences.

\section{Results}
\label{sec:results}
\subsection{Synthetic benchmark}
\label{sec:synthetic}
\paragraph{Monomial library: the role of the constraints}
Four ground truths are considered:
T1, the von Mises cylinder;
T2, the Burzy\'nski--Yagn paraboloid with $d=1.3$ ($\{q^2,I_1\}$);
T3, a trigonal, pressure-sensitive surface $\{q^3,r_3,I_1\}$ with $d=1.25$ ($\xi=(0.724,0.100,0.176)$, $k=1.11$, $b_Z=1.01$, closed in hydrostatic tension);
and T4, the Mohr--Coulomb criterion with $d=1.5$, which is not contained in the library.
T1--T3 are convex and unique (audit: no violations) and satisfy $\nu^\mathrm{in}_+\le1/2$.
Each data set contains the eight basic tests $Z,D,K,B_Z,B_D,I_Z,U_D,A_Z$ twice, 16 random plane-stress directions and up to six random three-dimensional directions, giving $N=38$ points. Radii are perturbed by multiplicative log-normal noise of 2\% or 5\%, with three realisations each.
We compare the constrained pipeline, the same pipeline without admissibility penalties (``unconstrained''), and a purely algebraic SINDy-type baseline. The baseline ranks supports by the algebraic residual of $\sum_k\xi_km_k=1$, takes the coefficients from that least-squares solution, and selects the size by AICc on the radial residual. For the synthetic study the support size is selected by AICc ($s\le4$) instead of cross-validation to limit cost. Surface errors are RMS log-radius errors over 2000 directions, evaluated either within $30^\circ$ of a data direction (``near data'') or over all directions where both surfaces are closed.

\begin{table}[!htbp]
\centering\small
\caption{Synthetic benchmark: median surface error near the data, and number of admissible fits (non-convex and re-entrant fractions below 0.5\% on the audit set), over three noise realisations.}
\label{tab:synth}
\begin{tabular}{llcccccc}\toprule
truth & noise & \multicolumn{3}{c}{median surface error (near data)} & \multicolumn{3}{c}{admissible fits} \\
\cmidrule(lr){3-5}\cmidrule(lr){6-8}
 & & constr. & unconstr. & algebraic & constr. & unconstr. & algebraic \\ \midrule
T1 von Mises & 2\,\% & 0.002 & 0.007 & 0.001 & 3/3 & 1/3 & 3/3 \\
T1 von Mises & 5\,\% & 0.004 & 0.018 & 0.001 & 3/3 & 1/3 & 3/3 \\
T2 paraboloid & 2\,\% & 0.002 & 0.002 & 0.002 & 3/3 & 3/3 & 3/3 \\
T2 paraboloid & 5\,\% & 0.005 & 0.005 & 0.005 & 3/3 & 3/3 & 3/3 \\
T3 trigonal & 2\,\% & 0.002 & 0.003 & 0.062 & 3/3 & 2/3 & 0/3 \\
T3 trigonal & 5\,\% & 0.006 & 0.008 & 0.062 & 3/3 & 1/3 & 0/3 \\
T4 Mohr--Coulomb & 2\,\% & 0.067 & 0.117 & 0.067 & 3/3 & 0/3 & 3/3 \\
T4 Mohr--Coulomb & 5\,\% & 0.067 & 0.100 & 0.068 & 3/3 & 0/3 & 3/3 \\
\bottomrule\end{tabular}
\end{table}

\begin{figure}[!htbp]
\centering
\includegraphics[width=\linewidth]{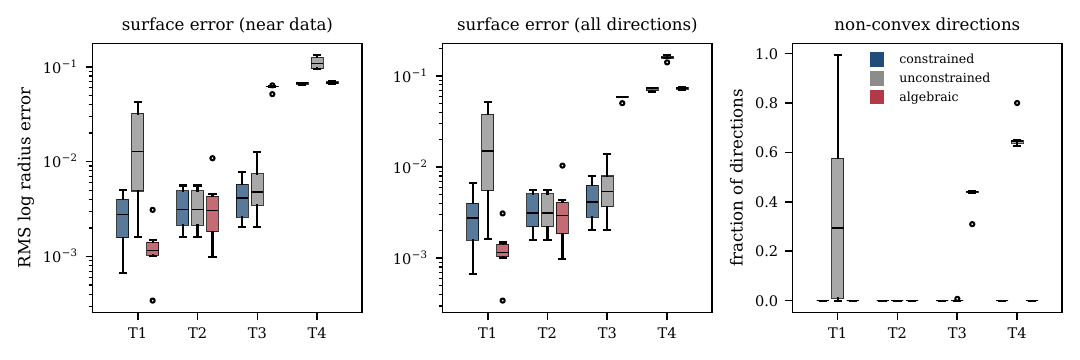}
\caption{Synthetic benchmark (all noise levels pooled): surface error near the data and over all directions, and the fraction of audit directions on which the identified surface is non-convex.}
\label{fig:synthetic}
\end{figure}

\Cref{tab:synth,fig:synthetic} summarise the results. All \SynN\ constrained fits are admissible on the audit set. Their median surface errors are of the order of the noise level divided by $\sqrt N$ for the in-library truths T1--T3. The true support is recovered for T2 and T3 in all realisations, and for T1 in five of six realisations, where $q$, $q^2$ and $q^3$ are equivalent single-term representations of the cylinder (Remark~1). The sixth realisation adds a small pressure term $I_1q$. The unconstrained variant fits the paraboloid equally well. For T1 and T3, however, it often selects additional terms that make the surface non-convex or re-entrant away from the data, so its errors are larger and only part of its fits are admissible. The algebraic baseline is accurate and admissible for T1, T2 and T4, but it misses the trigonal truth T3. Its ranking by the algebraic residual prefers a two-term support with about $6\%$ error that is non-convex on roughly 40\% of the directions. For the out-of-library truth T4 (Mohr--Coulomb), the constrained pipeline selects a Drucker--Prager-type cone, $\{q,I_1\}$, with about $7\%$ error, which is dominated by the corners of the hexagonal pyramid. The unconstrained pipeline adds terms that reduce the training residual but increase the surface error and make the surface non-convex on about two thirds of the directions. Over all runs, \SynCAdm\ of the constrained, \SynUAdm\ of the unconstrained and \SynAAdm\ of the algebraic fits are admissible.

\begin{figure}[!htbp]
\centering
\includegraphics[width=\linewidth]{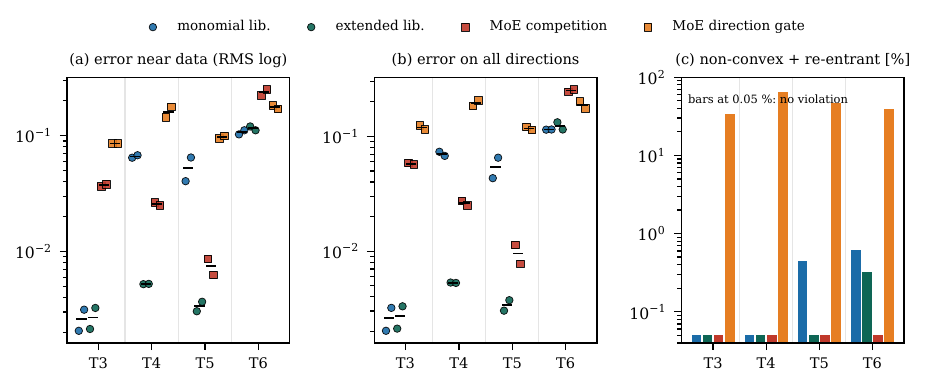}
\caption{Synthetic benchmark (two noise realisations at 2\%): surface error near the data and on all directions (markers: realisations, slightly offset for visibility; bars: median), and the mean fraction of non-convex or re-entrant audit directions. T3: trigonal (monomial library); T4: Mohr--Coulomb with $d=1.5$ (in no library); T5: PBP paraboloid (PBP atoms); T6: competition mixture of a Drucker--Prager cone and a mean-stress cap (competition MoE).}
\label{fig:synthetic_v2}
\end{figure}

\paragraph{Truths outside the monomial library}
Four truths test the extended model classes, with $2\%$ noise and two realisations each, the same design of $N=38$ directions, and AICc selection ($s\le4$, $E\le3$). The truths are T3 (trigonal, inside the monomial library), T4 (Mohr--Coulomb, $d=1.5$, which none of the libraries contains exactly), T5 (a paraboloid $0.769\,(q\Omega)^2+0.231\,I_1$ with the PBP section $\beta=0.6$, $\gamma=0.97$) and T6 (a competition mixture with $p=10$ of a Drucker--Prager cone, $(q+0.15I_1)/1.15$, and a mean-stress cap with hydrostatic strength $0.45$). \Cref{fig:synthetic_v2} compares the monomial library, the extended library, the competition MoE and the direction-gated MoE. To limit cost, the screening of the extended library in this study is exhaustive only up to $s=2$.

The extended library recovers T5 with the true support $\{I_1,(q\Omega)^2\}$ and shape parameters $(\beta,\gamma)=(0.60,0.98)$ and $(0.60,0.95)$. Its surface error is \SynTFiveExtEn, against \SynTFiveMonoEn\ for the monomial library, which needs three or four terms and remains rounded. For the Mohr--Coulomb truth T4 the extended library selects $\{I_1,q\Omega\}$ with $\gamma$ at its bound and $\beta\approx0.78$. This is the PBP representation of the irregular Mohr--Coulomb hexagon, and its error (\SynTFourExtEn) is an order of magnitude below that of the best monomial model, a Drucker--Prager cone (\SynTFourMonoEn). For T3 the extended library is as accurate as the monomial one (\SynTThreeExtEn\ vs.\ \SynTThreeMonoEn). It even finds a two-term surrogate, $\{I_1,(q\Omega)^3\}$ with $\beta=0$ and $\gamma\approx0.23$, that reproduces the trigonal section. The competition MoE is accurate whenever one expert describes the truth (T4: \SynTFourSmaxEn, T5: \SynTFiveSmaxEn). It is less accurate for T3, whose cubic meridian lies outside the expert family. All library and competition-MoE fits are admissible on the audit set up to at most $0.6\%$ of the directions. The direction-gated MoE is non-convex on 34--65\% of the directions for every truth and is never more accurate than the competition MoE.

The two-mechanism truth T6 is the instructive failure. No model class identifies it from 38 points, not even the competition MoE that contains it. That model reaches the noise level in the training residual, but its surface error near the data is \SynTSixSmaxEn. Its exponent is $p\approx27$--$32$ instead of $10$, and the deviatoric sections of its experts are far from circular. The true parameters fit the data equally well (RMS $0.021$) and give an error of $0.002$. The data thus do not determine the mixture: several pairs of experts pass through the same 38 points and differ in between. A weak prior on the deviatoric shape and the cap curvature, $10^{-2}\sum_e(\gamma_e^2+b_e^2)$, reduces the error to $0.07$ in both realisations. Mixtures with flexible experts therefore need mechanism-specific priors or designed tests near the intersections of the experts, a point we return to in \cref{sec:overall}.

\FloatBarrier
\subsection{Experimental data}
\label{sec:real}
\begin{figure}[!htbp]
\centering
\includegraphics[width=\linewidth]{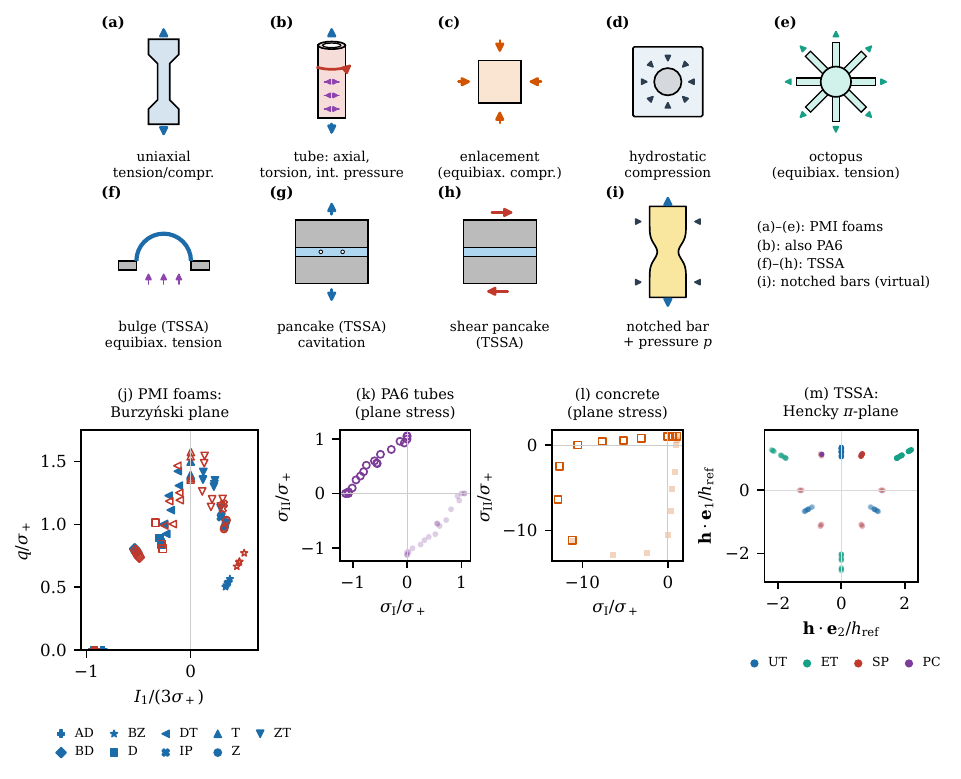}
\caption{Test specimens (schematic) and data. (a)--(e) Specimens of the PMI programme \citep[Ch.~14]{Kolupaev2018}; (b) also PA6; (f)--(h) TSSA tests; (i) virtual notched-bar tests of Section~\ref{sec:struct_results}. (j) PMI data in the Burzy\'nski plane (filled: 110IG, open: 71IG). (k, l) Plane-stress data of PA6 and concrete (mirrored about $\sigma_\mathrm I=\sigma_\mathrm{II}$, faint). (m) TSSA failure points in the Hencky $\pi$-plane with their sixfold symmetric replicas (faint); the cavitation datum $\operatorname{tr}\mathbf h/h_\mathrm{ref}=0.039$ lies outside this plane.}
\label{fig:specimens}
\end{figure}

\Cref{fig:specimens} gives an overview of the test specimens and of the stress or strain states they probe. The PMI foams Rohacell 71IG and 110IG were tested with nine specimen types (Table~\ref{tab:datasummary}): uniaxial tension and compression, thin-walled tubes under tension--torsion and internal pressure, enlacement (equibiaxial compression), hydrostatic compression, and an octopus specimen for equibiaxial tension. The PA6 and concrete data are plane-stress states from tubes and biaxially loaded plates. The TSSA data come from uniaxial tension, bulge, shear-pancake and pancake tests \citep{Drass2020coupled}.

\begin{figure}[!htbp]
\centering
\includegraphics[width=\linewidth]{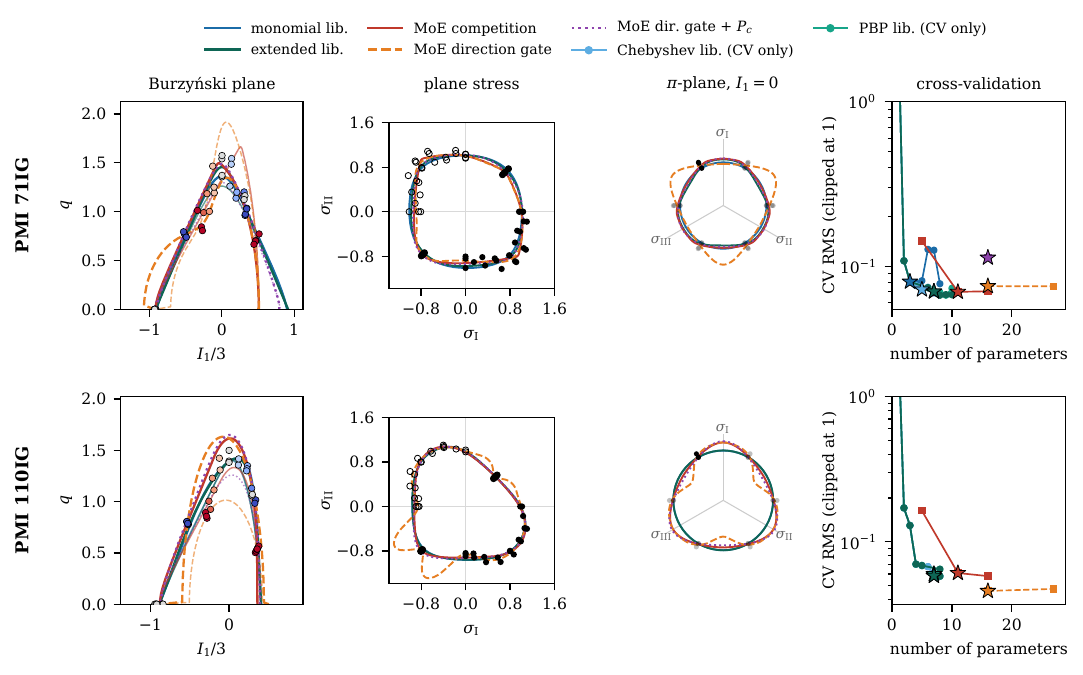}
\caption{PMI foams 71IG (top) and 110IG (bottom). From left: Burzy\'nski plane with the tension meridian (thick) and compression meridian (thin) of the selected models of each class, data coloured by Lode angle (blue $\theta=0$ to red $\theta=\pi/3$); plane-stress section (data mirrored, open symbols); $\pi$-plane at $I_1=0$ with the deviatoric projections of near-deviatoric data; training-path cross-validation errors versus number of parameters (stars: selected models).}
\label{fig:real_pmi}
\end{figure}

\begin{figure}[!htbp]
\centering
\includegraphics[width=\linewidth]{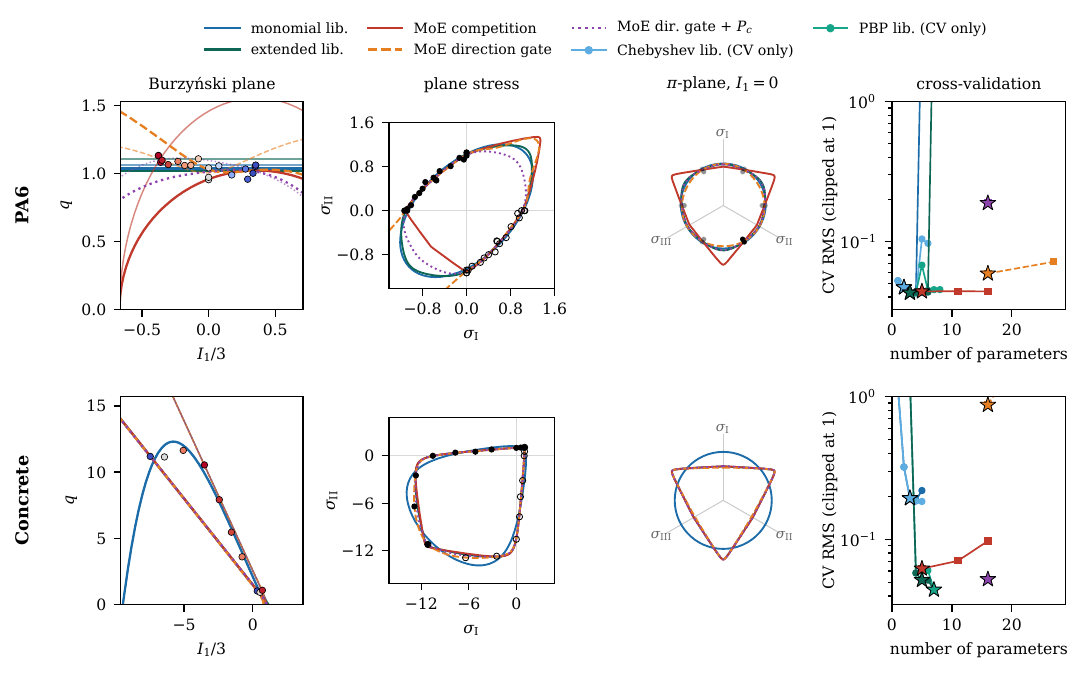}
\caption{Polyamide PA6 (top) and concrete (bottom); panels as in \cref{fig:real_pmi}.}
\label{fig:real_ps}
\end{figure}

\Cref{tab:v2overview} summarises all model classes on all data sets. \Cref{fig:real_pmi,fig:real_ps} show meridian, plane-stress and $\pi$-plane sections and the cross-validation curves of the stress-space data sets, and \cref{fig:tssa} the TSSA results. For the monomial library, \cref{tab:real} additionally compares the refined constrained models of a previous version (refitted with 1000 constraint directions) with the measured strength ratios and with Kolupaev's fits. All selected monomial supports of the current version of implementation coincide with those of the previous version.

\begin{table}[!htbp]
\centering\scriptsize
\caption{All model classes on all data sets: support size $s$ (libraries) or number of experts $E$ (MoE), number of parameters (including the PBP shape parameters and the exponent $p$), RMS radial log-residual, leakage-free cross-validated RMS, fractions of the 2000 audit directions with non-convex or re-entrant surface, and key extrapolations. Libraries and the penalised direction gate are fitted with the admissibility penalties; the competition MoE is admissible by construction.}
\label{tab:v2overview}
\setlength{\tabcolsep}{3pt}
\resizebox{\linewidth}{!}{\begin{tabular}{llrrrrrrl}
\toprule
Data set & Model & $s$/$E$ & $n_\mathrm{par}$ & RMS & CV & non-conv.\,[\%] & re-entr.\,[\%] & key extrapolations \\
\midrule
PMI 71IG & monomial lib. & 3 & 3 & 0.080 & 0.081 & 0.0 & 0.0 & $d$=1.00, $b_Z$=0.74, $a^\mathrm{hyd}_+$=0.89 \\
 & Chebyshev lib. & 5 & 5 & 0.070 & 0.072 & 1.1 & 0.0 & $d$=0.94, $b_Z$=0.71, $a^\mathrm{hyd}_+$=0.87 \\
 & PBP lib. & 5 & 7 & 0.069 & 0.070 & 0.2 & 0.0 & $d$=0.96, $b_Z$=0.72, $a^\mathrm{hyd}_+$=0.90 \\
 & extended lib. & 5 & 7 & 0.069 & 0.070 & 0.2 & 0.0 & $d$=0.96, $b_Z$=0.72, $a^\mathrm{hyd}_+$=0.90 \\
 & MoE competition & 2 & 11 & 0.059 & 0.070 & 0.0 & 0.0 & $d$=0.92, $b_Z$=0.71, $a^\mathrm{hyd}_+$=0.51 \\
 & MoE direction gate & 2 & 16 & 0.056 & 0.076 & 47.4 & 0.0 & $d$=0.88, $b_Z$=0.71, $a^\mathrm{hyd}_+$=0.50 \\
 & MoE dir. gate + $P_c$ & 2 & 16 & 0.063 & 0.113 & 3.3 & 0.0 & $d$=0.91, $b_Z$=0.70, $a^\mathrm{hyd}_+$=0.77 \\
\midrule
PMI 110IG & monomial lib. & 7 & 7 & 0.054 & 0.060 & 0.8 & 1.0 & $d$=0.98, $b_Z$=0.55, $a^\mathrm{hyd}_+$=0.42 \\
 & Chebyshev lib. & 7 & 7 & 0.051 & 0.058 & 1.2 & 0.0 & $d$=0.97, $b_Z$=0.54, $a^\mathrm{hyd}_+$=0.41 \\
 & PBP lib. & 7 & 7 & 0.054 & 0.060 & 0.8 & 1.0 & $d$=0.98, $b_Z$=0.55, $a^\mathrm{hyd}_+$=0.42 \\
 & extended lib. & 7 & 7 & 0.051 & 0.058 & 1.2 & 0.0 & $d$=0.97, $b_Z$=0.54, $a^\mathrm{hyd}_+$=0.41 \\
 & MoE competition & 2 & 11 & 0.045 & 0.061 & 0.0 & 0.0 & $d$=0.95, $b_Z$=0.55, $a^\mathrm{hyd}_+$=0.37 \\
 & MoE direction gate & 2 & 16 & 0.035 & 0.046 & 39.0 & 0.0 & $d$=0.86, $b_Z$=0.53, $a^\mathrm{hyd}_+$=0.50 \\
 & MoE dir. gate + $P_c$ & 2 & 16 & 0.048 & 1.160 & 3.8 & 0.0 & $d$=0.95, $b_Z$=0.55, $a^\mathrm{hyd}_+$=0.40 \\
\midrule
PA6 & monomial lib. & 2 & 2 & 0.045 & 0.047 & 0.0 & 0.1 & $d$=1.02, $b_Z$=1.02, $a^\mathrm{hyd}_+$=$\infty$ \\
 & Chebyshev lib. & 2 & 2 & 0.045 & 0.047 & 0.0 & 0.1 & $d$=1.02, $b_Z$=1.02, $a^\mathrm{hyd}_+$=$\infty$ \\
 & PBP lib. & 1 & 3 & 0.036 & 0.043 & 0.0 & 0.0 & $d$=1.09, $b_Z$=1.09, $a^\mathrm{hyd}_+$=$\infty$ \\
 & extended lib. & 1 & 3 & 0.036 & 0.043 & 0.0 & 0.0 & $d$=1.09, $b_Z$=1.09, $a^\mathrm{hyd}_+$=$\infty$ \\
 & MoE competition & 1 & 5 & 0.036 & 0.044 & 0.0 & 0.0 & $d$=1.08, $b_Z$=1.30, $a^\mathrm{hyd}_+$=1.36 \\
 & MoE direction gate & 2 & 16 & 0.035 & 0.059 & 71.1 & 0.0 & $d$=1.08, $b_Z$=1.25, $a^\mathrm{hyd}_+$=1.29 \\
 & MoE dir. gate + $P_c$ & 2 & 16 & 0.036 & 0.188 & 0.8 & 0.0 & $d$=1.08, $b_Z$=0.89, $a^\mathrm{hyd}_+$=1.10 \\
\midrule
Concrete & monomial lib. & 3 & 3 & 0.137 & 0.194 & 0.0 & 0.0 & $d$=7.45, $b_Z$=0.69, $a^\mathrm{hyd}_+$=0.76 \\
 & Chebyshev lib. & 3 & 3 & 0.137 & 0.194 & 0.0 & 0.0 & $d$=7.45, $b_Z$=0.69, $a^\mathrm{hyd}_+$=0.76 \\
 & PBP lib. & 5 & 7 & 0.033 & 0.044 & 0.0 & 0.0 & $d$=10.12, $b_Z$=0.99, $a^\mathrm{hyd}_+$=1.10 \\
 & extended lib. & 3 & 5 & 0.035 & 0.052 & 0.0 & 0.0 & $d$=10.21, $b_Z$=0.99, $a^\mathrm{hyd}_+$=1.10 \\
 & MoE competition & 1 & 5 & 0.038 & 0.062 & 0.0 & 0.0 & $d$=10.20, $b_Z$=0.98, $a^\mathrm{hyd}_+$=0.97 \\
 & MoE direction gate & 2 & 16 & 0.029 & 0.874 & 64.9 & 0.0 & $d$=10.44, $b_Z$=1.06, $a^\mathrm{hyd}_+$=0.84 \\
 & MoE dir. gate + $P_c$ & 2 & 16 & 0.035 & 0.052 & 2.6 & 0.0 & $d$=10.16, $b_Z$=0.98, $a^\mathrm{hyd}_+$=0.97 \\
\midrule
TSSA & monomial lib. & 3 & 3 & 0.093 & 0.094 & 1.0 & 0.0 & $e_\mathrm{ET}$=1.70, $e_\mathrm{PS}$=1.11, vol.\ compr.\ closed \\
 & Chebyshev lib. & 3 & 3 & 0.093 & 0.094 & 1.0 & 0.0 & $e_\mathrm{ET}$=1.70, $e_\mathrm{PS}$=1.11, vol.\ compr.\ closed \\
 & PBP lib. & 2 & 4 & 0.081 & 0.086 & 0.0 & 0.0 & $e_\mathrm{ET}$=1.87, $e_\mathrm{PS}$=1.15, vol.\ compr.\ open \\
 & extended lib. & 2 & 4 & 0.081 & 0.086 & 0.0 & 0.0 & $e_\mathrm{ET}$=1.87, $e_\mathrm{PS}$=1.15, vol.\ compr.\ open \\
 & MoE competition & 1 & 5 & 0.085 & 0.089 & 0.0 & 0.0 & $e_\mathrm{ET}$=1.87, $e_\mathrm{PS}$=1.15, vol.\ compr.\ open \\
 & MoE direction gate & 2 & 16 & 0.079 & 0.086 & 36.7 & 0.0 & $e_\mathrm{ET}$=1.84, $e_\mathrm{PS}$=1.06, vol.\ compr.\ open \\
 & MoE dir. gate + $P_c$ & 2 & 16 & 0.085 & 0.088 & 0.0 & 0.0 & $e_\mathrm{ET}$=1.87, $e_\mathrm{PS}$=1.15, vol.\ compr.\ open \\
\bottomrule
\end{tabular}}
\end{table}

\begin{table}[!htbp]
\centering\scriptsize
\caption{Monomial library, refined constrained models (previous version selection, refitted with 1000 constraint directions and penalty factors up to $10^3$; numerically audited on 2000 directions, not certified on the continuum). Characteristic values are normalised by the model's $\sigma_+$; brackets give 90\% bootstrap intervals conditional on the selected support (model-form uncertainty is not included); $\infty$ marks an open surface; $\dagger n$ marks the percentage of bootstrap models that are open in that direction. ``Measured'' values are computed directly from the data; Kolupaev's fits are from \citet[Ch.~14]{Kolupaev2018}. The CV row gives the leakage-free values of the current version.}
\label{tab:real}
\setlength{\tabcolsep}{3pt}
\resizebox{\linewidth}{!}{\begin{tabular}{lcccc}\toprule
 & PMI 71IG & PMI 110IG & PA6 & Concrete \\ \midrule
$N$ & 46 & 44 & 19 & 10 \\
$s^\star$ (selected terms) & 3 & 7 & 2 & 3 \\
terms & \scriptsize $q,\,I_1^{2},\,I_1^{2} q^{2}$ & \scriptsize $q,\,I_1,\,I_1 q,\,I_1^{2},\,I_1^{4} q,\,I_1^{2} r_3,\,I_1^{2} q^{4}$ & \scriptsize $q,\,q r_3$ & \scriptsize $q,\,I_1,\,I_1^{4}$ \\
RMS radial log-residual & 0.080 & 0.054 & 0.045 & 0.137 \\
CV RMS (leakage-free, v2) & 0.081 & 0.060 & 0.047 & 0.194 \\
model $\sigma_+/\sigma_\mathrm{ref}$ & 1.01 & 0.98 & 1.03 & 1.18 \\
\midrule
$d$ & 1.00 {\scriptsize [1.00, 1.00]} & 0.98 {\scriptsize [0.95, 1.01]} & 1.02 {\scriptsize [1.02, 1.02]} & 7.45 {\scriptsize [5.27, 9.81]} \\
$k$ & 1.35 {\scriptsize [1.33, 1.36]} & 1.42 {\scriptsize [1.37, 1.46]} & 1.01 {\scriptsize [1.01, 1.01]} & 1.79 {\scriptsize [1.68, 1.85]} \\
$b_Z$ & 0.74 {\scriptsize [0.74, 0.75]} & 0.55 {\scriptsize [0.55, 0.56]} & 1.02 {\scriptsize [1.02, 1.02]} & 0.69 {\scriptsize [0.69, 0.71]} \\
$b_D$ & 0.74 {\scriptsize [0.74, 0.75]} & 0.77 {\scriptsize [0.75, 0.79]} & 1.00 {\scriptsize [1.00, 1.00]} & 9.19 {\scriptsize [8.20, 25.96]} \\
$a^\mathrm{hyd}_+$ & 0.89 {\scriptsize [0.87, 0.91]} & 0.41 {\scriptsize [0.41, 0.43]} & $\infty$ {\scriptsize (open)} & 0.76 {\scriptsize [0.73, 0.82]} \\
$a^\mathrm{hyd}_-$ & 0.89 {\scriptsize [0.87, 0.91]} & 0.91 {\scriptsize [0.88, 0.92]} & $\infty$ {\scriptsize (open)} & 8.03 {\scriptsize [7.30, 10.10]}{\scriptsize$^{\dagger10}$} \\
$\nu^\mathrm{in}_+$ & -0.02 {\scriptsize [-0.03, -0.01]} & -0.58 {\scriptsize [-0.61, -0.53]} & 0.50 {\scriptsize [0.50, 0.50]} & -0.16 {\scriptsize [-0.19, -0.11]} \\
\midrule
measured $d$ & 0.89 & 0.86 & 1.07 & 10.54 \\
measured $k$ & 1.46 & 1.43 & 0.95 & -- \\
measured $b_Z$ & 0.71 & 0.53 & -- & -- \\
measured $b_D$ & 0.77 & 0.79 & -- & 11.19 \\
measured $a^\mathrm{hyd}_-$ & 0.92 & 0.89 & -- & -- \\
\midrule
Kolupaev fit $d$ & 0.98 & 1.10 & 1.10 & 10.96 \\
Kolupaev fit $k$ & 1.28 & 1.41 & 1.04 & 1.62 \\
Kolupaev fit $b_Z$ & 0.63 & 0.53 & 1.10 & 1.00 \\
Kolupaev fit $b_D$ & 0.78 & 0.72 & 1.00 & 11.19 \\
Kolupaev fit $a^\mathrm{hyd}_+$ & 0.47 & 0.35 & -- & 1.10 \\
\midrule
audit: non-convex / re-entrant & 0.0\,\% / 0.0\,\% & 0.0\,\% / 0.1\,\% & 0.0\,\% / 0.0\,\% & 0.0\,\% / 0.0\,\% \\
unconstr.: $s^\star$, CV RMS & 4, 0.067 & 5, 0.044 & 2, 0.043 & 3, 0.117 \\
unconstr.: non-convex / re-entrant & 73.9\,\% / 0.0\,\% & 45.1\,\% / 3.9\,\% & 0.0\,\% / 43.2\,\% & 81.2\,\% / 36.0\,\% \\
\bottomrule\end{tabular}}
\end{table}

\FloatBarrier
\paragraph{PMI foams: monomial library}
\emph{Data correction for 110IG.} The tabulated equibiaxial tensile strengths of 110IG in \citet[Table~14.10]{Kolupaev2018} ($2.76$--$3.11$\,MPa) are about twice the values that follow from the tabulated forces with the book's own force--stress relation, Eq.~(14.4), and twice the values plotted in the book's Figs.~14.33--14.34. The 71IG values are consistent. We use one half of the tabulated 110IG values (details in \cref{app:data}). The correction matters: with the uncorrected values the selection returns a six-term model with $b_Z=0.97$ and $a^\mathrm{hyd}_+=1.16$, almost three times the value obtained below, and a larger CV error ($0.068$ against $0.059$, both with the warm-started cross-validation of a previous version of this approach). We detected the inconsistency while cross-checking our fits against the book's figures. It should be confirmed with the original test records.

\begin{figure}[!htbp]
\centering
\includegraphics[width=\linewidth]{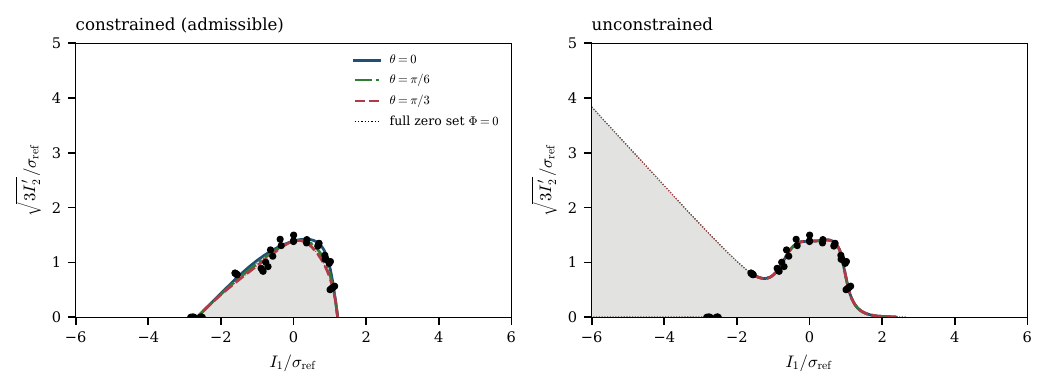}
\caption{PMI 110IG: complete zero set $\Phi=0$ (dotted) and first-root surface on the meridian planes $\theta=0,\pi/6,\pi/3$ for the refined constrained model (left) and the unconstrained model (right); shaded: $\Phi<0$. The unconstrained model is non-convex, open in hydrostatic compression, and reaches the hydrostatic compression data (on the axis) only through a needle-like feature that is not visible at this scale.}
\label{fig:branches}
\end{figure}

\begin{figure}[!htbp]
\centering
\includegraphics[width=\linewidth]{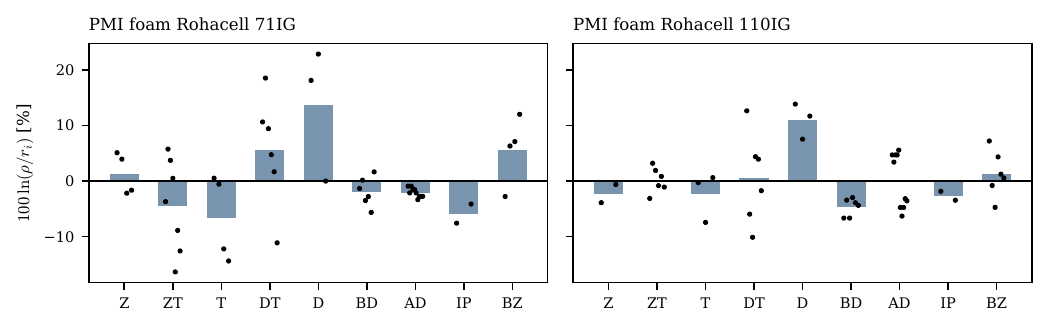}
\caption{PMI foams: radial log-residuals of the refined constrained models grouped by test type (bars: mean; dots: individual tests). Positive values mean that the model overestimates the strength.}
\label{fig:residuals}
\end{figure}

\begin{figure}[!htbp]
\centering
\includegraphics[width=\linewidth]{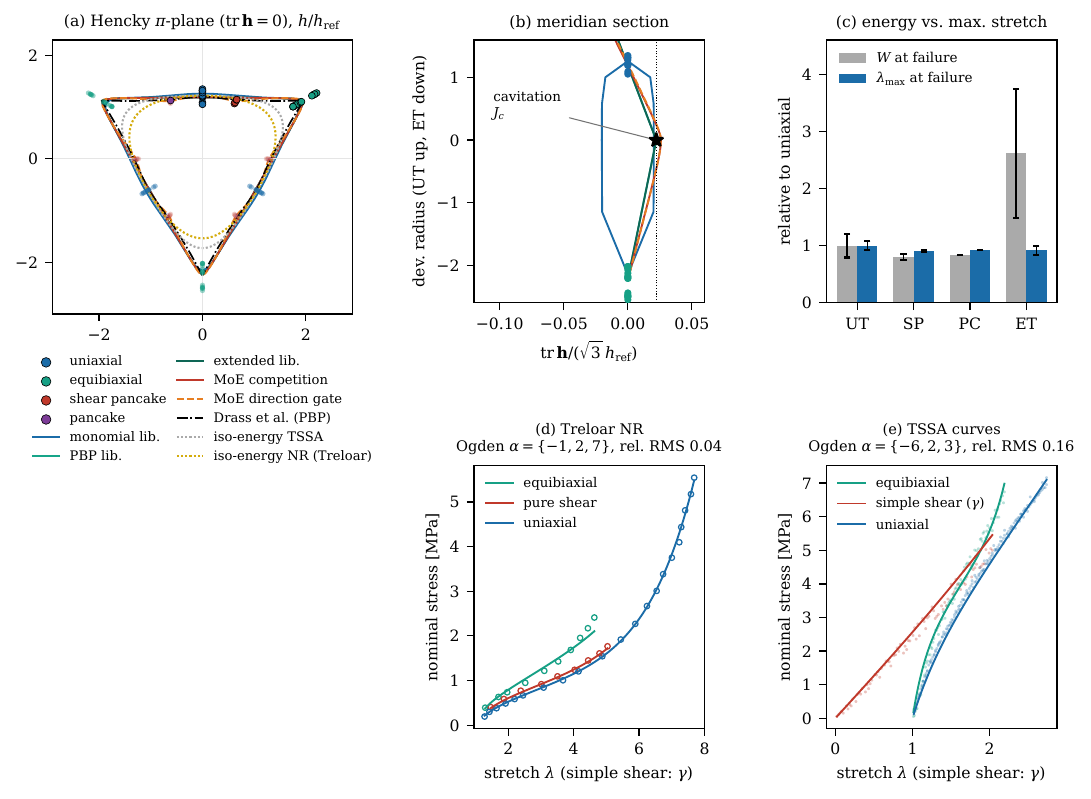}
\caption{TSSA. (a) Hencky $\pi$-plane: failure points (sixfold symmetric replicas faint), models of all classes, the PBP criterion of \citet{Drass2020coupled} mapped from the stretch $\pi$-plane, and iso-energy contours through the uniaxial failure point for TSSA and Treloar's natural rubber. (b) Meridian section through the uniaxial (up) and equibiaxial (down) directions with the cavitation datum (star); the monomial model closes the surface in volumetric compression. (c) Stored energy and largest principal stretch at failure relative to uniaxial tension (bars: coefficient of variation). (d, e) Sparse Ogden fits to Treloar's data and to the TSSA curves.}
\label{fig:tssa}
\end{figure}

For Rohacell 110IG the selected model has \PMIbSsel\ terms with an RMS radial residual of \PMIbRmse\ (leakage-free CV \VPMIbMonoCV). Its characteristic values $d=\PMIbd$, $k=\PMIbk$, $b_Z=\PMIbbZ$, $b_D=\PMIbbD$ and $a^\mathrm{hyd}_-=\PMIbahminus$ are close to the measured ratios and to Kolupaev's fits with the geometrical-mechanical criterion (\cref{tab:real}). The extrapolated hydrostatic tensile strength is $a^\mathrm{hyd}_+=\PMIbahplus$ with a 90\% bootstrap interval of $[\PMIbahplusLo,\PMIbahplusHi]$; Kolupaev obtained $0.35$--$0.36$. Grouped by test type (\cref{fig:residuals}), the residuals show that the model lies on average \PMIbResD\,\% above the three uniaxial compression tests, while the hydrostatic compression and equibiaxial compression series are matched within a few per cent. This supports Kolupaev's remark that the $D$ measurements on PMI do not fit the remaining data and should be repeated \citep[Sect.~14.5.2]{Kolupaev2018}. The hydrostatic series is visibly bimodal: specimens with a protective layer failed at pressures about 10\,\% higher than those without (Table~14.7 of \citealp{Kolupaev2018}), and the model passes between the two groups. The unconstrained monomial pipeline of an earlier version achieves a CV error of \PMIbUnconvCV\ (with warm-started folds) but is non-convex on \PMIbUnconvNC\ and re-entrant on \PMIbUnconvRE\ of the audit directions. \Cref{fig:branches} shows why. The unconstrained surface has a concave dent between $D$ and $B_D$ and is open in hydrostatic compression. It matches the ten hydrostatic compression tests only through a needle on the hydrostatic axis: exactly on the axis its radius is $1.53$, but already $0.5^\circ$ off the axis no root exists within the validity region. The low residual of the unconstrained model is thus bought with a physically meaningless feature, which the audit detects and visual inspection of meridian sections would miss. The constrained surface is closed and convex and passes through the same data with a residual of a few per cent. For 71IG the selected model has \PMIaSsel\ terms (RMS \PMIaRmse, CV \VPMIaMonoCV). This model is even in $I_1$, so $d=\PMIad$ by construction, whereas the measured ratio is $0.89$; the uniaxial compression tests scatter strongly (1.44--1.81\,MPa) and lie well below the model (\cref{fig:residuals}). The model captures the high torsional strength ($k=\PMIak$, measured $1.46$) and closes the surface in hydrostatic tension at $a^\mathrm{hyd}_+=\PMIaahplus$. Cross-validation does not support an additional odd term in $I_1$ (\cref{fig:real_pmi}), so the monomial library cannot resolve the tension--compression asymmetry of 71IG; the Lode terms of the extended libraries can (see below).

\paragraph{PMI foams: extended libraries and mixtures}
For 110IG the extended libraries add little. The Chebyshev and extended libraries replace $I_1^2r_3$ by $I_1^2q^2\cos3\theta$ (RMS \VPMIbExtRmse, CV \VPMIbExtCV\ against \VPMIbMonoCV), and all characteristic values change by at most $0.02$. For 71IG the Lode terms matter. A previous version of the approach had to leave the tension--compression asymmetry unresolved, because the selected model is even in $I_1$ and $d\equiv1$. The Chebyshev library adds $q\cos3\theta$ and $q\cos6\theta$, and the PBP library adds two atoms with $\beta=\VPMIaPbpbeta$, $\gamma=\VPMIaPbpgamma$. Both lower the CV error from \VPMIaMonoCV\ to \VPMIaChebCV\ and \VPMIaPbpCV, respectively, and yield $d=\VPMIaPbpd$ (measured $0.89$). The asymmetry is thus produced by the Lode dependence: the compression meridian lies inside the tension meridian although the model is even in $I_1$ (\cref{fig:real_pmi}).

\begin{figure}[!htbp]
\centering
\includegraphics[width=\linewidth]{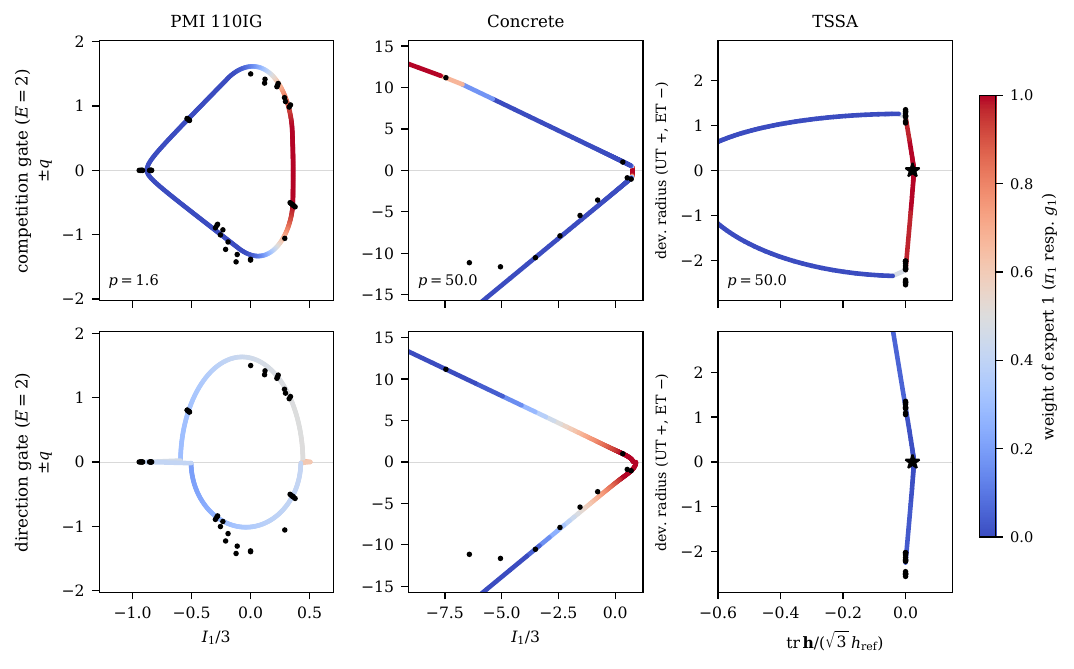}
\caption{Gates of two-expert models in meridian sections ($+$: tension meridian $\theta=0$, $-$: compression meridian $\theta=\pi/3$; for TSSA uniaxial up, equibiaxial down). Colour: responsibility $\pi_1$ (competition gate, top) or gate weight $g_1$ (direction gate, bottom) of expert~1. Black: data, plotted on the tension side for $\theta<\pi/6$ and on the compression side otherwise. For concrete and TSSA, cross-validation selects one expert; the two-expert models are shown for comparison.}
\label{fig:gates}
\end{figure}

The competition MoE needs two experts for both foams. A single expert, a closed shifted ellipsoid, misses the data (CV \VPMIbSmaxCV\ with $E=2$ against $0.16$ with $E=1$ for 110IG), and the hard-concrete relaxation keeps two to three experts for $\lambda_0\le3\cdot10^{-3}$ and collapses to one with a large error for $\lambda_0=10^{-2}$ (Table~\ref{tab:v2sparsity}). The two 110IG experts have a clear physical reading (\cref{fig:gates}). Expert~1 is dominated by the mean stress ($c_1=0.61$) with a small Rankine-type deviatoric part ($\beta_1\approx0$). It is responsible for the flat tensile cut-off between $Z$, $B_Z$ and $A_Z$, i.e.\ for tensile fracture of the cell walls. Expert~2 is a cone that opens towards compression ($c_2=-0.35$) with an inverted-triangle section ($\beta_2=2$). It carries uniaxial, biaxial and hydrostatic compression, i.e.\ cell collapse. Its prediction of the hydrostatic tensile strength, $a^\mathrm{hyd}_+=\VPMIbSmaxahp$ with the bootstrap interval $[\VPMIbSmaxahpLo,\VPMIbSmaxahpHi]$, does \emph{not} overlap with the interval of the monomial model, $[\VPMIbMonoahpLo,\VPMIbMonoahpHi]$. Both models are admissible and cross-validate equally well. Kolupaev's fits gave $0.35$--$0.36$. The fitted exponent is small ($p=\VPMIbSmaxp$), so the transition between the mechanisms is broad. In the competing-risk reading this corresponds to a Weibull modulus far below the one estimated from the scatter of replicate tests (Section~\ref{sec:overall}).

The direction-gated MoE attains the lowest CV error for 110IG (\VPMIbGateCV), but it is non-convex on \VPMIbGateNC\,\% of the directions. Like the unconstrained monomial model of \cref{fig:branches}, it reaches the hydrostatic compression data only through a narrow feature on the axis (its meridian section in \cref{fig:gates} is open there). With the convexity penalty it becomes nearly convex (\VPMIbGatePenNC\,\%), but its cross-validation breaks down (CV \VPMIbGatePenCV) because single folds converge to degenerate gates. Its weights are diffuse (0.3--0.6 for both experts in \cref{fig:gates}), so they offer no mechanism reading.

\paragraph{Polyamide PA6}
The PA6 data cover only the plane-stress region $Z$--$K$--$D$ of thin-walled tubes. The monomial library selects $\{q,\ q\,r_3\}$ (RMS \VPAsixMonoRmse, CV \VPAsixMonoCV); the Chebyshev terms are not used. A single PBP atom $q\Omega$ with $\beta=\VPAsixPbpbeta$, $\gamma=\VPAsixPbpgamma$ fits better (RMS \VPAsixPbpRmse, CV \VPAsixPbpCV) with three parameters. This is a pressure-insensitive criterion with a mildly hexagonal section, $d=\VPAsixPbpd$ and $k=\VPAsixPbpk$ (measured $1.07$ and $0.95$). The competition MoE selects one expert and reaches the same accuracy (CV \VPAsixSmaxCV). It closes the surface, however, along the hydrostatic axis and predicts $b_Z=\VPAsixSmaxbZ$, $b_D=\VPAsixSmaxbD$ and $a^\mathrm{hyd}_+=\VPAsixSmaxahp$. Both libraries leave the surface open in hydrostatic tension and compression with $b_Z\approx b_D\approx1$. None of these values is constrained by tube tests. The spread between model classes, from an open surface to $a^\mathrm{hyd}_+=1.4$ and from $b_D=0.6$ to $1.0$, shows how far the extrapolation of plane-stress tube data is a property of the model class rather than of the material. The direction gate reaches the lowest training error but a higher CV error than every library and the competition MoE (\VPAsixGateCV) and is non-convex on \VPAsixGateNC\,\% of the directions. With the convexity penalty it becomes almost convex but generalises even worse (CV \VPAsixGatePenCV).

\paragraph{Concrete}
The ten plane-stress points of \citet{Tasuji1978} span strengths from $1$ (tension) to about $13$ (biaxial compression). With leave-one-out cross-validation the monomial model is unchanged: $\{q,I_1,I_1^4\}$, RMS \VConcMonoRmse, CV \VConcMonoCV, $d=\VConcMonod$ against the measured $10.54$. The PBP atom removes this limitation. Already the two-term model $\{I_1,\ q\Omega\}$ reaches an RMS of $0.036$ (CV $0.058$) with $\beta\approx0.1$ and $\gamma$ at its bound. This is a cone with a nearly Rankine-type triangular section, close to the Mohr--Coulomb and Rankine-type forms that Kolupaev uses for concrete \citep[Ch.~10--11]{Kolupaev2018}. The selected five-term PBP model (RMS \VConcPbpRmse, CV \VConcPbpCV) gives $d=\VConcPbpd$, $b_D=\VConcPbpbD$, $k=\VConcPbpk$, $b_Z=\VConcPbpbZ$ and $a^\mathrm{hyd}_+=\VConcPbpahp$. These values agree with the measured ratios ($10.54$, $11.19$) and with Kolupaev's fits ($d=10.96$, $k=1.62$, $b_Z=1.00$, $b_D=11.19$, $a^\mathrm{hyd}_+=1.10$; \cref{tab:real}) to within a few per cent. The surface is open in hydrostatic compression, as the data neither require nor exclude a cap. The competition MoE selects a single expert of the same kind, a cone with $\beta=0.10$ and $\gamma=0.99$ (RMS \VConcSmaxRmse, CV \VConcSmaxCV). It predicts a slightly lower hydrostatic tensile strength ($a^\mathrm{hyd}_+=\VConcSmaxahp$, bootstrap $[\VConcSmaxahpLo,\VConcSmaxahpHi]$ against $[\VConcExtahpLo,\VConcExtahpHi]$ for the extended library). The decisive feature for concrete is therefore the sharp, triangular deviatoric section, not the meridian. With ten points the direction-gated MoE overfits catastrophically (CV \VConcGateCV) and is non-convex on \VConcGateNC\,\% of the directions; the convexity penalty restores a reasonable model (CV \VConcGatePenCV).

\FloatBarrier
\paragraph{Structural silicone TSSA in Hencky space}
\label{sec:tssa_res}
The TSSA data comprise 32 deviatoric failure points and the cavitation datum (Section~\ref{sec:hencky}, \cref{fig:specimens}m). \Cref{fig:tssa} shows the models of all classes in the Hencky $\pi$-plane and in the meridian section through the uniaxial and equibiaxial directions. The deviatoric section is a rounded triangle with flat sides normal to the uniaxial directions. The equibiaxial failure strain lies $\approx1.84$ times further from the origin than the uniaxial one, and the planar points (shear pancake and pancake) lie at $1.05$. The PBP and extended libraries select the two-term model $\{I_1,\ q\Omega\}$ with $\beta=\VTSSAPbpbeta$ and $\gamma=\VTSSAPbpgamma$ (at its bound), RMS \VTSSAPbpRmse\ and CV \VTSSAPbpCV. For $\beta=0$ and $\gamma\to1$, $q\Omega$ is proportional to the largest principal deviatoric strain. The discovered surface is therefore a \emph{maximum-principal-stretch criterion with a volumetric (cavitation) term},
\begin{equation}
  \kappa_1\,\bigl(h_\mathrm I-\tfrac13\operatorname{tr}\mathbf h\bigr)+\kappa_2\operatorname{tr}\mathbf h=1,
  \label{eq:tssa_disc}
\end{equation}
with $\kappa_1$ and $\kappa_2$ following from the coefficients in Table~\ref{tab:v2terms}. On incompressible paths it predicts failure at a constant largest principal stretch. Its equibiaxial ratio $e_\mathrm{ET}=\VTSSAPbpeET$ is close to the measured $1.84$. The planar ratio $e_\mathrm{PS}=\VTSSAPbpePS$ lies about 9\% above the measurements, which fail somewhat earlier than a pure maximum-stretch criterion predicts. This is a systematic Lode correction that the 32 points support only weakly. The residual scatter is of the size of the replicate scatter (coefficients of variation of $8\%$ for the uniaxial and $9\%$ for the equibiaxial radii). The monomial (and identical Chebyshev) model $\{I_1^2,q^6,q^3r_3\}$ fits worse (RMS \VTSSAMonoRmse, CV \VTSSAMonoCV), needs a sixth-degree polynomial to approximate the triangle, is non-convex on \VTSSAMonoNC\,\% of the directions, and, because $I_1^2$ is even, closes the surface in volumetric compression at the same distance as the cavitation point. The latter is physically meaningless. The Drass et al.\ PBP criterion, fitted in the stretch $\pi$-plane and mapped to Hencky space, is almost indistinguishable from the discovered surface in the deviatoric plane (\cref{fig:tssa}a). The same data are thus a nearly Tresca-shaped hexagon in the stretch $\pi$-plane and a nearly Rankine-shaped triangle in Hencky space; the logarithmic space makes the maximum-stretch structure explicit.

The competition MoE selects a single expert (all $E$ have the same CV error, \VTSSASmaxCV), $\sqrt{a(q\Omega)^2+b\,(\operatorname{tr}\mathbf h)^2}+c\operatorname{tr}\mathbf h$ with $\beta\approx0$ and $\gamma$ at its bound, i.e.\ the same maximum-stretch section with a volumetric cone. With two experts (\cref{fig:gates}), one expert takes the cavitation corner and the other the distortional part. The gate then reads as the share of cavitation versus distortional failure, but the data cannot resolve the intersection, and the hard-concrete $L_0$ relaxation keeps one expert for all penalty weights (Table~\ref{tab:v2sparsity}). The direction-gated MoE fits slightly better (RMS \VTSSAGateRmse), is non-convex on \VTSSAGateNC\,\% of the directions, and is not more accurate in cross-validation. All admissible models agree on the deviatoric section to within the scatter. Where they disagree is the volumetric side, where only one model-based datum exists.

\paragraph{Energetic reading of TSSA}
The Ogden fits (\cref{fig:tssa}d,e) select $\alpha=\{\TrelAlphas\}$ for Treloar's natural rubber (relative RMS \TrelRelRms\,\%) and $\alpha=\{\TSSAhypAlphas\}$ for TSSA (\TSSAhypRelRms\,\%; the larger error reflects specimen scatter, Mullins softening and the strain-rate history of the curves). At the failure points the stored energy of TSSA differs strongly between test types: the equibiaxial value is $\TSSAWratio$ times the planar and $\TSSAWratioUT$ times the uniaxial one. The largest principal stretch varies only between \TSSAlmaxMin\ and \TSSAlmaxMax\ (\cref{fig:tssa}c). The iso-energy contours through the uniaxial failure point, for TSSA and for Treloar's rubber alike, are much rounder than the data and underestimate the equibiaxial failure strain by $23\%$ (TSSA) and $32\%$ (natural rubber) (\cref{fig:tssa}a). A critical stored energy is therefore not a consistent failure criterion for TSSA, whereas a critical largest principal stretch nearly is. This agrees with the molecular picture of failure by the limited extensibility of the network chains that are aligned with the largest principal stretch. The equibiaxial energies rely on an extrapolation of the Ogden model beyond the largest stretch of the bulge curves ($2.2$ against $2.6$ at failure). Any further stiffening would increase them, so the conclusion is conservative.

\FloatBarrier
\subsection{Structural identification from notched bars}
\label{sec:struct_results}
To test end-to-end identification, we take as ground truth the admissible four-term model from the 110IG selection path ($s=4$, RMS residual \StrTruthRmse\ on the real data, no audit violations, $a^\mathrm{hyd}_+=\StrTruthAh$). It is more compact than the selected seven-term model, which keeps the identification well conditioned. Nine notched-bar tests are simulated with notch widths $w\in\{1.5,3,6\}$ (root radii $0.56$, $2.25$, $9$) and superposed pressures $p/\sigma_\mathrm{ref}\in\{0,0.3,0.6\}$. Their first-failure loads are perturbed with 4\% log-normal noise. The point data are regenerated in the directions of the 44 real 110IG tests with 4\% noise. \Cref{tab:struct} lists the failure loads and the critical stress states of the truth. The critical triaxiality ranges from \StrTriaxMin\ to \StrTriaxMax, and $I_1/\sigma_\mathrm{ref}$ reaches \StrIoneMax. The notched bars thus probe the tensile, triaxial region between $Z$ and $A_Z$. Among the homogeneous tests, only the equibiaxial tension test reaches this region. In all nine tests the critical Gauss point lies at the notch root, although the triaxiality of the unit-load field is highest at the centre of the net section (\cref{fig:structural}a).

On the support of the truth, three identifications are compared: points only; points plus notched bars mapped to homogeneous net-section states $(P_f-p,-p,-p)$ (``naive''); and points plus notched bars through the differentiable FE model (``end-to-end''). Two point-data scenarios are considered. In scenario~A all 44 test directions of the 110IG programme are used (\StrNseed\ noise realisations). In scenario~B only the 20 tubular plane-stress tests ($Z$, $ZT$, $T$, $DT$, $D$) are used (\StrTubeN\ realisations), which is the typical programme of a tension--torsion machine without biaxial, hydrostatic or pressure tests.

In scenario~A the point data alone already determine the four coefficients well: the median surface error in the half-space $I_1>0$ is \StrPErrT, and the RMS error of $a^\mathrm{hyd}_+$ is \StrPAhRMSE. Adding the notched bars end-to-end leaves the surface error essentially unchanged (\StrJErrT) and halves the RMS error of $a^\mathrm{hyd}_+$ (\StrJAhRMSE). Nine further loads with the same 4\% scatter add little to 44 well-distributed tests. The naive mapping, in contrast, biases the surface (median error \StrNErrT, RMS error of $a^\mathrm{hyd}_+$ \StrNAhRMSE). The net-section nominal stress underestimates the local stress at the critical point by the stress concentration factor, and it misrepresents the triaxiality.

In scenario~B the tube tests leave the tensile triaxial region unconstrained. The median surface error for $I_1>0$ is \StrTubePErrT\ with points only, \StrTubeJErrT\ end-to-end and \StrTubeNErrT\ with the naive mapping; the RMS errors of $a^\mathrm{hyd}_+$ are \StrTubePAhRMSE, \StrTubeJAhRMSE\ and \StrTubeNAhRMSE. Here the naive mapping is clearly harmful, while end-to-end identification uses the notched bars as intended.

Two idealisations of this study are tested separately. First, generating the synthetic failure loads on a finer mesh ($24\times54$) than used for identification ($16\times36$) avoids an ``inverse crime''. In scenario~A over \StrFineXN\ realisations, this changes the median surface errors to \StrFineXPErrT\ (points), \StrFineXJErrT\ (end-to-end) and \StrFineXNErrT\ (naive). Second, pre-failure linear elasticity is retained. Because the truth and the identification share the same forward model, the study isolates the effect of the stress mapping. It does not quantify the model error of a linear-elastic forward model for real foams, which is discussed in Section~\ref{sec:disc}. The structural results are therefore a proof of concept.

\begin{table}[!htbp]
\centering\small
\caption{Simulated notched-bar tests (truth: four-term admissible 110IG model): first-failure load and critical stress state.}
\label{tab:struct}
\begin{tabular}{rrrrrrr}\toprule
$w$ & $\rho_n$ & $p/\sigma_\mathrm{ref}$ & $P_f/\sigma_\mathrm{ref}$ & triaxiality & $I_1/\sigma_\mathrm{ref}$ & $q/\sigma_\mathrm{ref}$ \\ \midrule
1.5 & 0.56 & 0.0 & 0.282 & 0.65 & 1.09 & 0.56 \\
1.5 & 0.56 & 0.3 & 0.476 & 0.33 & 0.94 & 0.95 \\
1.5 & 0.56 & 0.6 & 0.641 & 0.18 & 0.68 & 1.28 \\
3.0 & 2.25 & 0.0 & 0.498 & 0.50 & 1.05 & 0.70 \\
3.0 & 2.25 & 0.3 & 0.815 & 0.24 & 0.82 & 1.15 \\
3.0 & 2.25 & 0.6 & 1.001 & 0.07 & 0.31 & 1.41 \\
6.0 & 9.00 & 0.0 & 0.748 & 0.40 & 1.00 & 0.84 \\
6.0 & 9.00 & 0.3 & 1.164 & 0.17 & 0.65 & 1.30 \\
6.0 & 9.00 & 0.6 & 1.191 & -0.05 & -0.21 & 1.33 \\
\bottomrule\end{tabular}
\end{table}

\subsection{Gradient-based specimen design}
\label{sec:oed_results}
\begin{figure}[!htbp]
\centering
\includegraphics[width=\linewidth]{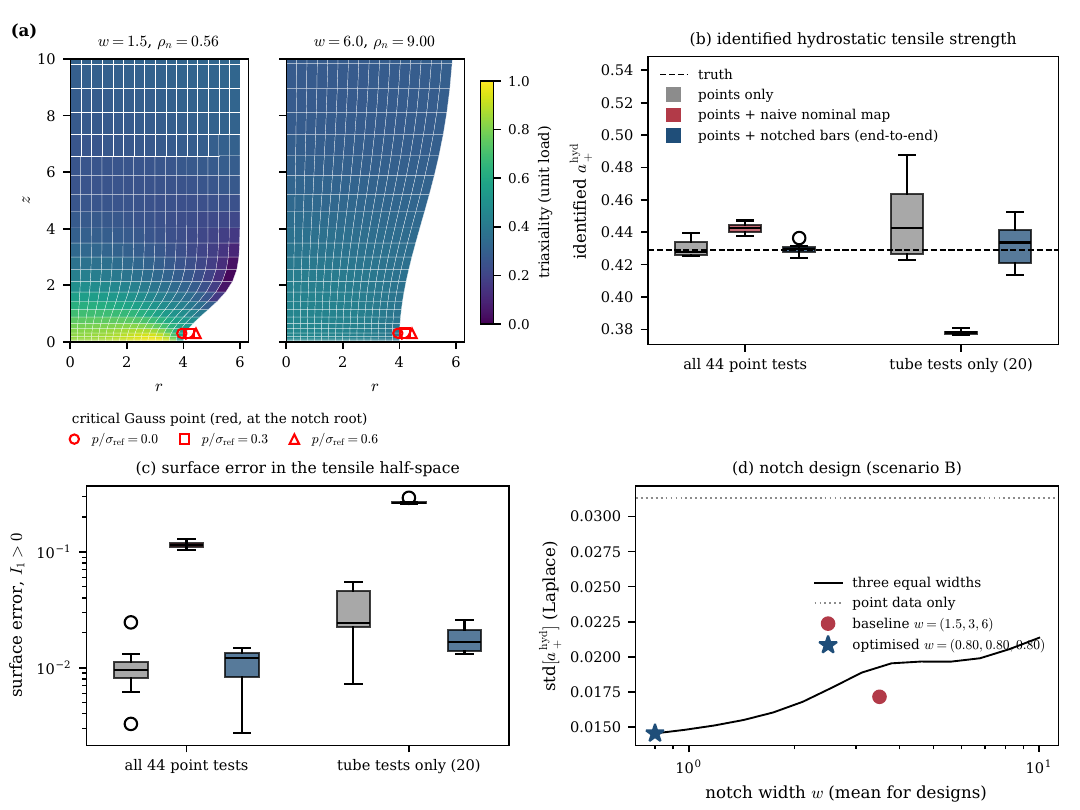}
\caption{Structural identification. (a) Notched-bar meshes for $w=1.5$ and $w=6$, coloured by the triaxiality of the unit-load field, with the critical Gauss points of the truth for three superposed pressures. (b) Identified $a^\mathrm{hyd}_+$ (truth dashed) and (c) surface error for $I_1>0$ for scenarios A (all point tests) and B (tube tests only). (d) Scenario B: Laplace standard deviation of $a^\mathrm{hyd}_+$ for three equal notch widths (line), the baseline design, the optimised design, and point data only.}
\label{fig:structural}
\end{figure}

For the same truth and noise model, the Laplace standard deviation of $a^\mathrm{hyd}_+$ is evaluated for three specimens, each tested at $p/\sigma_\mathrm{ref}\in\{0,0.3,0.6\}$. The notch widths are optimised by L-BFGS-B with gradients through the FE model, the mesh mapping and the implicit roots. The design gradients agree with central finite differences to a relative deviation of at most \OedGradErr\ (scenario~A) and \OedTGradErr\ (scenario~B). In scenario~A the standard deviation is \OedStdPts\ from the point data alone, \OedStdBase\ with the baseline design $w=(1.5,3,6)$ and \OedStdOpt\ with the optimised design $w=(\OedWopt)$, i.e.\ root radii $(\OedRhoOpt)$. In scenario~B the corresponding values are \OedTStdPts, \OedTStdBase\ and \OedTStdOpt, with optimised $w=(\OedTWopt)$. Relative to the baseline design, the optimisation reduces the standard deviation by \OedRed\,\% (A) and \OedTRed\,\% (B); relative to the point data alone, the reductions are \OedRedPts\,\% and \OedTRedPts\,\% (\cref{fig:structural}d). The design problem is non-convex, so we used three starting designs. Several optimal widths lie at the bounds of the admissible range $w\in[0.8,10]$. In scenario~B, three equally sharp notches are best, because they produce the highest triaxiality. A practical design would therefore have to include manufacturing limits and the notch-sensitivity of the material, which the present first-failure model ignores.

\FloatBarrier
\subsection{Analysis across all examples}
\label{sec:overall}
\begin{figure}[!htbp]
\centering
\includegraphics[width=\linewidth]{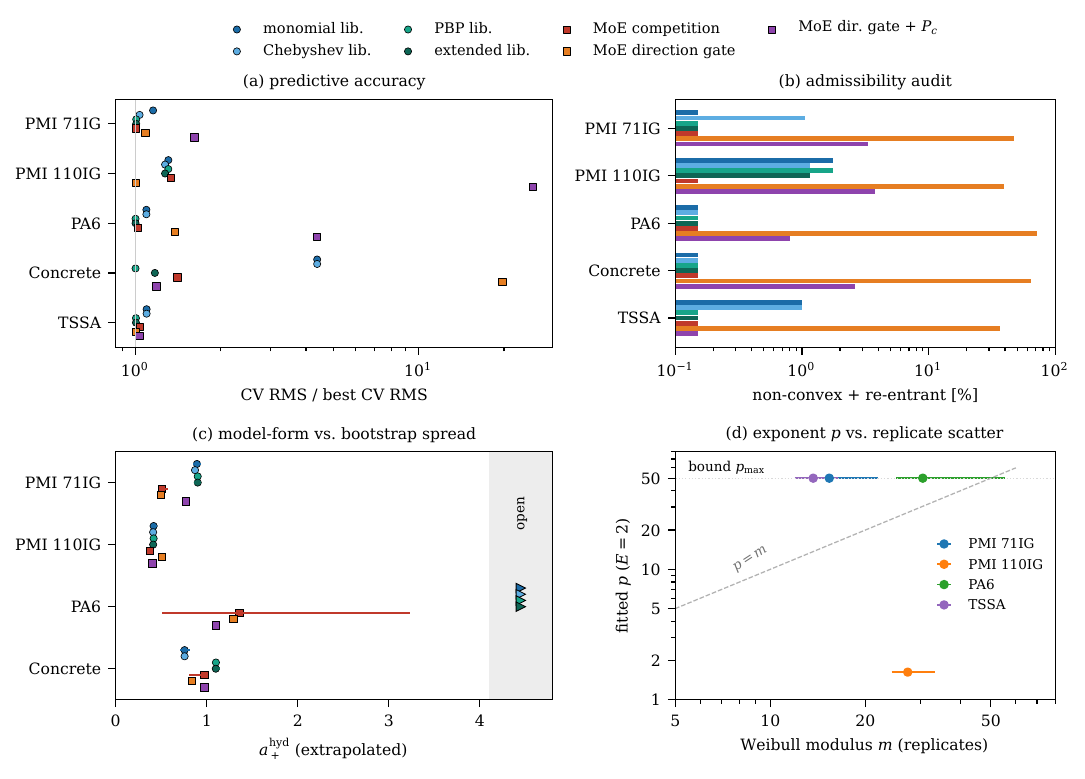}
\caption{Analysis across all examples. (a) Cross-validated error relative to the best class per data set. (b) Fraction of non-convex or re-entrant audit directions (bars at $0.15\%$ indicate none). (c) Extrapolated hydrostatic tensile strength of the selected model of each class with 90\% bootstrap intervals (lines) for the monomial, extended and competition-MoE models; triangles in the grey band: open surface, $a^\mathrm{hyd}_+=\infty$. (d) Exponent $p$ of the two-expert competition MoE against the Weibull modulus of the replicate scatter (90\% bootstrap intervals); dotted: upper bound of $p$.}
\label{fig:overall}
\end{figure}

\Cref{fig:overall} and \cref{tab:v2overview,tab:v2spread} collect the results of all model classes on all five data sets. Four observations stand out.

\paragraph{1. Accuracy: sharp Lode functions matter more than more terms}
The PBP or extended library is the most accurate admissible class on all five data sets; for 71IG it shares this place with the competition MoE (\cref{fig:overall}a). Its advantage over the monomial library ranges from $3\%$ (110IG, where only a Chebyshev term and no atom is selected) through $9$--$14\%$ (PA6, TSSA, 71IG) to a factor of four (concrete, CV \VConcPbpCV\ against \VConcMonoCV). The Chebyshev terms help only for 71IG. The competition MoE stays within $1.0$--$1.4$ times the best CV error, with five to eleven parameters. The direction-gated MoE reaches the lowest \emph{training} errors everywhere and the lowest CV error for 110IG and TSSA. It is, however, non-convex on 37--71\% of the directions, and it produces the three largest CV failures (concrete without the penalty, 110IG and PA6 with it). Penalising its convexity reduces this to below 4\%, but its cross-validation becomes erratic (CV $1.16$ for 110IG, $0.19$ for PA6). The gate has enough freedom to degenerate in single folds.

\paragraph{2. Admissibility: by penalty, by construction, or not at all}
The penalised libraries leave at most $1.8\%$ non-convex or re-entrant audit directions even without the refit with 1000 constraint directions used an earlier version of the approach (\cref{fig:overall}b). These fractions are results of the discrete audit on 2000 directions; they bound the violations on the sampled directions, not on the continuum (Section~\ref{sec:convex_audit}). The competition MoE has none by construction. The unconstrained direction gate is the only class whose violations are systematic. This confirms the conclusion of an earlier version with a model class that has nothing to do with polynomials: flexibility without admissibility constraints buys training accuracy with physically meaningless surfaces.

\paragraph{3. What the data determine and what the model class decides}
\Cref{tab:v2spread} compares, for every characteristic value, the range over all admissible classes that cross-validate within a factor $1.5$ of the best with the 90\% bootstrap width of a single class. For values backed by tests (e.g.\ $d$, $k$, $b_Z$, $b_D$, $a^\mathrm{hyd}_-$ of the foams, or $e_\mathrm{ET}$, $e_\mathrm{PS}$ of TSSA), the median relative range is \SpreadMeas\,\%, comparable to or smaller than the bootstrap width. For extrapolated values it is \SpreadExtr\,\%, and individual cases are far larger (\cref{fig:overall}c).
\begin{itemize}
\item \emph{71IG:} the hydrostatic tensile strength ranges from $0.51$ (competition MoE) to $0.90$ (libraries), against bootstrap widths of about $0.06$. The libraries and the competition MoE cross-validate identically (CV $0.070$). Kolupaev's fit gave $0.47$.
\item \emph{110IG:} the libraries give $a^\mathrm{hyd}_+\approx0.41$, the competition MoE gives $0.37$, and the bootstrap intervals do not overlap.
\item \emph{PA6:} $b_Z$ ranges from $1.02$ to $1.30$ and $b_D$ from $0.64$ to $1.00$, and four of five classes leave the surface open in hydrostatic tension.
\end{itemize}
Bootstrap intervals conditional on one model class thus systematically understate the uncertainty of extrapolated values. The spread over admissible model classes is a cheap and more honest estimate of model-form uncertainty. Where it is large, it identifies the tests that would be most informative. For the foams this is the triaxial tensile region, which the notched-bar design of Section~\ref{sec:oed} targets.

\paragraph{4. Mechanisms and the gate}
The number of experts selected by cross-validation, and confirmed by the hard-concrete relaxation (Table~\ref{tab:v2sparsity}), is two for both PMI foams and one for PA6, concrete and TSSA. For the foams the two experts separate tensile cell-wall fracture from compressive cell collapse (\cref{fig:gates}). For concrete a single cone with a nearly Rankine-type section suffices once the deviatoric shape can be sharp, and the same holds for TSSA in Hencky space with a cavitation term. The Gumbel-softmax routing of the direction gate becomes hard (84--100\% of the data assigned to one expert), but this does not make it more accurate or interpretable (Table~\ref{tab:v2sparsity}).

The quantitative prediction of the competing-risk reading is not confirmed by these data. The fitted exponents of the two-expert models lie at the bound $p_\mathrm{max}=50$ for 71IG, PA6 and TSSA and at $p=1.6$ for 110IG. The replicate scatter, in contrast, gives Weibull moduli of $14$--$30$ (\cref{fig:overall}d). With the published data, $p$ is determined only by the few states near the intersection of two experts. A test series designed along such an intersection, for example at intermediate triaxiality for the foams, would make $p$ identifiable. Comparing it with the scatter-based modulus would then be a direct test of the interpretation. Until such a test is made, the exponent is best read as a regime-transition parameter. The \emph{qualitative} reading, in contrast, is consistent across materials. Where two mechanisms are needed, they are the ones that the materials science of these materials suggests (tensile fracture versus collapse of the foam cells), and where one suffices, its form (a maximum-stretch surface for the elastomer, a Rankine-type cone for concrete) matches the established engineering criteria.

\begin{table}[!htbp]
\centering\small
\caption{Model-form versus parameter uncertainty: range of characteristic values over the admissible model classes ($\le5\%$ audit violations, CV within $1.5\times$ the best class), median 90\% bootstrap width of the monomial, extended and competition-MoE models, and the number of classes with an open surface in that direction. ``measured'': a test probes this direction; ``extrapolated'': no test in this direction.}
\label{tab:v2spread}
\setlength{\tabcolsep}{4pt}
\resizebox{\linewidth}{!}{\begin{tabular}{llcccr}
\toprule
Data set & quantity & status & range over admissible classes & bootstrap 90\% width & open \\
\midrule
PMI 71IG & $d$ & measured & 0.92--1.00 & 0.04 & 0/5 \\
 & $k$ & measured & 1.31--1.43 & 0.10 & 0/5 \\
 & $b_Z$ & measured & 0.71--0.74 & 0.02 & 0/5 \\
 & $b_D$ & measured & 0.73--0.77 & 0.01 & 0/5 \\
 & $a^\mathrm{hyd}_+$ & extrapolated & 0.51--0.90 & 0.06 & 0/5 \\
 & $a^\mathrm{hyd}_-$ & measured & 0.87--0.91 & 0.05 & 0/5 \\
\midrule
PMI 110IG & $d$ & measured & 0.95--0.98 & 0.06 & 0/5 \\
 & $k$ & measured & 1.41--1.45 & 0.11 & 0/5 \\
 & $b_Z$ & measured & 0.54--0.55 & 0.02 & 0/5 \\
 & $b_D$ & measured & 0.77--0.82 & 0.04 & 0/5 \\
 & $a^\mathrm{hyd}_+$ & extrapolated & 0.37--0.42 & 0.03 & 0/5 \\
 & $a^\mathrm{hyd}_-$ & measured & 0.89--0.90 & 0.05 & 0/5 \\
\midrule
PA6 & $d$ & measured & 1.02--1.09 & 0.05 & 0/5 \\
 & $k$ & measured & 1.00--1.01 & 0.09 & 0/5 \\
 & $b_Z$ & extrapolated & 1.02--1.30 & 0.05 & 0/5 \\
 & $b_D$ & extrapolated & 0.64--1.00 & 0.00 & 0/5 \\
 & $a^\mathrm{hyd}_+$ & extrapolated & 1.36--1.36 & 2.94 & 4/5 \\
 & $a^\mathrm{hyd}_-$ & extrapolated & 0.66--0.66 & 1.18 & 4/5 \\
\midrule
Concrete & $d$ & measured & 10.12--10.21 & 2.99 & 0/4 \\
 & $k$ & extrapolated & 1.61--1.62 & 0.04 & 0/4 \\
 & $b_Z$ & extrapolated & 0.98--0.99 & 0.05 & 0/4 \\
 & $b_D$ & measured & 10.91--11.09 & 3.82 & 0/4 \\
 & $a^\mathrm{hyd}_+$ & extrapolated & 0.97--1.10 & 0.08 & 0/4 \\
 & $a^\mathrm{hyd}_-$ & extrapolated & -- & 2.98 & 4/4 \\
\midrule
TSSA & $e_\mathrm{ET}$ & measured & 1.70--1.87 & 0.12 & 0/6 \\
 & $e_\mathrm{PS}$ & measured & 1.11--1.15 & 0.03 & 0/6 \\
 & $e_\mathrm{HT}$ & measured & 0.02--0.02 & 0.02 & 0/6 \\
\bottomrule
\end{tabular}}
\end{table}

\section{Discussion}
\label{sec:disc}
\paragraph{What admissibility buys, and what it costs}
Without penalties, sparse library models and direction-gated mixtures reach cross-validated errors similar to or lower than the constrained models, but they are non-convex or re-entrant on large parts of the stress space (\cref{tab:real,tab:v2overview}). The constraints act as a strong prior. They exclude surfaces that the data alone would prefer, typically because a few test series disagree at the level of a few per cent. We consider this the correct behaviour for a strength criterion that will be extrapolated in design. A lower cross-validated error obtained with a non-convex surface is not evidence against convexity; it reflects scatter or systematic differences between specimen types. The residuals grouped by test type (\cref{fig:residuals}) make such differences visible and turn them into statements about which test series should be repeated, as Kolupaev recommends for the compression tests on PMI \citep[Sect.~14.5.2]{Kolupaev2018}. The factor-of-two inconsistency in the tabulated equibiaxial tensile strengths of 110IG (\cref{app:data}) is an error that a purely data-driven fit would have absorbed silently. The competition-gated mixture is the only model class here in which admissibility costs nothing: convexity and uniqueness hold by construction (\cref{sec:moe}).

\paragraph{Richer libraries: what helps and what does not}
The Chebyshev terms answer a natural question, whether trigonometric functions of the Lode angle should be in the library. Isotropy admits only $\cos3k\theta$, and these are rational invariants with negative powers of $q$ (Proposition~\ref{prop:cos}). On the present data they help only for 71IG, where $q\cos3\theta$ and $q\cos6\theta$ resolve the tension--compression asymmetry (\cref{tab:v2overview}). The other data sets do not resolve Lode harmonics beyond those already contained in $r_3$ and $r_3^2$. The PBP atom helps where the Lode dependence has corners, and it does so with very few terms. It is decisive for concrete (a fourfold lower CV error and characteristic values within a few per cent of the measured ones) and for TSSA, whose Hencky-space section is a rounded triangle. It also helps for PA6 and for the synthetic Mohr--Coulomb and PBP truths. Two costs come with larger libraries. First, the heuristic search becomes harder. Without the nested candidates of the smaller libraries, the union library occasionally returned worse supports than the monomial library at the same size. Seeding the search with the nested paths makes the training error of a larger library non-increasing in the library, but it does not make the search exhaustive. Second, overcomplete unions contain near-collinear terms, so the selected \emph{support} is less stable than the selected \emph{surface}. The PBP atom also leaves Kolupaev's class of rational criteria, condition (v). We regard this as acceptable because the atom has a closed form, is convex by construction and is used in the literature on rubber, rock and concrete \citep{Podgorski1985,Bigoni2004,Rosendahl2019}.

\paragraph{Libraries versus mixtures of experts}
The two approaches are complementary rather than competing. A sparse library model is one smooth equation whose terms have no individual meaning; its strength is the systematic search over a large, structured space and the direct link to Kolupaev's catalogue. A mixture model is a small set of mechanisms with a rule for their competition. With the competition gate it is admissible by construction, it stays within $1.0$--$1.4$ times the best cross-validated error on all data sets with five to eleven parameters (\cref{fig:overall}), and its gate selects between physically motivated mechanisms (\cref{sec:gate_meaning}). Its weaknesses are identifiability and optimisation. The expert family has five parameters per mechanism, the loss is non-convex, and different local optima can fit the data equally well while differing far from the data. The synthetic two-mechanism truth T6 shows this drastically. The fitted competition MoE reproduces the data at the noise level, yet its surface error near the data is $0.22$, whereas the true parameters give $0.002$. A weak shape prior, $10^{-2}\sum_e(\gamma_e^2+b_e^2)$, reduces the error to $0.07$ (\cref{app:impl2}). Mixtures therefore need either well-distributed data or mechanism-based priors, for example on the deviatoric shape of a mechanism. The direction gate adds flexibility, and with it the largest cross-validation failures and non-convexity on more than a third of the directions, but no interpretability. Penalising convexity removes most of the non-convexity, but it makes the fits unstable across folds. A few per cent of non-convex directions remain, because the gate gradient keeps entering the curvature.

\paragraph{Testing the gate}
The interpretation of the competition gate makes two predictions that can be falsified. First, the exponent $p$ should be of the order of the Weibull modulus of the strength scatter. \Cref{fig:overall}d compares the fitted $p$ with moduli estimated from replicate tests. There is no quantitative agreement: $p$ sits at its upper bound for three data sets and is $1.6$ for 110IG, whereas the replicate moduli are $14$--$30$. This is not surprising, for two reasons. The replicate scatter mixes material variability with specimen and test-machine effects, and the data constrain $p$ only through the few states where two mechanisms are comparable. A designed test series near an expert intersection, such as combined deviatoric and volumetric loading of TSSA, would identify $p$ directly. Second, the responsibilities predict the fraction of fracture surfaces of each type, e.g.\ cohesive tearing versus cavitation-dominated fracture in TSSA, which can be checked by fractography. Neither check is possible with the published data. We therefore use the competition gate as a phenomenological regime selector, whose experts and transitions are physically motivated, and regard the statistical and dissipative readings as hypotheses that the model makes testable, not as results of this paper.

\paragraph{Strain space and the energetic reading of TSSA}
The TSSA results show the value of choosing the space in which the surface is simple. In Hencky space the uniaxial, planar and equibiaxial tests lie on the three characteristic meridians of the deviatoric plane, and the surface is close to a maximum-principal-stretch criterion with a slight Lode correction. The same data in the stretch $\pi$-plane need the nearly Tresca-shaped PBP criterion of \citet{Drass2020coupled}. Our analysis supports a stretch-based rather than an energy-based reading. The stored energy at failure differs by a factor of about $\TSSAWratio$ between equibiaxial and planar tension, whereas the maximal principal stretch varies by only about ten per cent (\TSSAlmaxMin--\TSSAlmaxMax). This statement depends on the extrapolation of the Ogden model beyond the largest stretch of the equibiaxial curves (about $2.2$ against $2.6$ at failure). Its direction is robust, however, because any stiffening model increases the equibiaxial energy further. Treloar's natural rubber gives iso-energy contours of the same shape, so the conclusion does not hinge on the TSSA fit. The volumetric part of the surface, in contrast, rests on a single model-based value, $J_c$. The monomial model closes the surface in volumetric compression through its even term $(\operatorname{tr}\mathbf h)^2$, which is unphysical; the PBP model and the competition MoE leave the compression side open. Coupled deviatoric--volumetric tests, e.g.\ pancakes of different aspect ratio \citep{Gent1959,LopezPamies2011}, are needed before the shape of the cavitation cap can be discovered rather than assumed.

\paragraph{Search and selection}
The screened beam search is a heuristic and is not guaranteed to find the best support of a given size. Because supports are chosen with all data, the reported CV errors are slightly optimistic; a nested procedure would remove this bias at higher cost. Cross-validation must, moreover, start every fold fit independently of the full-data fit: warm starts leak information into coefficients that the training fold does not constrain (\cref{app:impl2}). The bootstrap intervals are conditional on the selected support. The spread of extrapolated values \emph{across} model classes (\cref{fig:overall}c) is a more honest indicator of model-form uncertainty. It is often larger than the bootstrap intervals, which is the central quantitative lesson of the cross-material analysis (\cref{sec:overall}).

\paragraph{Discrete enforcement of constraints}
Convexity and uniqueness of the library models are enforced on a finite set of directions and checked on a denser audit set. All audit fractions in this paper are results of such discrete checks (Section~\ref{sec:convex_audit}). Refitting with $1000$ constraint directions reduces the residual violations on the sampled directions, but no discrete check is a certificate on the continuum. Because the even-$m$ sub-library is polynomial in the stresses, sum-of-squares relaxations could certify convexity exactly. The competition MoE is a second route to certified admissibility. It gives up the library's systematic search, and as in input-convex neural networks \citep{Fuhg2022yield}, convexity becomes structural.

\paragraph{Structural identification}
The structural study is synthetic, and its most restrictive assumption is linear-elastic pre-failure behaviour. For PMI foams this assumption is questionable for three reasons. First, in compression the foams deform non-linearly well before the strength is reached (onset of cell-wall bending and buckling), and the superposed pressures of the study, up to $p/\sigma_\mathrm{ref}=0.6$, amount to about two thirds of the hydrostatic compressive strength of 110IG ($a^\mathrm{hyd}_-=\PMIbahminus$). The material in the specimen is thus pre-loaded into a range where a secant stiffness smaller than the initial one, and a different volumetric response, must be expected. Second, local non-linearity at the notch root redistributes stress. It lowers the peak stress and changes the triaxiality at the root relative to the elastic solution. A linear-elastic forward model then overestimates the local stress at failure, and the identified surface is biased towards lower strengths in exactly the triaxial region that the notched bars are meant to probe. Third, the elastic triaxiality at the notch root depends on the Poisson ratio, which for foams differs from the assumed $\nu=0.3$ and may differ between tension and compression. In the synthetic study the truth and the identification share the forward model, so these errors cancel by construction; with real specimens they enter as a model bias that the noise model does not contain. Methodologically, a non-linear pre-failure model removes the affine structure of Proposition~\ref{prop:affine}. The first failure load must then be found by event detection along an incrementally solved load path, and the gradients follow from implicit (adjoint) differentiation of the equilibrium path. This is feasible in the same JAX framework, at the cost of one non-linear solve per load step. Before real notched-bar tests are used, the forward-model error should be bounded, for example by comparing linear-elastic and elastic--plastic (crushable-foam) analyses of the tested geometries. Real notched-bar tests on foams would therefore require (i) such a pre-failure model, (ii) a treatment of size and gradient effects, for example through critical distances \citep{Taylor2007} or weakest-link statistics (which the competing-risk reading of the gate makes natural), and (iii) a converged discretisation at the notch root (\cref{app:mesh}). All three can be added to the differentiable pipeline. The design study is \emph{locally} optimal: the Laplace variance is evaluated at the (here known) true surface. In practice the design must be computed for a provisional surface and updated sequentially, or averaged over the bootstrap and model-class ensemble in a Bayesian design criterion \citep{Chaloner1995}. With these caveats, end-to-end identification removes the bias of the nominal-stress mapping, structural tests pay off mainly when the homogeneous programme leaves the triaxial tensile region uncovered, and the specimen geometry can be optimised by gradients through the FE solve.

\paragraph{Scope and relation to EUCLID}
The framework is restricted to isotropic, rate-independent limit surfaces in stress or strain space. Anisotropic materials require structural tensors in the library and in the experts. EUCLID \citep{Flaschel2022,Xu2025} discovers yield surfaces from full-field data. Its Lode-angle Fourier series is closely related to our Chebyshev library, and Proposition~\ref{prop:cos} explains why only cosine modes are needed there as well. The present work targets the complementary situation of sparse failure states and structural failure loads, and it makes Kolupaev's admissibility conditions, including the unique-definition condition, explicit and differentiable. Both directions can be combined: the libraries, penalties and gated experts could be inserted into EUCLID's force-balance loss.

\section{Conclusions}
\label{sec:concl}
We have recast the equivalent stress concept of \citet{Kolupaev2018} as a sparse, admissibility-constrained and differentiable identification problem, and we have extended it in two directions. First, the invariant library now goes beyond monomials. Isotropy restricts trigonometric Lode terms to cosine modes, which are rational invariants with negative powers of $q$, and parametric PBP atoms provide sharp deviatoric shapes with two learnable parameters. Second, mixtures of convex gauge experts serve as baselines. Their competition gate is convex and unique by construction and acts as a regime selector between mechanisms. Under stated assumptions its responsibilities are also the share of failure probability in competing-risk fracture statistics, the share of dissipation in Koiter's multi-surface plasticity, and a Boltzmann weight in the Gibbs variational principle. On the present data the qualitative regime reading is consistent across materials, whereas the quantitative identification of the exponent with a Weibull modulus is not supported. A free direction gate has none of these meanings and overfits.

On five experimental data sets (two PMI foams, polyamide, concrete and a structural silicone in Hencky-strain space) and a synthetic benchmark, we draw four conclusions. (i) Admissibility constraints or admissible-by-construction representations are indispensable; unconstrained models with equal or lower cross-validated error are non-convex or re-entrant on large parts of stress space. (ii) A sharp Lode function matters more than many terms. Monomials suffice for the foams, but PBP atoms are decisive for the silicone and for concrete, whose data the monomial library could not resolve; with two to five terms they reproduce the measured strength ratios of concrete within a few per cent. (iii) Values that the tests do not constrain, above all the hydrostatic tensile strength, differ more between model classes than within bootstrap intervals. Model-form uncertainty must therefore be reported. (iv) The structure of a surface becomes interpretable when the space and the model class are chosen well: the TSSA surface is a maximum-stretch surface, not a critical-energy surface. End-to-end identification from notched-bar failure loads and gradient-based specimen design, finally, connect the discovered surfaces to structural testing. Future work will address certified convexity, sequential and Bayesian design over the model-class ensemble, fractographic tests of the gate, coupled deviatoric--volumetric tests on elastomers, and anisotropy.

\section*{Data and code availability}
The transcribed experimental data (from Tables 14.3, 14.6--14.10 and 14.13 of \citealp{Kolupaev2018}), the extracted TSSA failure stretches, the TSSA stress--strain curves, Treloar's data and the JAX implementation used for all results are provided as supplementary material.

\section*{Declaration of generative AI in the writing process}
During the preparation of this work the author used AI, specifically Anthropic Claude Opus 5.5, openAI gpt-6-astra in its ChatGPT version as well as Google gemini to draft code, text and figures. The author reviewed, iterated and edited all content and takes full responsibility for the publication.

\appendix
\section{Transcribed data}
\label{app:data}
All data were transcribed from \citet{Kolupaev2018}. For the PMI foams, the principal stresses of Table~14.9 were checked for consistency with the tabulated $(\sigma_{11},\tau_{12})$ (maximum deviation below $0.03$\,MPa). The supplementary tests comprise the enlacement test (equibiaxial compression $B_D$, Table~14.6, raw stress column), the hydrostatic compression test ($A_D$, Table~14.7, specimens with and without protective layer), the internal pressure test on tubes ($\sigma_t,\sigma_a,\sigma_r$ at failure, Table~14.8) and the equibiaxial tension test on the octopus specimen ($B_Z$, Table~14.10). For 110IG, the tabulated equibiaxial stresses ($2.76$--$3.11$\,MPa) are about twice the values implied by Eq.~(14.4), $\sigma_\mathrm I=\tfrac12F\,s_\mathrm{oct}$, with the tabulated forces and the scaling factors of Fig.~14.29. They are also twice the values plotted by \citet{Kolupaev2018} himself in Figs.~14.33--14.34 ($b_Z\approx0.50$--$0.56$). The 71IG values, in contrast, agree with Eq.~(14.4) and with Fig.~14.32. We therefore use one half of the tabulated 110IG values. We found the inconsistency while cross-checking our 110IG fits against the book's figures. The decision matters. With the unhalved values, the selected constrained model has six terms, an RMS radial residual of $0.064$ (CV $0.068$, against $0.054$ and $0.059$ with the corrected data), $b_Z=0.97$, $k=1.28$, and an extrapolated hydrostatic tensile strength $a^\mathrm{hyd}_+=1.16$, almost three times the value obtained with the corrected data. The correction should be confirmed with the original test records. PMI data are normalised by the mean of the uniaxial tensile tests in Table~14.9 ($\sigma_\mathrm{ref}=\PMIaSref$\,MPa for 71IG and $\PMIbSref$\,MPa for 110IG). PA6 and concrete data are used in the normalisation of the source ($64.55$\,MPa and $418$\,psi). Table~\ref{tab:datasummary} summarises the data sets.

\begin{table}[!htbp]
\centering\small
\caption{Data sets used in Section~\ref{sec:real}.}
\label{tab:datasummary}
\begin{tabular}{p{0.22\linewidth}p{0.17\linewidth}rp{0.42\linewidth}}
\toprule
Material & Source & $N$ & Test types \\
\midrule
PMI Rohacell 71IG & \citet{Kolupaev2018}, Tab.~14.6--14.10 & 46 & $Z$, $D$, $T$, $ZT$, $DT$ (tube), $B_D$, $A_D$, internal pressure, $B_Z$ \\
PMI Rohacell 110IG & Tab.~14.6--14.10 & 44 & as above \\
Polyamide PA6 (Dillenberger) & Tab.~14.3 & 19 & plane stress, $Z$--$K$--$D$ region (tube) \\
Concrete (Tasuji et al.) & Tab.~14.13 & 10 & plane stress, biaxial \\
Silicone TSSA (Hencky space) & \citet{Drass2020coupled}, Fig.~5 & 33 & uniaxial (10), bulge (16), shear pancake (5), pancake (mean), cavitation $J_c$ \\
\bottomrule
\end{tabular}
\end{table}

\section{Finite element model: mesh study}
\label{app:mesh}
Table~\ref{tab:mesh} reports the maximum Gauss-point principal stress (net-section stress concentration) and the maximum stress triaxiality under unit net-section load for three meshes. The $16\times36$ mesh used in the paper is within $3\%$ of the finest mesh for the sharpest notch and within $1\%$ for the others. Since the synthetic structural data in Section~\ref{sec:struct_results} are generated and identified with the same discretisation, the discretisation error does not bias the comparison. For experimental structural data, a converged mesh, or a critical-distance averaging, would be required.

\begin{table}[!htbp]
\centering\small
\caption{Mesh study for the notched bar (Gauss-point values, unit net-section stress, $\nu=0.3$).}
\label{tab:mesh}
\begin{tabular}{lccc}
\toprule
mesh ($n_r\times n_z$) & $w=1.5$ ($\rho_n=0.56$) & $w=3$ ($\rho_n=2.25$) & $w=6$ ($\rho_n=9$) \\
\midrule
$12\times28$ & 2.526 & 1.621 & 1.187 \\
$16\times36$ & 2.601 & 1.631 & 1.188 \\
$24\times54$ & 2.673 & 1.640 & 1.189 \\
\bottomrule
\end{tabular}
\end{table}

\section{Sensitivity to penalty weights and constraint directions}
\label{app:sens}
Table~\ref{tab:sens} refits the selected 110IG support with penalty weights $\mu_c=\mu_u=\mu_\nu\in\{1,10,100\}$ and with $240$ or $1000$ constraint directions. The characteristic values change by less than $1\%$, and the RMS residual by less than $0.001$. The audit violations depend mainly on the number of constraint directions: with $240$ directions, $0.7$--$1.3\%$ of the audit directions are non-convex and about $1\%$ re-entrant; with $1000$ directions and $\mu\ge10$, at most $0.1\%$ are. This motivates the refinement step with $1000$ directions. Like every discrete check, it reduces violations on the sampled directions but does not certify convexity on the continuum (Section~\ref{sec:convex_audit}).

\begin{table}[!htbp]
\centering\small
\caption{Sensitivity of the 110IG model (selected support) to penalty weights and constraint directions.}
\label{tab:sens}
\begin{tabular}{rrccccccc}\toprule
$\mu$ & $N_d$ & RMS & $d$ & $k$ & $b_Z$ & $a^\mathrm{hyd}_+$ & non-convex & re-entrant \\ \midrule
1 & 240 & 0.0532 & 0.977 & 1.415 & 0.550 & 0.416 & 1.3\,\% & 1.0\,\% \\
1 & 1000 & 0.0538 & 0.982 & 1.420 & 0.551 & 0.414 & 0.7\,\% & 0.1\,\% \\
10 & 240 & 0.0535 & 0.980 & 1.417 & 0.551 & 0.417 & 0.8\,\% & 1.0\,\% \\
10 & 1000 & 0.0541 & 0.985 & 1.423 & 0.553 & 0.415 & 0.1\,\% & 0.1\,\% \\
100 & 240 & 0.0536 & 0.980 & 1.417 & 0.551 & 0.417 & 0.7\,\% & 1.0\,\% \\
100 & 1000 & 0.0542 & 0.984 & 1.422 & 0.553 & 0.415 & 0.0\,\% & 0.1\,\% \\
\bottomrule\end{tabular}
\end{table}

\section{Implementation details}
\label{app:impl}
The library has $K=38$ terms ($n=6$). Ray roots use a quadratic grid of $256$ points on $[0,t_\mathrm{max}]$, $t_\mathrm{max}=3\max_i r_i$, followed by $40$ bisection steps. Convexity and uniqueness penalties use $240$ Fibonacci directions; uniqueness is enforced for $t\in[1.1\rho,t_\mathrm{max}]$ with margin $0.02$. Penalty weights are $\mu_c=\mu_u=\mu_\nu=10$ with continuation factors $\{1,100\}$ and ridge $\alpha=10^{-6}$ on scaled coefficients. Beam search uses exhaustive screening up to $s=3$, beam width $60$ and $8$ refits per size, plus warm-started nested refits. Cross-validation is $5$-fold (PMI, PA6) or leave-one-out (concrete) with the one-standard-error rule. The bootstrap uses $40$ resamples. The audit uses $2000$ directions: a direction counts as non-convex if the smallest scaled tangential curvature is below $-10^{-3}$, as re-entrant if $\Phi<-10^{-6}$ at any of $200$ geometrically spaced radii in $[1.01\rho,t_\mathrm{max}]$, and as weakly unique if $\Phi<10^{-3}$ somewhere in $[1.1\rho,t_\mathrm{max}]$. The structural first-failure load uses a smooth minimum with temperature $\tau=2\times10^{-3}$ over all Gauss points, $\tilde P_f=-\tau\ln\sum_x\exp(-t^\star(x)/\tau)$, which satisfies $P_f-\tau\ln N_x\le\tilde P_f\le P_f$; the same operator is used for data generation and identification. Joint fits are polished by $400$ Adam iterations after L-BFGS-B, because the first-root and minimum operators make the objective only piecewise smooth. All computations run in double precision on a two-core CPU.

\section{TSSA and rubber data}
\label{app:tssa}
\paragraph{Failure stretches}
The ultimate stretches of DOWSIL\texttrademark\ TSSA were extracted from the vector graphics of Fig.~5 of the manuscript of \citet{Drass2020coupled} (author's archive). The markers were read directly from the PDF drawing operators; there is no digitisation error beyond the plotting precision. Tick marks fix the scale, and grey markers (individual tests) are separated from white markers (means). The figure replicates every test threefold by rotation in the stretch $\pi$-plane. We keep the replica in the sector $[-60^\circ,60^\circ]$ around the uniaxial axis. This gives $10$ uniaxial, $16$ equibiaxial (bulge) and $5$ shear-pancake tests and the mean of the pancake tests, whose individual values are not plotted. Principal stretches follow from the $\pi$-plane coordinates and incompressibility, $\lambda_1\lambda_2\lambda_3=1$. A least-squares refit of the PBP criterion in the stretch $\pi$-plane to the extracted points gives $\lambda_c=1.644$, $\beta=0.827$ and $\gamma=1$ (bound). The published values are $1.6432$, $0.8175$ and $0.9999$ \citep{Drass2020coupled}, so the extraction is validated. The shear-pancake and pancake points lie at a stretch-space angle of about $15^\circ$, which is the image of planar deformation $(\lambda,1,1/\lambda)$ at large stretch. In Hencky space they lie on the planar meridian $\theta=\pi/6$. The critical cavitation volume ratio $J_c=1.041$ was obtained by inverse analysis of pancake tests \citep{Drass2020coupled,Drass2020pec}. It is a model-based value, not a direct measurement, and it is always kept in the training set (Section~\ref{sec:tssa_res}).

\paragraph{Stress--strain curves}
The TSSA curves (10 uniaxial tension, 9 uniaxial compression, 5 equibiaxial bulge and 5 shear-pancake tests) are engineering stress--strain curves resampled to a common strain grid. Each series ends at the smallest failure strain of that series, so the curves carry no failure information. The Treloar data comprise 21 uniaxial, 10 equibiaxial and 9 pure-shear points \citep{Treloar1944,HyperDataJl}. The selected Ogden models are listed in Section~\ref{sec:tssa_res}. Uniaxial compression was excluded from the TSSA fit, because friction affects these curves and because uniaxial compression is kinematically equivalent to equibiaxial tension, so it does not add a deformation mode.

\section{Selected models}
\label{app:v2models}
\begin{table}[!htbp]
\centering\scriptsize
\caption{Selected library models: active terms with coefficients (on normalised data) and PBP shape parameters.}
\label{tab:v2terms}
\begin{tabular}{llp{0.62\linewidth}}
\toprule
Data set & Library & selected terms (coefficients), shape parameters \\
\midrule
PMI 71IG & mono & $q$ (+0.733), $I_1^{2}$ (+0.136), $I_1^{2} q^{2}$ (+0.112) \\
PMI 71IG & cheb & $q$ (+0.743), $I_1^{2}$ (+0.136), $I_1^{2} q^{2}$ (+0.104), $q \cos 3\theta$ (-0.0484), $q \cos 6\theta$ (+0.0104) \\
PMI 71IG & pbp & $q$ (+1.22), $I_1^{2}$ (+0.132), $I_1^{2} q^{2}$ (+0.00536), $(q\Omega)^{2}$ (-0.372), $(q\Omega)^{6}$ (+0.0012); $\beta=0.74$, $\gamma=0.451$ \\
PMI 71IG & ext & $q$ (+1.22), $I_1^{2}$ (+0.132), $I_1^{2} q^{2}$ (+0.00536), $(q\Omega)^{2}$ (-0.372), $(q\Omega)^{6}$ (+0.0012); $\beta=0.74$, $\gamma=0.451$ \\
\midrule
PMI 110IG & mono & $q$ (+0.72), $I_1$ (+0.439), $I_1 q$ (-0.406), $I_1^{2}$ (+0.307), $I_1^{4} q$ (+0.00102), $I_1^{2} r_3$ (-0.0667), $I_1^{2} q^{4}$ (+0.0203) \\
PMI 110IG & cheb & $q$ (+0.718), $I_1$ (+0.442), $I_1 q$ (-0.413), $I_1^{2}$ (+0.309), $I_1^{4} q$ (-0.000559), $I_1^{2} q^{4}$ (+0.0218), $I_1^{2} q^{2} \cos 3\theta$ (-0.0684) \\
PMI 110IG & pbp & $q$ (+0.72), $I_1$ (+0.439), $I_1 q$ (-0.406), $I_1^{2}$ (+0.307), $I_1^{4} q$ (+0.00102), $I_1^{2} r_3$ (-0.0666), $I_1^{2} q^{4}$ (+0.0203) \\
PMI 110IG & ext & $q$ (+0.718), $I_1$ (+0.442), $I_1 q$ (-0.413), $I_1^{2}$ (+0.309), $I_1^{4} q$ (-0.000559), $I_1^{2} q^{4}$ (+0.0218), $I_1^{2} q^{2} \cos 3\theta$ (-0.0684) \\
\midrule
PA6 & mono & $q$ (+0.956), $q r_3$ (+0.01) \\
PA6 & cheb & $q$ (+0.956), $q r_3$ (+0.01) \\
PA6 & pbp & $(q\Omega)$ (+0.984); $\beta=0.72$, $\gamma=0.734$ \\
PA6 & ext & $(q\Omega)$ (+0.984); $\beta=0.72$, $\gamma=0.734$ \\
\midrule
Concrete & mono & $q$ (+0.476), $I_1$ (+0.374), $I_1^{4}$ (+1.8e-05) \\
Concrete & cheb & $q$ (+0.476), $I_1$ (+0.374), $I_1^{4}$ (+1.8e-05) \\
Concrete & pbp & $q$ (+0.0408), $I_1$ (+0.295), $(q\Omega)$ (+0.639), $I_1 (q\Omega)$ (+0.000152), $(q\Omega)^{2}$ (+0.000346); $\beta=0.05$, $\gamma=0.995$ \\
Concrete & ext & $I_1$ (+0.296), $q \cos 6\theta$ (+0.000643), $(q\Omega)$ (+0.68); $\beta=0.10$, $\gamma=0.994$ \\
\midrule
TSSA & mono & $I_1^{2}$ (+810), $q^{6}$ (+0.0382), $q^{3} r_3$ (+0.0351) \\
TSSA & cheb & $I_1^{2}$ (+810), $q^{6}$ (+0.0382), $q^{3} r_3$ (+0.0351) \\
TSSA & pbp & $I_1$ (+26), $(q\Omega)$ (+0.685); $\beta=0.01$, $\gamma=0.995$ \\
TSSA & ext & $I_1$ (+26), $(q\Omega)$ (+0.685); $\beta=0.01$, $\gamma=0.995$ \\
\bottomrule
\end{tabular}
\end{table}

\begin{table}[!htbp]
\centering\small
\caption{Selected competition-gate MoE models: exponent $p$ and expert parameters \eqref{eq:expert}.}
\label{tab:v2moe}
\begin{tabular}{lrrrrrrrr}
\toprule
Data set & $E$ & $p$ & expert & $a_e$ & $b_e$ & $c_e$ & $\beta_e$ & $\gamma_e$ \\
\midrule
PMI 71IG & 2 & 50.0 & 1 & 0.390 & 0.001 & -0.332 & 2.00 & 0.449 \\
 & & & 2 & 0.457 & 0.240 & +0.162 & 0.28 & 0.971 \\
\midrule
PMI 110IG & 2 & 1.6 & 1 & 0.038 & 0.100 & +0.607 & 0.00 & 0.930 \\
 & & & 2 & 0.313 & 0.001 & -0.348 & 2.00 & 0.648 \\
\midrule
PA6 & 1 & 20.3 & 1 & 1.093 & 0.137 & -0.129 & 0.29 & 0.995 \\
\midrule
Concrete & 1 & 16.3 & 1 & 0.459 & 0.001 & +0.297 & 0.10 & 0.994 \\
\midrule
TSSA & 1 & 7.8 & 1 & 0.470 & 4.646 & +20.204 & 0.01 & 0.995 \\
\bottomrule
\end{tabular}
\end{table}

\begin{table}[!htbp]
\centering\small
\caption{Sparsity-enforcing relaxations: hard-concrete $L_0$ gates for the competition MoE ($E_\mathrm{max}=4$) at four penalty weights, and Gumbel-softmax routing for the direction-gated MoE ($E=3$).}
\label{tab:v2sparsity}
\begin{tabular}{lrcrr}
\toprule
Data set & $\lambda_0$ & deterministic gates $z_e$ & active & RMS \\
\midrule
PMI 71IG & 1e-04 & (1.00, 0.55, 1.00, 0.00) & 3 & 0.059 \\
 & 1e-03 & (1.00, 0.00, 1.00, 0.00) & 2 & 0.059 \\
 & 3e-03 & (0.07, 0.00, 1.00, 0.00) & 2 & 0.059 \\
 & 1e-02 & (0.00, 0.00, 1.00, 0.00) & 1 & 0.135 \\
 & \multicolumn{4}{l}{Gumbel routing ($E=3$): RMS 0.137, hard assignments 100\,\%, non-convex 0.0\,\%} \\
\midrule
PMI 110IG & 1e-04 & (1.00, 1.00, 1.00, 1.00) & 4 & 0.044 \\
 & 1e-03 & (0.00, 1.00, 1.00, 1.00) & 3 & 0.044 \\
 & 3e-03 & (0.00, 0.43, 0.81, 1.00) & 3 & 0.044 \\
 & 1e-02 & (0.00, 0.00, 0.00, 0.83) & 1 & 0.141 \\
 & \multicolumn{4}{l}{Gumbel routing ($E=3$): RMS 0.137, hard assignments 84\,\%, non-convex 13.1\,\%} \\
\midrule
PA6 & 1e-04 & (0.59, 1.00, 0.00, 0.00) & 2 & 0.036 \\
 & 1e-03 & (0.00, 1.00, 0.00, 0.00) & 1 & 0.036 \\
 & 3e-03 & (0.00, 1.00, 0.00, 0.00) & 1 & 0.036 \\
 & 1e-02 & (0.00, 0.94, 0.00, 0.00) & 1 & 0.036 \\
 & \multicolumn{4}{l}{Gumbel routing ($E=3$): RMS 0.035, hard assignments 84\,\%, non-convex 18.0\,\%} \\
\midrule
Concrete & 1e-04 & (1.00, 0.00, 1.00, 0.00) & 2 & 0.035 \\
 & 1e-03 & (1.00, 0.00, 0.00, 0.00) & 1 & 0.038 \\
 & 3e-03 & (0.97, 0.00, 0.33, 0.00) & 2 & 0.035 \\
 & 1e-02 & (1.00, 0.00, 0.00, 0.00) & 1 & 0.038 \\
 & \multicolumn{4}{l}{Gumbel routing ($E=3$): RMS 0.267, hard assignments 100\,\%, non-convex 0.0\,\%} \\
\midrule
TSSA & 1e-04 & (0.00, 0.00, 1.00, 0.00) & 1 & 0.085 \\
 & 1e-03 & (0.00, 0.00, 1.00, 0.00) & 1 & 0.085 \\
 & 3e-03 & (0.00, 0.00, 1.00, 0.00) & 1 & 0.085 \\
 & 1e-02 & (0.00, 0.00, 0.00, 0.91) & 1 & 0.085 \\
 & \multicolumn{4}{l}{Gumbel routing ($E=3$): RMS 0.081, hard assignments 100\,\%, non-convex 0.0\,\%} \\
\bottomrule
\end{tabular}
\end{table}

\section{Implementation details}
\label{app:impl2}
The Chebyshev features use $\cos3\theta=r_3/q^3$ with $q=\sqrt{3I_2'+10^{-12}}$, which keeps the features bounded and differentiable away from the hydrostatic axis. PBP atoms use $\gamma\le0.995$. Library refits optimise the active coefficients and, if a PBP atom is active, $(\beta,\gamma)$ with bounded L-BFGS-B; the penalty settings are those of \ref{app:impl}. Cross-validation refits every fold from the algebraic least-squares initialisation on the training fold. A first version of the approach warm-started the fold refits from the full-data fit. We found that such warm starts leak information into coefficients that a training fold does not constrain. For TSSA, the volumetric coefficients are fixed by the single cavitation datum, and with warm starts the cross-validation error at $s=4$ was $0.084$, although the training data of the fold that held out the cavitation datum contained no volumetric information at all. All cross-validation results of this implementation are therefore leakage-free. The cavitation datum is always kept in training, and it enters the CV score with its fold-averaged training residual. Without this, the one-standard-error rule chose a model without any volumetric term. The MoE models are fitted from $12$ random starts (CV folds: six fresh starts). Larger $E$ is additionally warm-started from the optimum with $E-1$ experts plus a negligible new expert, which makes the training error non-increasing in $E$. The penalised direction gate is initialised from the unconstrained one. The hard-concrete runs use $\beta_\mathrm{hc}=2/3$, $(\zeta,\gamma_\mathrm{hc})=(1.1,-0.1)$, Adam with step $0.02$ and $2500$ iterations, followed by a deterministic refit of the surviving experts. The Gumbel runs anneal $T$ from $2$ to $0.1$ over $2000$ Adam iterations. Weibull moduli of the replicate scatter are maximum-likelihood estimates of the pooled radii, each normalised by the mean of its replicate group (same test label and direction within $2^\circ$, at least three replicates). They are corrected by $\sqrt{\sum(n_g-1)/\sum n_g}$ for the variance removed by the normalisation, and their 90\% intervals are obtained by bootstrap.

\FloatBarrier
\bibliographystyle{elsarticle-harv}
\bibliography{references}

@book{Kolupaev2018,
  author = {Kolupaev, Vladimir A.},
  title = {Equivalent Stress Concept for Limit State Analysis},
  series = {Advanced Structured Materials},
  volume = {86},
  publisher = {Springer},
  address = {Cham},
  year = {2018},
  doi = {10.1007/978-3-319-73049-3}
}

@incollection{Altenbach2014,
  author = {Altenbach, Holm and Bolchoun, Alexandre and Kolupaev, Vladimir A.},
  title = {Phenomenological yield and failure criteria},
  booktitle = {Plasticity of Pressure-Sensitive Materials},
  editor = {Altenbach, Holm and {\"O}chsner, Andreas},
  publisher = {Springer},
  address = {Berlin},
  pages = {49--152},
  year = {2014}
}

@article{Yu2002,
  author = {Yu, Mao-Hong},
  title = {Advances in strength theories for materials under complex stress state in the 20th century},
  journal = {Applied Mechanics Reviews},
  volume = {55},
  number = {5},
  pages = {169--218},
  year = {2002}
}

@article{Podgorski1985,
  author = {Podg{\'o}rski, Jerzy},
  title = {General failure criterion for isotropic media},
  journal = {Journal of Engineering Mechanics},
  volume = {111},
  number = {2},
  pages = {188--201},
  year = {1985}
}

@article{Bigoni2004,
  author = {Bigoni, Davide and Piccolroaz, Andrea},
  title = {Yield criteria for quasibrittle and frictional materials},
  journal = {International Journal of Solids and Structures},
  volume = {41},
  number = {11--12},
  pages = {2855--2878},
  year = {2004}
}

@article{Tasuji1978,
  author = {Tasuji, M. Ehsani and Slate, Floyd O. and Nilson, Arthur H.},
  title = {Stress-strain response and fracture of concrete in biaxial loading},
  journal = {ACI Journal Proceedings},
  volume = {75},
  number = {7},
  pages = {306--312},
  year = {1978},
  doi = {10.14359/10944}
}

@article{Brunton2016,
  author = {Brunton, Steven L. and Proctor, Joshua L. and Kutz, J. Nathan},
  title = {Discovering governing equations from data by sparse identification of nonlinear dynamical systems},
  journal = {Proceedings of the National Academy of Sciences},
  volume = {113},
  number = {15},
  pages = {3932--3937},
  year = {2016}
}

@article{Kaheman2020,
  author = {Kaheman, Kadierdan and Kutz, J. Nathan and Brunton, Steven L.},
  title = {{SINDy-PI}: a robust algorithm for parallel implicit sparse identification of nonlinear dynamics},
  journal = {Proceedings of the Royal Society A},
  volume = {476},
  pages = {20200279},
  year = {2020},
  doi = {10.1098/rspa.2020.0279}
}

@article{Rudy2017,
  author = {Rudy, Samuel H. and Brunton, Steven L. and Proctor, Joshua L. and Kutz, J. Nathan},
  title = {Data-driven discovery of partial differential equations},
  journal = {Science Advances},
  volume = {3},
  number = {4},
  pages = {e1602614},
  year = {2017}
}

@article{Flaschel2021,
  author = {Flaschel, Moritz and Kumar, Siddhant and De Lorenzis, Laura},
  title = {Unsupervised discovery of interpretable hyperelastic constitutive laws},
  journal = {Computer Methods in Applied Mechanics and Engineering},
  volume = {381},
  pages = {113852},
  year = {2021},
  doi = {10.1016/j.cma.2021.113852}
}

@article{Flaschel2022,
  author = {Flaschel, Moritz and Kumar, Siddhant and De Lorenzis, Laura},
  title = {Discovering plasticity models without stress data},
  journal = {npj Computational Materials},
  volume = {8},
  pages = {91},
  year = {2022},
  doi = {10.1038/s41524-022-00752-4}
}

@article{Flaschel2023,
  author = {Flaschel, Moritz and Kumar, Siddhant and De Lorenzis, Laura},
  title = {Automated discovery of generalized standard material models with {EUCLID}},
  journal = {Computer Methods in Applied Mechanics and Engineering},
  volume = {405},
  pages = {115867},
  year = {2023},
  doi = {10.1016/j.cma.2022.115867}
}

@article{Xu2025,
  author = {Xu, Haotian and Flaschel, Moritz and De Lorenzis, Laura},
  title = {Discovering non-associated pressure-sensitive plasticity models with {EUCLID}},
  journal = {Advanced Modeling and Simulation in Engineering Sciences},
  year = {2025},
  doi = {10.1186/s40323-024-00281-3}
}

@article{Fuhg2024review,
  author = {Fuhg, Jan N. and others},
  title = {A review on data-driven constitutive laws for solids},
  journal = {Archives of Computational Methods in Engineering},
  volume = {32},
  pages = {1841--1883},
  year = {2025},
  doi = {10.1007/s11831-024-10196-2}
}

@article{Fuhg2022yield,
  author = {Fuhg, Jan N. and others},
  title = {Machine-learning convex and texture-dependent macroscopic yield from crystal plasticity simulations},
  journal = {Materialia},
  volume = {23},
  pages = {101446},
  year = {2022},
  doi = {10.1016/j.mtla.2022.101446}
}

@article{Vlassis2021,
  author = {Vlassis, Nikolaos N. and Sun, WaiChing},
  title = {Sobolev training of thermodynamic-informed neural networks for interpretable elasto-plasticity models with level set hardening},
  journal = {Computer Methods in Applied Mechanics and Engineering},
  volume = {377},
  pages = {113695},
  year = {2021},
  doi = {10.1016/j.cma.2021.113695}
}

@article{Linka2023,
  author = {Linka, Kevin and Kuhl, Ellen},
  title = {A new family of constitutive artificial neural networks towards automated model discovery},
  journal = {Computer Methods in Applied Mechanics and Engineering},
  volume = {403},
  pages = {115731},
  year = {2023},
  doi = {10.1016/j.cma.2022.115731}
}

@misc{jax2018,
  author = {Bradbury, James and Frostig, Roy and Hawkins, Peter and Johnson, Matthew James and Leary, Chris and Maclaurin, Dougal and Necula, George and Paszke, Adam and VanderPlas, Jake and Wanderman-Milne, Skye and Zhang, Qiao},
  title = {{JAX}: composable transformations of {Python+NumPy} programs},
  year = {2018},
  howpublished = {\url{http://github.com/jax-ml/jax}}
}

@article{Lewis1996,
  author = {Lewis, Adrian S.},
  title = {Convex analysis on the {H}ermitian matrices},
  journal = {SIAM Journal on Optimization},
  volume = {6},
  number = {1},
  pages = {164--177},
  year = {1996},
  doi = {10.1137/0806009}
}

@article{Byrd1995,
  author = {Byrd, Richard H. and Lu, Peihuang and Nocedal, Jorge and Zhu, Ciyou},
  title = {A limited memory algorithm for bound constrained optimization},
  journal = {SIAM Journal on Scientific Computing},
  volume = {16},
  number = {5},
  pages = {1190--1208},
  year = {1995}
}

@book{Hastie2009,
  author = {Hastie, Trevor and Tibshirani, Robert and Friedman, Jerome},
  title = {The Elements of Statistical Learning},
  edition = {2},
  publisher = {Springer},
  address = {New York},
  year = {2009}
}

@book{Efron1993,
  author = {Efron, Bradley and Tibshirani, Robert J.},
  title = {An Introduction to the Bootstrap},
  publisher = {Chapman \& Hall},
  address = {New York},
  year = {1993}
}

@article{Chaloner1995,
  author = {Chaloner, Kathryn and Verdinelli, Isabella},
  title = {Bayesian experimental design: a review},
  journal = {Statistical Science},
  volume = {10},
  number = {3},
  pages = {273--304},
  year = {1995}
}

@book{Hill1950,
  author = {Hill, Rodney},
  title = {The Mathematical Theory of Plasticity},
  publisher = {Clarendon Press},
  address = {Oxford},
  year = {1950}
}

@book{Taylor2007,
  author = {Taylor, David},
  title = {The Theory of Critical Distances: A New Perspective in Fracture Mechanics},
  publisher = {Elsevier},
  address = {Oxford},
  year = {2007}
}

@article{Drass2020pec,
  author = {Drass, Michael and Du Bois, Paul A. and Schneider, Jens and Kolling, Stefan},
  title = {Pseudo-elastic cavitation model: part {I}---finite element analyses on thin silicone adhesives in fa{\c{c}}ades},
  journal = {Glass Structures \& Engineering},
  volume = {5},
  pages = {41--65},
  year = {2020},
  doi = {10.1007/s40940-019-00115-4}
}

@article{Drass2020coupled,
  author = {Drass, Michael and Rosendahl, Philipp L. and Kraus, Michael A.},
  title = {Coupled distortional-dilatational failure mode concept for silicone adhesives},
  journal = {Mechanics of Materials},
  pages = {103511},
  year = {2020},
  doi = {10.1016/j.mechmat.2020.103511}
}

@article{Rosendahl2019,
  author = {Rosendahl, Philipp L. and Drass, Michael and Felger, Julian and Schneider, Jens and Becker, Wilfried},
  title = {Equivalent strain failure criterion for multiaxially loaded incompressible hyperelastic elastomers},
  journal = {International Journal of Solids and Structures},
  volume = {166},
  pages = {32--46},
  year = {2019},
  doi = {10.1016/j.ijsolstr.2019.01.030}
}

@article{Jacobs1991,
  author = {Jacobs, Robert A. and Jordan, Michael I. and Nowlan, Steven J. and Hinton, Geoffrey E.},
  title = {Adaptive mixtures of local experts},
  journal = {Neural Computation},
  volume = {3},
  number = {1},
  pages = {79--87},
  year = {1991}
}

@inproceedings{Shazeer2017,
  author = {Shazeer, Noam and Mirhoseini, Azalia and Maziarz, Krzysztof and Davis, Andy and Le, Quoc and Hinton, Geoffrey and Dean, Jeff},
  title = {Outrageously large neural networks: the sparsely-gated mixture-of-experts layer},
  booktitle = {International Conference on Learning Representations (ICLR)},
  year = {2017}
}

@article{Bischof2022moe,
  author = {Bischof, Rafael and Kraus, Michael A.},
  title = {Mixture-of-experts-ensemble meta-learning for physics-informed neural networks},
  journal = {Proceedings of 33. Forum Bauinformatik},
  year = {2022}
}

@inproceedings{Louizos2018,
  author = {Louizos, Christos and Welling, Max and Kingma, Diederik P.},
  title = {Learning sparse neural networks through {$L_0$} regularization},
  booktitle = {International Conference on Learning Representations (ICLR)},
  year = {2018}
}

@inproceedings{Jang2017,
  author = {Jang, Eric and Gu, Shixiang and Poole, Ben},
  title = {Categorical reparameterization with {G}umbel-softmax},
  booktitle = {International Conference on Learning Representations (ICLR)},
  year = {2017}
}

@inproceedings{Maddison2017,
  author = {Maddison, Chris J. and Mnih, Andriy and Teh, Yee Whye},
  title = {The concrete distribution: a continuous relaxation of discrete random variables},
  booktitle = {International Conference on Learning Representations (ICLR)},
  year = {2017}
}

@article{Blei2003,
  author = {Blei, David M. and Ng, Andrew Y. and Jordan, Michael I.},
  title = {Latent {D}irichlet allocation},
  journal = {Journal of Machine Learning Research},
  volume = {3},
  pages = {993--1022},
  year = {2003}
}

@article{Weibull1951,
  author = {Weibull, Waloddi},
  title = {A statistical distribution function of wide applicability},
  journal = {Journal of Applied Mechanics},
  volume = {18},
  pages = {293--297},
  year = {1951}
}

@article{Koiter1953,
  author = {Koiter, Warner T.},
  title = {Stress-strain relations, uniqueness and variational theorems for elastic-plastic materials with a singular yield surface},
  journal = {Quarterly of Applied Mathematics},
  volume = {11},
  number = {3},
  pages = {350--354},
  year = {1953}
}

@article{Treloar1944,
  author = {Treloar, Leslie R. G.},
  title = {Stress-strain data for vulcanised rubber under various types of deformation},
  journal = {Transactions of the Faraday Society},
  volume = {40},
  pages = {59--70},
  year = {1944}
}

@article{Ogden1972,
  author = {Ogden, Raymond W.},
  title = {Large deformation isotropic elasticity -- on the correlation of theory and experiment for incompressible rubberlike solids},
  journal = {Proceedings of the Royal Society of London A},
  volume = {326},
  pages = {565--584},
  year = {1972}
}

@article{Hamdi2006,
  author = {Hamdi, Adel and Na{\"\i}t Abdelaziz, Mo{\"\i}ssa and A{\"\i}t Hocine, Na{\"\i}m and Heuillet, Patrick and Benseddiq, Noureddine},
  title = {A fracture criterion of rubber-like materials under plane stress conditions},
  journal = {Polymer Testing},
  volume = {25},
  number = {8},
  pages = {994--1005},
  year = {2006}
}

@misc{HyperDataJl,
  author = {{Aminofa70}},
  title = {{HyperData.jl}: experimental data for hyperelastic materials},
  howpublished = {\url{https://github.com/Aminofa70/HyperData.jl}},
  note = {accessed 7 November 2026},
  year = {2024}
}

@article{LopezPamies2011,
  author = {Lopez-Pamies, Oscar and Idiart, Mart{\'\i}n I. and Nakamura, Toshio},
  title = {Cavitation in elastomeric solids: {I}---a defect-growth theory},
  journal = {Journal of the Mechanics and Physics of Solids},
  volume = {59},
  number = {8},
  pages = {1464--1487},
  year = {2011}
}

@article{Gent1959,
  author = {Gent, Alan N. and Lindley, P. B.},
  title = {Internal rupture of bonded rubber cylinders in tension},
  journal = {Proceedings of the Royal Society of London A},
  volume = {249},
  pages = {195--205},
  year = {1959}
}
\end{document}